\documentclass{raa_twocolumn}
\usepackage{natbib}
\usepackage{amssymb,amsmath}
\usepackage{upgreek}
\bibpunct{(}{)}{;}{a}{}{,}
\usepackage{graphicx,times}
\usepackage[morefloats = 38]{morefloats}
\usepackage[figuresright]{rotating}
\usepackage[flushleft]{threeparttable}

\begin{document}

\title{FAST pulsar database: I. Polarization profiles of 682 pulsars}

 \volnopage{ {\bf 2023} Vol.\ {\bf 23} No. {\bf 03}, A02}
   \setcounter{page}{1}

   \author{ P. F. Wang\inst{1,2}, J. L. Han\inst{1,2}, J. Xu\inst{1,2}, C. Wang\inst{1,2}, Y. Yan\inst{1,2}, W. C. Jing\inst{1,2},
     W. Q. Su\inst{1,2}, D. J. Zhou\inst{1,2}, \and T. Wang\inst{1,2}
   }

   \institute{
     National Astronomical Observatories, Chinese Academy of Sciences, 20A Datun Road,
     Chaoyang District, Beijing 100012, China;  {\it pfwang@nao.cas.cn; hjl@nao.cas.cn}\\
     \and
     School of Astronomy, University of Chinese Academy of Sciences,
     Beijing 100049, China; \\
    \vs \no
   {\small Received 2022 XXX; accepted 2022 XXX}
}

   \abstract{Pulsar polarization profiles are very basic database for
     understanding the emission processes in pulsar
     magnetosphere. After careful polarization calibration of the
     19-beam L-band receiver and verification of beam-offset
     observation results, we obtain polarization profiles of 682
     pulsars from observations by the Five-hundred-meter Aperture
     Spherical radio Telescope (FAST) during the survey tests for the
     Galactic Plan Pulsar Snapshot (GPPS) survey and other normal FAST
     projects. Among them, polarization profiles of about 460 pulsars
     are observed for the first time. The profiles exhibit diverse
     features. Some pulsars have a polarization position angle curve
     with a good S-shaped swing, and some with orthogonal modes; some
     have components with highly linearly components or strong
     circularly polarized components; some have a very wide profile,
     coming from an aligned rotator, and some have an interpulse from
     a perpendicular rotator; some wide profiles are caused by
     interstellar scattering. We derive geometry parameters for 190
     pulsars from the S-shaped position angle curves or with
     orthogonal modes. We find that the linear and circular
     polarization or the widths of pulse profiles have various
     frequency dependencies. Pulsars with large fraction of linear
     polarization are more likely to have a large  Edot
     .
   }

   \authorrunning{P. F. Wang et al. }            
   \titlerunning{FAST pulsar database: I. Polarization profiles}  
   \maketitle

%
\section{Introduction}           
\label{sect:intro}

Pulsars are highly polarized radio sources in the universe. Its highly
polarized emission was first noticed about 50 years ago soon after the
pulsar discovery \citep{ls68}. Since then, great efforts have been
made to uncover polarization features at a wide range of frequencies
by many radio telescopes. The big projects produce large
databases. For example at 1.4~GHz, 300 pulsars have been observed with
the Lovel telescope \citep{gl98}, 98 pulsars by Arecibo
\citep{wcl+99}, initially 45 \citep{wmlq93} and then 66 \citep{mhq98}
and recently 600 pulsars \citep{jk18} by the Parkes telescope. At
higher frequencies, 32 pulsars have been observed at 4.9~GHz by the
Effelsberg radio telescope \citep{vkk98}, and 48 pulsars at 3.1~GHz by
the Parkes telescope \citep{kjm05}. At lower frequencies, 100 pulsars
were observed at 774~MHz by the Green Bank telescope \citep{hdvl09},
57 pulsars at 430~MHz and the lower frequencies by Arecibo
\citep{hr10}, 123 pulsars at 333~MHz and/or 618~MHz by GMRT
  \citep{mbm+16}, and 20 pulsars below 200~MHz by LOFAR
\citep{nsk+15}. Recently, \citet{sjd+21} extended the polarimetric
observations to the ultra wide frequency band, from 704 to 4032~MHz,
for 40 pulsars by using Parkes. \citet{pkj+23}measured the
  polarimetry of 1170 pulsars using MeerKAT in a frequency band from
  856 to 1712~MHz.

\begin{table*}
  \centering
  \caption{FAST observations for pulsar polarization profiles}
  \label{table:obs}
  \begin{tabular}{cccccr}
    \hline
    \hline
   Projects   &  PI    &  Obs. Mode    & Beams    & Obs. Dates     & Obs. Time  \\
\hline
GPPS survey   & J.~L. Han    &  snapshot    & M01--M19 & 201904--      &  5 min     \\
              &        &  tracking    & M01--M19 &               & 15 min     \\
              &        &  swiftcalibration & M01--M19 &          & 15 min     \\
PT2020$\_$0087& P.~F. Wang   &  tracking    &   M01    & 202104-202105 & 30 min     \\
PT2020$\_$0161& P.~F. Wang   &  tracking    &   M01    & 202009-202009 & 180 min    \\
PT2020$\_$0164& J. Xu     &  tracking    &   M01    & 202010-202104 & 10 min     \\
PT2021$\_$0051& J. Xu     &  snapshotdec & M01--M19 & 202109-202111 &  5 min     \\
PT2022$\_$0169& J. Xu     &  tracking    &   M01    & 202208-202211 & 15 min     \\

\hline
    \end{tabular}
\end{table*}

Pulsar polarization profiles exhibit rich features
\citep[e.g.][]{lm88,rsw89,rr03}. Pulsar polarization is predominantly
linear. Some pulsars are highly linearly polarized for the whole or
part of the profiles \citep[e.g.][]{gl98,kjm05,wh16}, like PSRs
B1259-63, B0355+54 and B1931+24. The degree of linear polarization
generally decreases with the increase of observing frequency
\citep[e.g.][]{man71, xkj+96, wwh15}. The circular polarization is
usually weak, but can be extremely strong for some pulsars like PSR
B1702-19 or for some component such as PSR J1920+2650
\citep{hdvl09}. The circular polarization has a single sign, or a sign
reversal for core or conal components \citep[e.g.][]{ran83,lm88,
  hmxq98}. The polarization position angle (PA) curves ideally have an
S-shape, such as PSRs B2045-16, J1420-6048, and B0833-45, but not for
most pulsars. Some pulsars exhibits the polarization angle jumps of 90
degrees \citep[e.g.][]{scr+84,ms00}, which is called as the orthogonal
modes. The polarization features are poorly understood,
although polarization profiles of about 1500 pulsars were
  reported in the literatures.


To understand these diverse polarization features, many theoretical
models have been developed, based on pulsar emission geometry,
emission mechanisms and propagation effects. The geometry of pulsar
emission region can be described by the rotating magnetosphere with a
dipole magnetic field, called the rotating vector model
\citep[RVM,][]{rc69}.
%
Pulsar radio emission is believed to be coherent radiation
generated by relativistic particles streaming along the curved dipole
magnetic field lines in the magnetosphere. Curvature radiation is one
of the most probable mechanisms, and the radiation of which is highly
linearly polarized \citep[e.g.][]{gan10, wwh12}. Circular polarization
produced by relativistic particles can be of one sign or have a sign
reversal, depending on the density gradients of the relativistic
particles and the geometry of line of sight to the emission magnetic
field lines \citep{wwh12}. The inverse Compton scattering of low
frequency waves by relativistical particles in pulsar magnetosphere
can also produce similar polarization features \citep{xlhq00}.
%
In addition to the emission processes, propagation effects can also
significantly affect the polarization states. The plasma in the pulsar
magnetosphere has two orthogonally polarized modes, the ordinary mode
(O-mode) and the extra-ordinary mode (X-mode) \citep[e.g.][]{ms77,
  ab86, wl07}. The polarized radiation produced by any emission
processes has to be coupled to these transverse modes to propagate out
in the magnetosphere. During propagation, it will experience
`adiabatic walking', `wave-mode coupling' and `cyclotron absorption'
\citep[e.g.][]{cr79, wlh10, bp12}, and the additional refraction for
the O-mode wave \citep{ba86, lp98}.
Moreover, the pulsar rotation
not only causes the phase lag between the PA curve and the pulse
profile \citep[e.g.][]{bcw91}, but also leads to the
separation of the O and X modes \citep[e.g.][]{wwh14}.
Therefore, polarization profiles at different frequencies can be
understood only if the dipole geometry, curvature emission process and
propagation effects within pulsar magnetosphere are jointly considered
\citep{wwh14}. Up to now, the origins of orthogonal modes, highly
linearly polarized emission and its frequency dependency can be
naturally explained \citep{wwh15,wh16}. To further test these
theories, much more sensitive observations are needed.

The Five-hundred-meter Aperture Spherical radio Telescope (FAST) gives
us a great opportunity to get more details of polarization profiles and
more profiles for fainter pulsars, because of its unprecedented sensitivity.
In this paper, we present the polarization profiles for 682 pulsars
observed by the FAST via several projects, including the test
observations during the Galactic Plan Pulsar Snapshot survey
(GPPS). The paper is organized as follows. In Section 2, we briefly
describe the observations, data reduction procedures and the
capability for polarization measurements of the 19 beams of the L-band
19-beam receiver. Observation results with classifications of
polarization features are presented in Section 3. Further
understanding of the polarization profiles are discussed in Section
4. Conclusions are given in Sections 5.

\begin{figure}
  \centering
  \includegraphics[width = 0.45\textwidth] {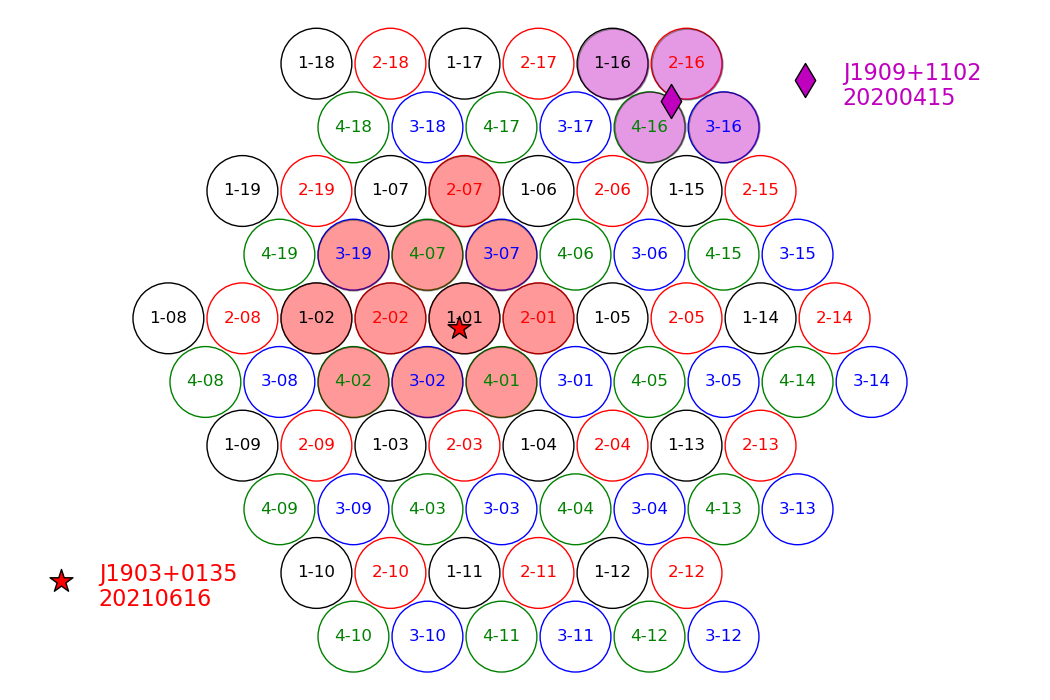}
  \caption{FAST has a 19-beam L-band receiver, which has been used for
    a snapshot mode in the GPPS survey \citep{hww+21}. With 4 nearby
    pointing, each with one color for beams, a sky coverage can be
    fully covered. The beam number and pointing number are indicated
    inside each beam (e.g. 2-07 is the beam M07 in the 2nd
    pointing). A bright pulsar can be detected by a few nearby beams,
    as illustrated for PSRs J1903+0135 and J1909+1102, which were
    observed at two different sessions with dates indicated. The
    polarization results of the targeted pulsar from these beams are
    surprising consistent (see results below in 
    Figure~\ref{fig:Pol_comp-snap} and listed in
    Table~\ref{table:Pol_comp}), which implies the excellent
    polarization characteristics of the 19-beam L-band receiver.}
  \label{fig:Pol_comp-snap_beam}
\end{figure}

\section{Observations and Data reduction}

\subsection{Observations}

The data for pulsar polarization profiles presented in this paper are
obtained by the FAST observations from 2019 to 2022 via six projects
(see Table~\ref{table:obs}).

The FAST GPPS survey is one of the five FAST key projects with project
numbers as ZD2020\_2, ZD2021\_2, and ZD2022\_2, which is carried out
for discovering new pulsars within the Galactic latitude of $\pm
10^\circ$ by employing all the 19 beams with the newly invented
snapshot observation mode \citep{hww+21}. The observations of nearby 4
pointing can cover a hexagonal sky coverage of about $28'$ in size
(see Figure~\ref{fig:Pol_comp-snap_beam}) which is called a cover. Any
cover with one or more known pulsars in any beam, not necessary in the
beam center (see also Figure~\ref{fig:Pol_comp-snap_beam} and
Table~\ref{table:Pol_comp}), was observed as the survey tests with
polarization data recorded, which provide a good database for this
work but a careful polarization calibration is really desired.

In the last three years, we carried out targeted observations for a
dedicated study of polarization features of known pulsars (FAST
projects: PT2020$\_$0087 and PT2020$\_$0161), for the general studies
of high Galactic latitude pulsars without polarimetry before (FAST
projects: PT2020$\_$0164, PT2021$\_$0051 and PT2022$\_$0169).

All these observations have been made with the 19-beam L-band receiver
of the FAST. These 19-beam receivers cover 1.0 -- 1.5 GHz band
\citep{jth+20}. Radio signals from two orthogonal linear polarization
channels, X and Y, of each beam are sampled. They are channelized for
4096 or 2048 channels, and then correlated for the XX, YY,
Re[X$^\ast$Y] and Im[X$^\ast$Y] polarization products in a digital
backend \citep[see][for details]{hww+21}. The data are accumulated to
a time resolution of 49.152 $\mu s$, and then written to fits
files. Observations often take 5, 10, 15, 30 minutes in general,
depending on objects and projects.

\begin{figure*}
  \centering
  \includegraphics[angle=0,width = 0.24\textwidth] {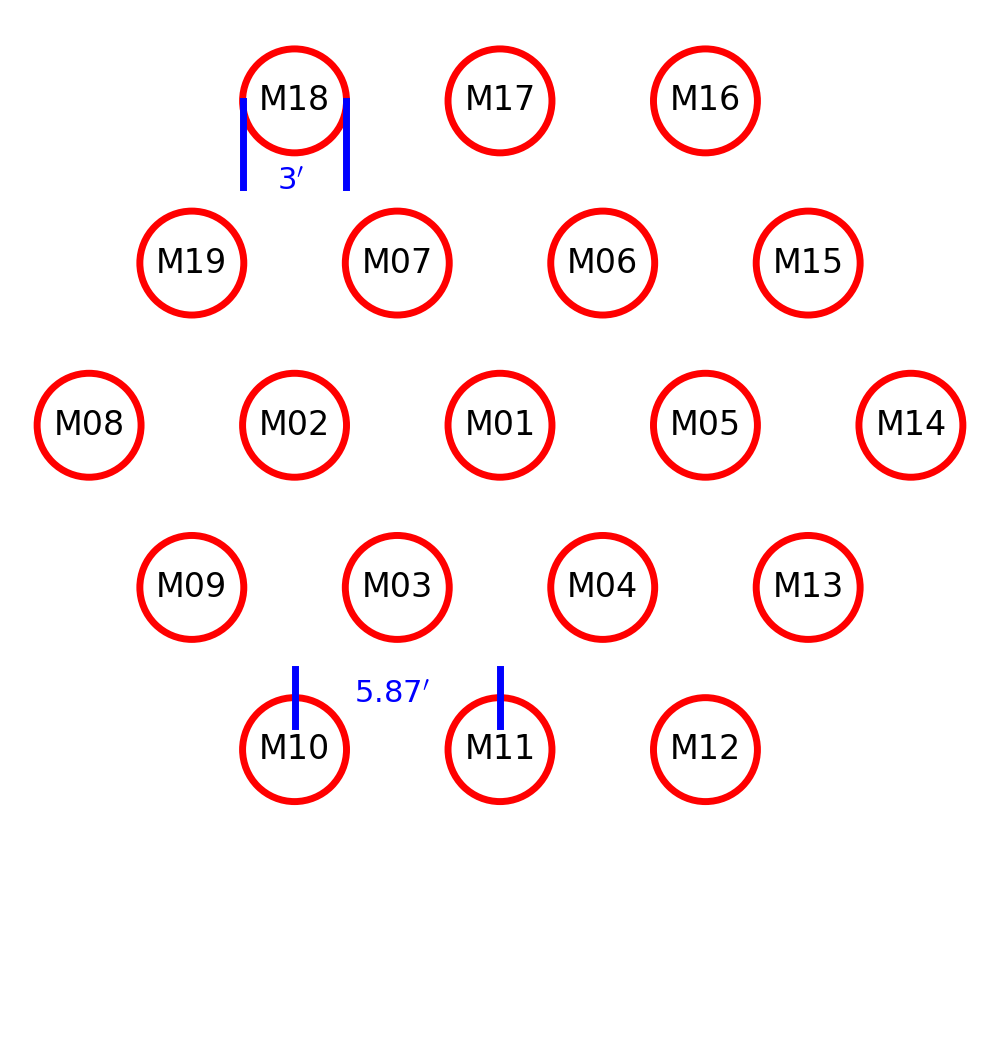}
  \includegraphics[angle=0,width = 0.24\textwidth] {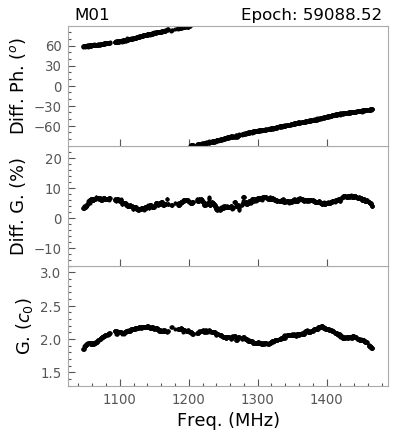}
  \includegraphics[angle=0,width = 0.24\textwidth] {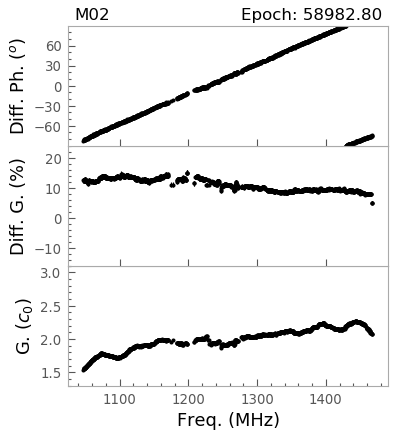}
  \includegraphics[angle=0,width = 0.24\textwidth] {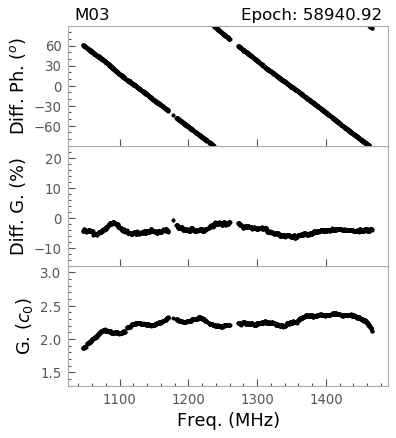}\\
  \includegraphics[angle=0,width = 0.24\textwidth] {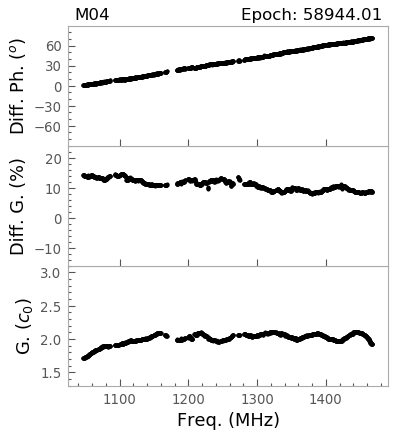}
  \includegraphics[angle=0,width = 0.24\textwidth] {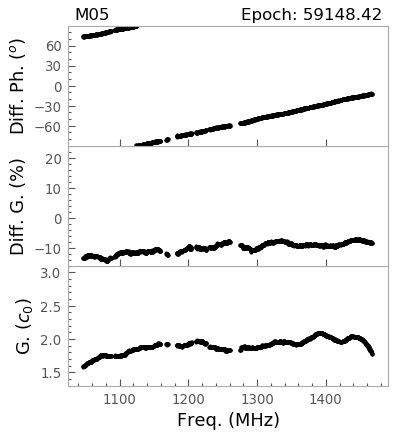}
  \includegraphics[angle=0,width = 0.24\textwidth] {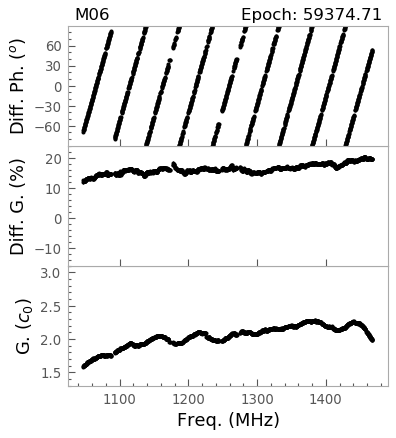}
  \includegraphics[angle=0,width = 0.24\textwidth] {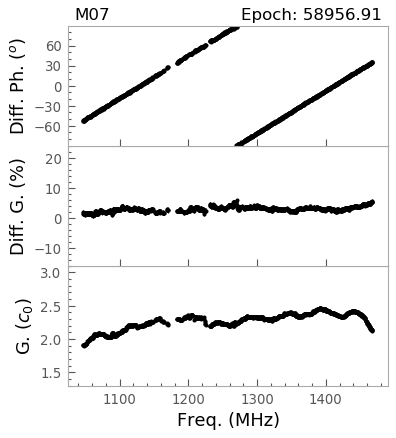}\\
  \includegraphics[angle=0,width = 0.24\textwidth] {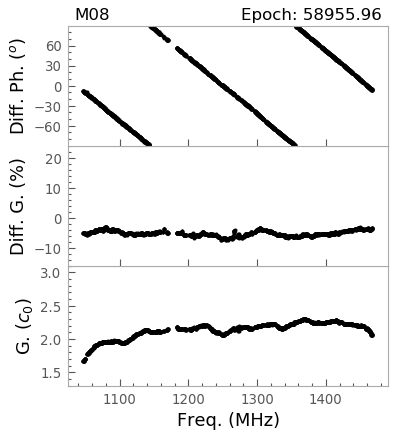}
  \includegraphics[angle=0,width = 0.24\textwidth] {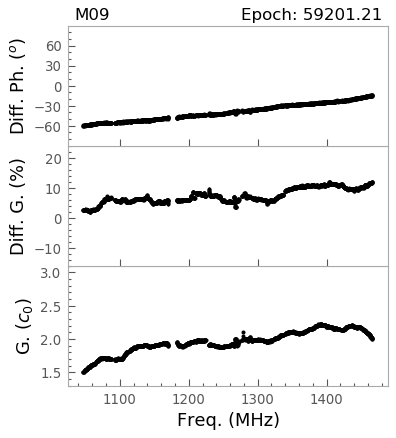}
  \includegraphics[angle=0,width = 0.24\textwidth] {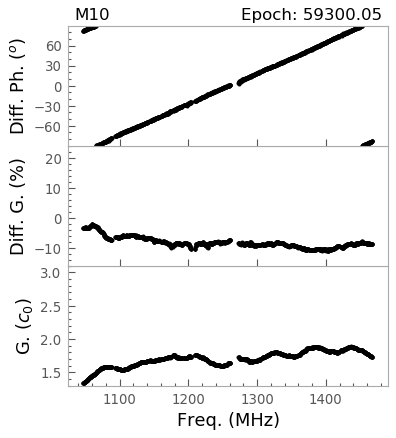}
  \includegraphics[angle=0,width = 0.24\textwidth] {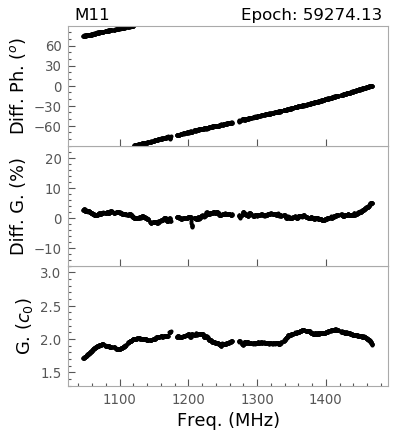}\\
  \includegraphics[angle=0,width = 0.24\textwidth] {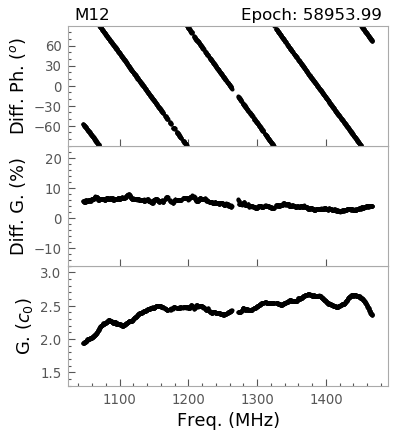}
  \includegraphics[angle=0,width = 0.24\textwidth] {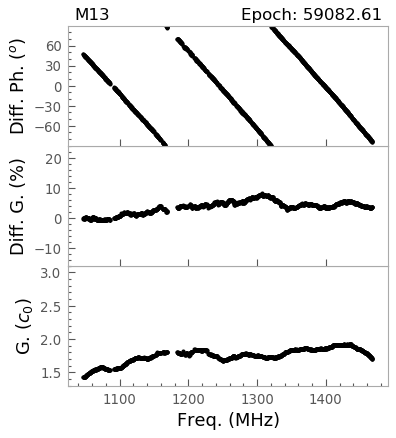}
  \includegraphics[angle=0,width = 0.24\textwidth] {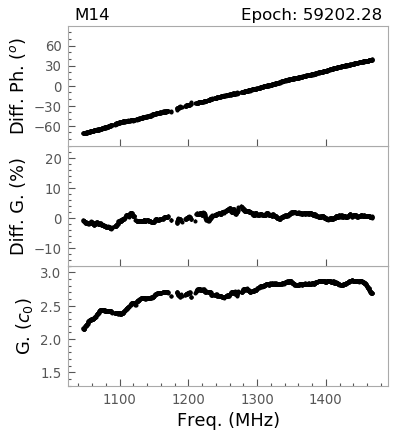}
  \includegraphics[angle=0,width = 0.24\textwidth] {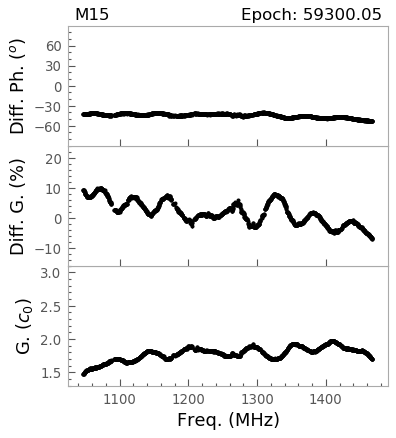}\\
  \includegraphics[angle=0,width = 0.24\textwidth] {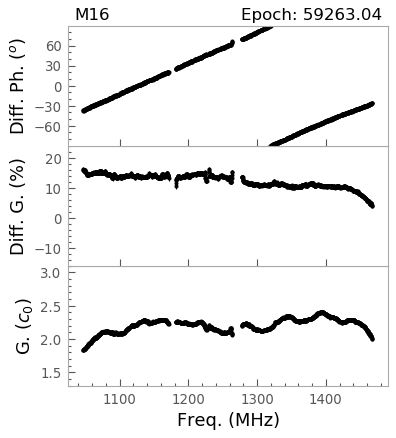}
  \includegraphics[angle=0,width = 0.24\textwidth] {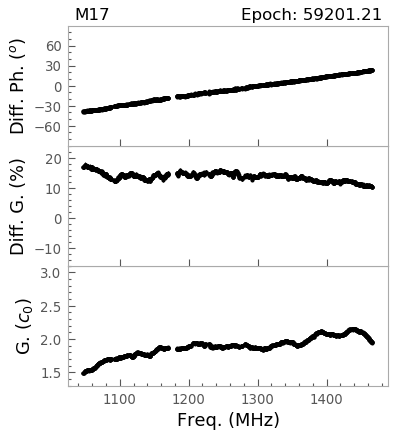}
  \includegraphics[angle=0,width = 0.24\textwidth] {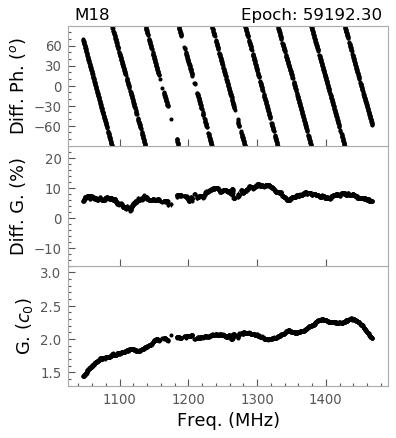}
  \includegraphics[angle=0,width = 0.24\textwidth] {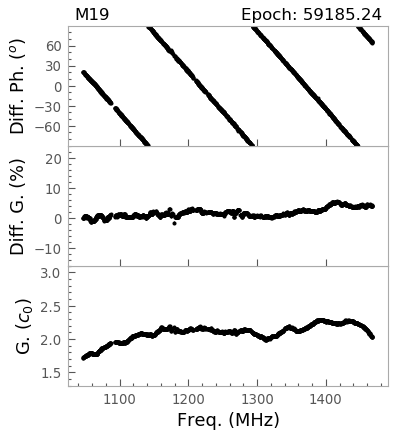}\\
  \caption{Instrumental characteristics as a function of observing
    frequency calculated for 19 beams of the FAST 19-beam L-band
    receiver. The sky distribution of the 19 beams is shown in the top
    left panel for beams M01 to M19. For each beam, the gain,
    differential gain and differential phase are plotted in three
    sub-panels. }
  \label{fig:pol_para}
\end{figure*}

\begin{table*}
  \centering
  \caption{No significant differences on polarization measurements
    observed at or near the center of beam M01 and at a very offset
    position in other beams.}
  \label{table:Pol_comp}
  \begin{tabular}{lrrrrrrrrrrrr}
\hline
\hline
PSR        & Obs. Date & Beam & Offset  & Int.   & S/N & RM$_{\rm ISM}$  & \multicolumn{2}{r}{$\rm \Delta PA$($^\circ$)} & \multicolumn{2}{r}{$\rm \Delta L/I$(\%)} & \multicolumn{2}{r}{$\rm \Delta V/I(\%)$} \\
           &           &      &($^\prime$)& (min)  &  &($\rm rad/m^2$)& ave.  & std. & ave. & std. & ave. & std.   \\
\hline
J2113+4644 & 20220308 & P1M01 &  0.0 & 120 & 52096.6 & -222.4(3)&  0.0 &     &  0.0 &     & 0.0 &     \\
           & 20200826 & P1M01 &  3.0 &  15 &  7802.4 & -221.6(3)& -2.9 & 2.1 &  1.8 & 2.8 & 0.3 & 1.1 \\
           & 20200917 & P1M12 &  3.0 &  15 &  7835.7 & -221.2(2)& -1.0 & 1.4 &  1.9 & 3.0 &-1.3 & 1.1 \\
J0631+1036 & 20210501 & P1M01 &  0.0 &  30 &  2638.8 &  143.6(2)&  0.0 &     &  0.0 &     & 0.0 &     \\
           & 20200816 & P1M02 &  2.2 &  15 &   440.9 &  145.2(3)& -6.4 & 1.7 & -1.5 &10.2 &-0.2 & 2.5 \\
J1915+1606 & 20191028 & P1M01 &  0.0 &  30 &  1472.1 &  357.2(2)&  0.0 &     &  0.0 &     & 0.0 &     \\
           & 20210422 &   M03 &  0.6 &   5 &   585.3 &  356.5(3)&  4.3 & 2.0 & -0.4 & 3.9 &-0.3 & 3.1 \\
J0711+0931 & 20210111 & P1M01 &  0.0 &  10 &  1135.5 &   60.9(7)&  0.0 &     &  0.0 &     & 0.0 &     \\
           & 20210526 &   M04 &  0.5 &   5 &   578.7 &   63.0(5)&  2.7 & 5.4 &  0.7 & 4.8 &-2.5 & 3.1 \\
J1844+1454 & 20210430 & P1M01 &  0.0 &  30 &  5171.3 &  119.1(3)&  0.0 &     &  0.0 &     & 0.0 &     \\
           & 20210606 &   M10 &  1.0 &   5 &  2824.5 &  118.1(3)&  0.7 & 1.8 & -0.6 & 3.1 & 0.1 & 0.9 \\
J1935+1616 & 20190919 & P1M01 &  0.0 &  50 &430977.0 &   -2.8(2)&  0.0 &     &  0.0 &     & 0.0 &     \\
           & 20210119 &   M11 &  2.5 &   5 & 13216.0 &   -2.6(3)&  3.5 & 2.6 &  0.7 & 4.4 & 0.0 & 1.3 \\
J1828+0625 & 20210118 & P1M01 &  0.0 &  10 &   223.0 &    24(2) &  0.0 &     &  0.0 &     & 0.0 &     \\
           & 20210615 &   M17 &  1.6 &   5 &   166.8 &   16(11) &  2.6 & 6.6 &  0.6 & 6.5 & -   &  -  \\
J1842+1332 & 20210115 & P1M01 &  0.0 &  10 &  1612.2 &  130.3(2)&  0.0 &     &  0.0 &     & 0.0 &    \\
           & 20210605 &   M19 &  1.2 &   5 &  1050.8 &  129.6(2)& -2.3 & 3.0 &  2.1 & 4.2 & 0.8 & 0.3 \\
\hline
J1903+0135 & 20210616 & P1M01 & 0.4 &  5 & 13125.7 &  73.0(4) &  0.0 &     &  0.0 &     & 0.0  &     \\
           &          & P3M02 & 2.5 &  5 &  2226.4 &  72.9(5) &  1.1 & 5.4 &  1.8 & 2.3 &  0.4 & 1.4 \\
           &          & P2M02 & 2.7 &  5 &  1901.0 &  72.2(8) &  3.5 & 4.3 &  0.6 & 2.5 &  0.2 & 1.0 \\
           &          & P4M01 & 2.8 &  5 &  1955.3 &  70.8(8) &  1.1 & 4.2 &  1.4 & 3.1 & -0.6 & 1.4 \\
           &          & P2M01 & 3.2 &  5 &  1138.0 &  75.3(11)& -0.5 &13.7 &  0.5 & 2.8 & -1.0 & 2.8 \\
           &          & P4M07 & 3.2 &  5 &   757.8 &  70.1(22)& -1.8 & 7.7 &  1.5 & 2.9 &  1.5 & 2.2 \\
           &          & P3M07 & 3.4 &  5 &   662.1 &  69.6(14)&  5.2 & 6.2 &  2.7 & 4.0 & -1.3 & 1.9 \\
           &          & P4M02 & 4.7 &  5 &    27.0 &  77(4)   &  9.9 & 6.5 & 11.8 & 3.8 &  -   & -   \\
           &          & P3M19 & 5.1 &  5 &    90.8 &  75.2(14)&  3.2 & 8.5 &  1.2 & 4.5 & -0.5 & 2.5 \\
           &          & P2M07 & 5.5 &  5 &    42.9 &  70(5)   & 11.4 &17.6 &  4.5 & 5.2 &  0.7 & 3.7 \\
           &          & P1M02 & 5.7 &  5 &    37.8 &  57(9)   & 25.1 & 3.4 &  1.1 & 1.5 & -2.1 & 2.3 \\
J1909+1102 & 20200414 & P4M16 & 1.4 &  5 &  2450.4 & 553.3(4) &  0.0 &     &  0.0 &     & 0.0  &     \\
           &          & P2M16 & 1.6 &  5 &  2733.1 & 554.0(4) & -0.5 & 2.8 & -1.0 & 5.7 & -1.2 & 2.0 \\
           &          & P3M16 & 2.3 &  5 &   767.7 & 552.4(7) &  0.5 & 3.2 & -0.2 & 5.0 & -0.4 & 1.7 \\
           &          & P1M16 & 2.8 &  5 &   644.2 & 553.7(7) &  0.7 & 4.6 &  0.7 & 8.2 &  2.4 & 3.3 \\

\hline
    \end{tabular}
\end{table*}

\begin{figure*}[t]
  \centering
  \includegraphics[angle=0,width = 0.29\textwidth] {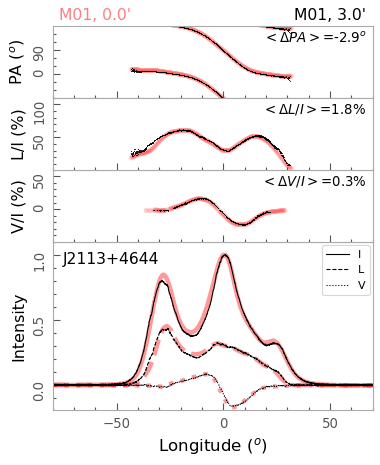}
  \includegraphics[angle=0,width = 0.29\textwidth] {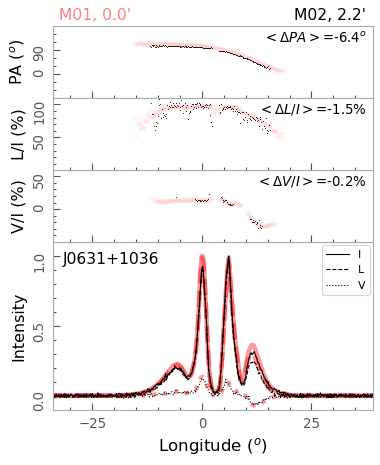}
  \includegraphics[angle=0,width = 0.29\textwidth] {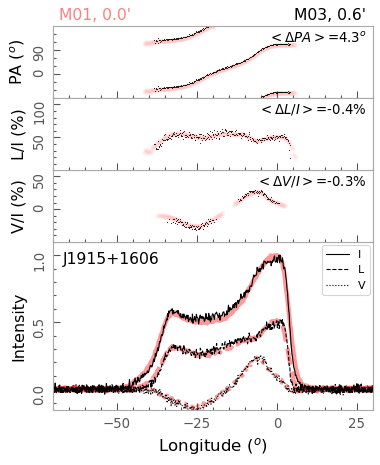}\\
  \includegraphics[angle=0,width = 0.29\textwidth] {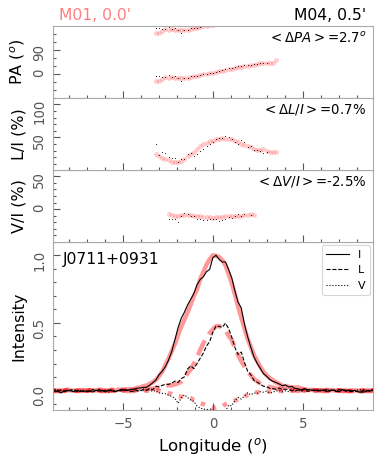}
  \includegraphics[angle=0,width = 0.29\textwidth] {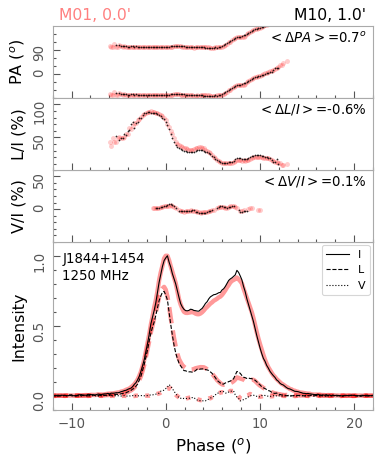}
  \includegraphics[angle=0,width = 0.29\textwidth] {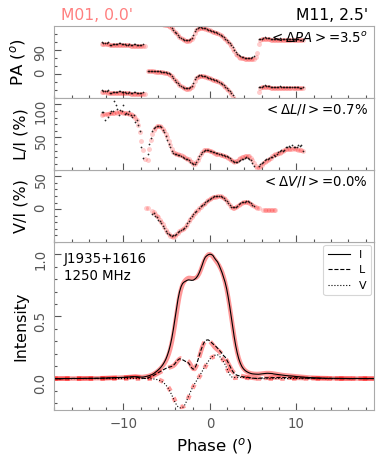}\\
  \includegraphics[angle=0,width = 0.29\textwidth] {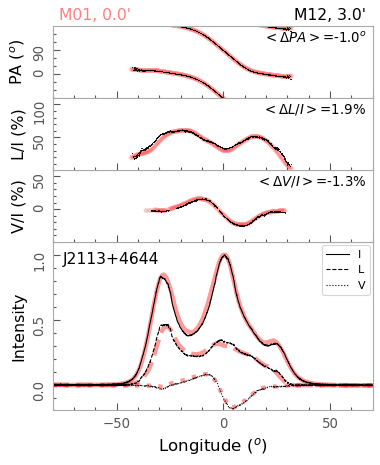}
  \includegraphics[angle=0,width = 0.29\textwidth] {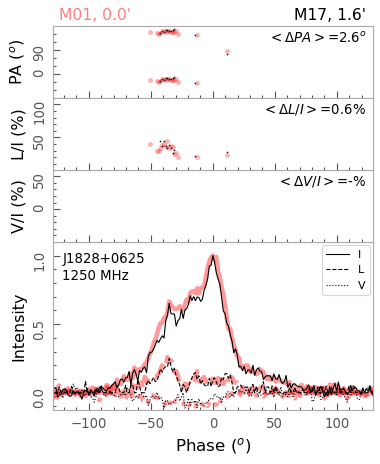}
  \includegraphics[angle=0,width = 0.29\textwidth] {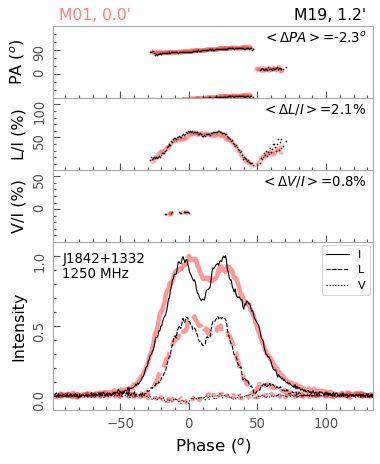}\\
  \caption{Comparisons for polarization profiles (thin lines) obtained
    from data recorded by the outer beams of the 19-beam L-band
    receiver when a pulsar is very offset from the beam center with
    the results (thick background lines) observed by the central beam
    M01 without offset. For each pulsar, in the bottom sub-panel are
    the total intensity, linear and circular polarization represented
    by solid, dashed and doted lines; and other sub-panels are the
    fractional circular and linear polarization and position angle
    curves of linear polarization with the linear polarized
    intensities exceeding 5$\sigma$, here $\sigma$ is obtained from
    the off pulse phase-bins. In general, results from the outer beams
    are very consistent with that from the central beam M01 in the
    thick background line, with a small derivations for $\langle
    \Delta PA \rangle$, $\langle \Delta L/I \rangle$ and $ \langle
    \Delta V/I \rangle$ as indicated in each sub-panel.  }
  \label{fig:Pol_comp}
\end{figure*}

\subsection{Data reduction}

Offline data processing has many steps.
The FAST data in fits files were recorded in the search mode
\citep[see e.g.][]{hww+21}, and have to be first folded for each
frequency channel and de-dispersed according to the ephemeris of a
given pulsar in the pulsar catalog \citep{mhth05} by using the
software tool DSPSR \citep{vb11}. The frequency channels at two edges
of the frequency band are  exercised with PSRCHIVE
\citep{hvm04} because of the low gain (see Figure~\ref{fig:pol_para}),
so that the frequency coverage is reduced from 1000-1500~MHz to
1048-1452~MHz. Radio frequency interference (RFI) is then exercised
both in frequency and time domains.

In general, the calibration signals are recorded for 2 minutes in
every observation session, which are didoes signals of a temperature
of 1.1~K injected to the feed at an angle of 45$^o$ with respect to
the two linear feed receptors. The signals are on for 1~s on and then
off for the other 1~s. The data recorded in the 4 polarization
channels, are folded to the period of 2~s to get a square wave, so
that the gain, differential gain and differential phase of the 19-beam
receiver can be obtained, as shown in Figure~\ref{fig:pol_para}. By
employing these system characteristic parameters obtained from the
nearest calibration observation session, we get the observed pulsar
signals calibrated.

The data for XX, YY, Re[X$^\ast$Y] and Im[X$^\ast$Y] are transformed
to the Stokes parameters, through $I=\rm XX+YY$, $Q=\rm XX-YY$, $U=2
\rm Re[X^\ast Y]$ and $V=2 \rm Im[X^\ast Y]$. They are
 integrated over frequency to form 8
frequency subbands with Faraday rotation corrected, and
 integrated in time so that the
polarization profiles can be analyzed for frequency dependence. The
final integrated pulse profiles are further
integrated from these 8 frequency subbands. The linear
polarization for phase bin $i$ is calculated as
$L_i=\sqrt{Q_i^2+U_i^2}$ and its associated PA as $\psi_i=1/2
\arctan(U_i/Q_i))$. The baseline bias of $L_i$ caused by the
  quadrature summation is corrected as, $L_{c,i}=
  \sqrt{|L_i^2-(\sigma_Q^2+\sigma_U^2)|}$ if $L_i^2\ge
  (\sigma_Q^2+\sigma_U^2)$, otherwise,
  $L_{c,i}=-\sqrt{|L_i^2-(\sigma_Q^2+\sigma_U^2)|}$. Here, $\sigma_Q$
  and $\sigma_U$ are the standard deviation of the off pulse Q and U
  data. The bias for the absolute circulation $|V_i|$ is also
  corrected, $|V_{c,i}|= \sqrt{|V_i^2-\sigma_V^2|}$ if $|V_i| \ge
  \sigma_V$, otherwise, $|V_{c,i}|=-\sqrt{|V_i^2-\sigma_V^2|}$, and
  $\sigma_V$ is estimated from the off-pulse V data. The degrees of
linear, circular and absolute circular polarization are
calculated as $L/I$, $V/I$, and $|V|/I$ with $I=\sum{I_i}$,
  $L=\sum{L_{c,i}}$, $V=\sum{V_i}$ and $|V|=\sum{|V_{c,i}|}$ for $N_d$
  phase bins with $I_i \ge 3\sigma_I$. Their uncertainties are
  estimated as, $\sigma_{L/I,m}=\sqrt{N_d
    [(\sigma_L/I)^2+(L\sigma_I/I^2)^2]}$, $\sigma_{V/I,m}=\sqrt{N_d
    [(\sigma_V/I)^2+(V\sigma_I/I^2)^2]}$ and
  $\sigma_{|V|/I,m}=\sqrt{N_d
    [(\sigma_{|V|}/I)^2+(|V|\sigma_I/I^2)^2]}$. In general,
  $\sigma_{L/I,m}\ge\sigma_{V/I,m}\ge\sigma_{|V|/I,m}$. Their total
  uncertainties are calculated as the quadrature sum of these
  measurement uncertainties and the estimated systematic uncertainty
  of 3 percent.

The polarization is presented in normal pulsar convention
\citep{vmjr10}. It reads, the PAs are with respect to the North
Celestial Pole, they are defined to increase counter-clockwise on the
sky, and the left-hand circular polarization is positive. This is set
for all FAST observations presented in this paper, though the data are
taken through the tracking, swiftcalibration and snapshotdec
modes. Note that FAST observations with the snapshot mode are made
with respect to the North Galactic pole, i.e., $\rm PA_{snap}$. A
correction of the PA should be made via $\rm PA= PA_{snap}- PA_c$, and
\begin{equation}
  \rm PA_c = \tan^{-1} \left[
    \frac{ \sin(\Delta RA) }
         { \cos \delta \cot 62.9^{\circ} - \sin \delta  \cos(\Delta RA) }
    \right].
  \label{eq:PAcorrect}
\end{equation}
Here, $\Delta RA = \rm RA_{psr} - \rm RA_{NGP}$, and $\rm RA_{psr}$
and $\rm RA_{NGP}$ are the right ascensions of a pulsar and the north
Galactic pole, $\delta$ is the declination of the pulsar, and the
inclination angle of the North Galactic pole to the North Celestial
pole is 62.9$^\circ$. After these corrections, the position
  angles are finally referred to the same frame.

The plane of linear polarization of pulsar radio emission has
  been rotated during their propagation through the magnetized
  interstellar medium.  Faraday rotation of
linearly polarized signals across the observational frequency band has
to be corrected to the central frequency 1250~MHz by using the
observed rotation measure (RM) of an observation, $\rm RM_{\rm
  obs}$. The initial value of RM is first found by searching for the
maximum of linearly polarized pulse emission as a function of many
trial RMs. The more accurate value is then obtained by iteratively
refining the differential PAs of the two halves of the total bandwidth
\citep[see][]{hmvd18}. These calculations are achieved by employing
the program RMFIT in the package of PSRCHIVE. The so-obtained $\rm
RM_{\rm obs}$ includes the RM contribution from the interstellar
medium, $\rm RM_{\rm ISM}$, and the variable RM contribution from the
earth ionosphere, $\rm RM_{\rm ion}$. The $\rm RM_{\rm ion}$ is
modeled by using the vertical total electron content maps of the
ionosphere, CODE\footnote{ftp://ftp.aiub.unibe.ch/CODE/}, and the
International Geomagnetic Reference Field
(IGRF-13)\footnote{https://www.ngdc.noaa.gov/IAGA/vmod/igrf.html} with
an updated code of
IONFR\footnote{https://sourceforge.net/projects/ionfarrot/}
\citep{ssh+13}. The variable RM contribution should be discounted, and
then the published RM value is purely from the interstellar medium and
is $\rm RM_{obs}-RM_{ion}$.

\begin{figure*}
  \centering

  \includegraphics[angle=0,width = 0.23\textwidth] {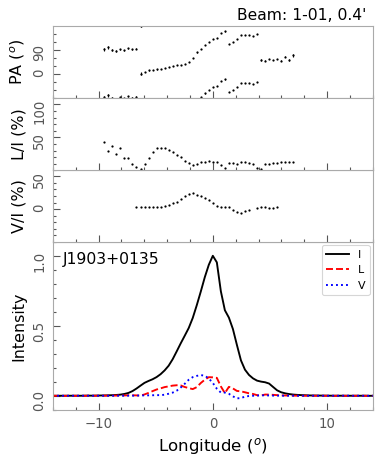}
  \includegraphics[angle=0,width = 0.23\textwidth] {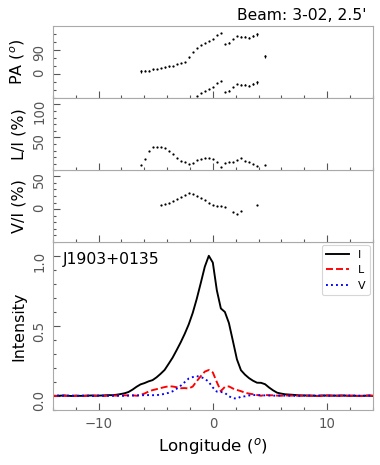}
  \includegraphics[angle=0,width = 0.23\textwidth] {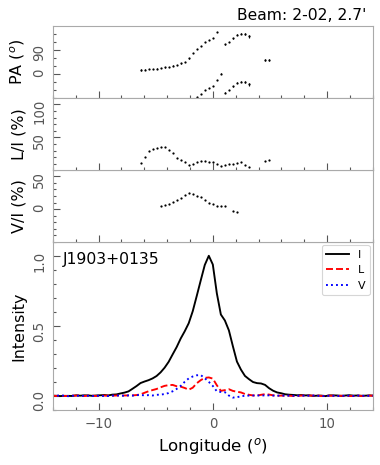}
  \includegraphics[angle=0,width = 0.23\textwidth] {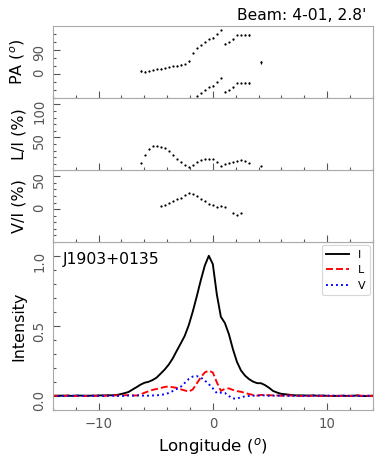}\\
  \includegraphics[angle=0,width = 0.23\textwidth] {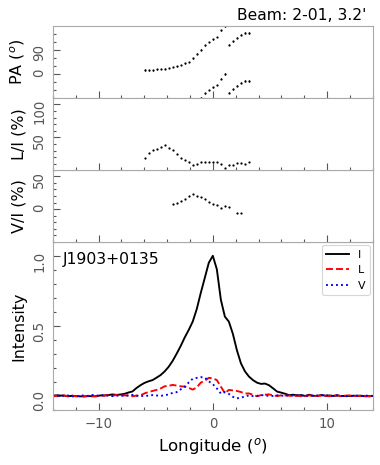}
  \includegraphics[angle=0,width = 0.23\textwidth] {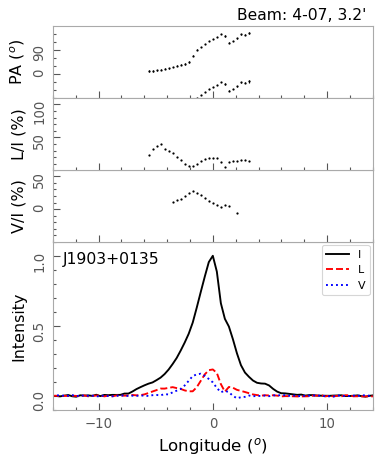}
  \includegraphics[angle=0,width = 0.23\textwidth] {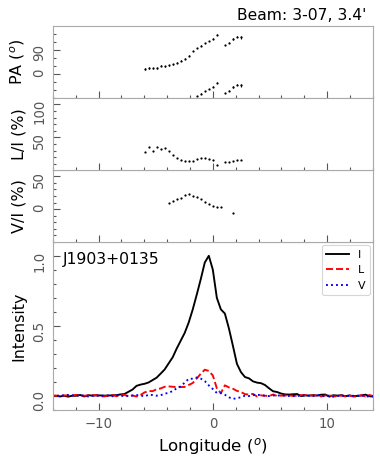}
  \includegraphics[angle=0,width = 0.23\textwidth] {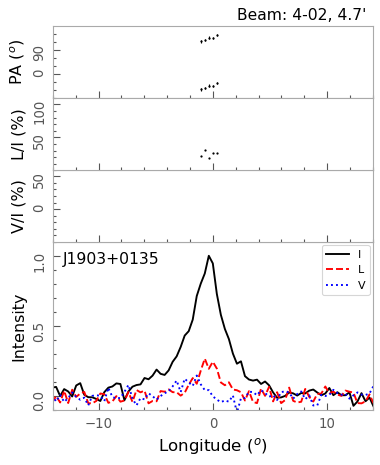}\\
  \includegraphics[angle=0,width = 0.23\textwidth] {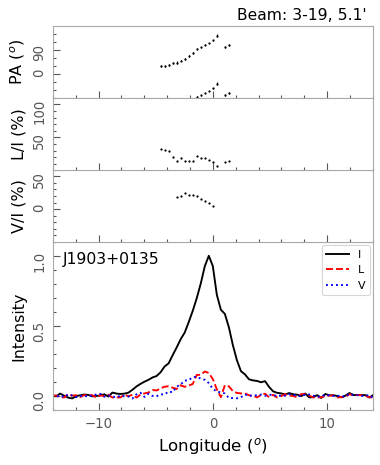}
  \includegraphics[angle=0,width = 0.23\textwidth] {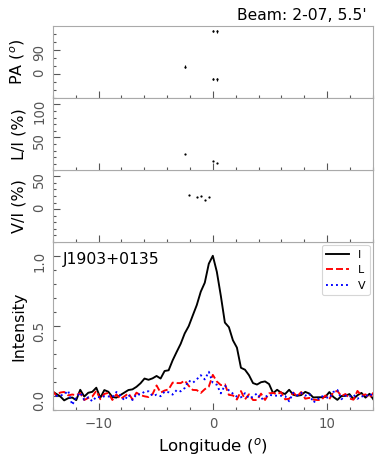}
  \includegraphics[angle=0,width = 0.23\textwidth] {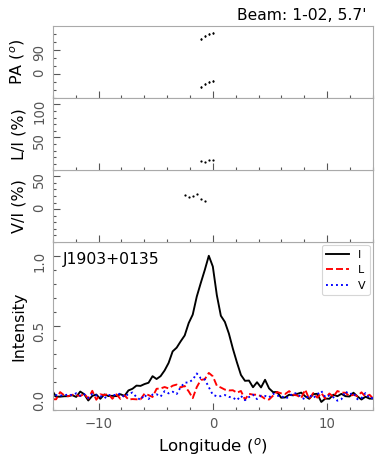}\\
  \includegraphics[angle=0,width = 0.23\textwidth] {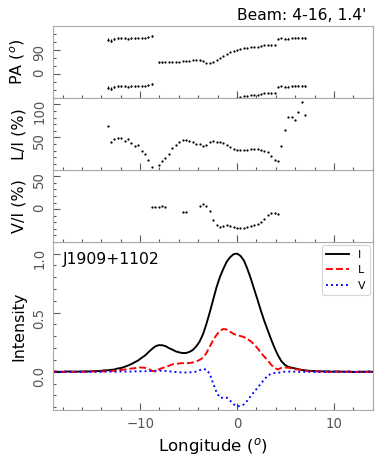}
  \includegraphics[angle=0,width = 0.23\textwidth] {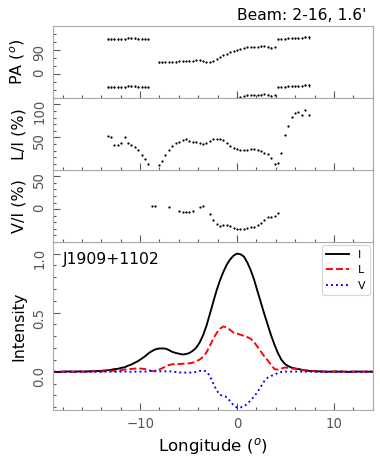}
  \includegraphics[angle=0,width = 0.23\textwidth] {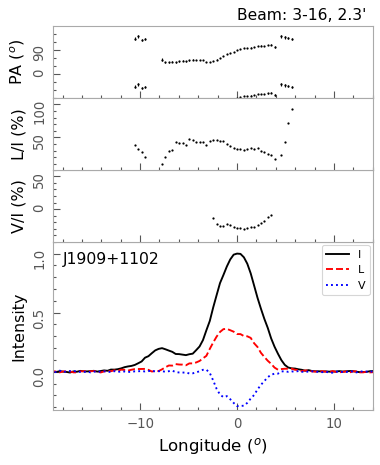}
  \includegraphics[angle=0,width = 0.23\textwidth] {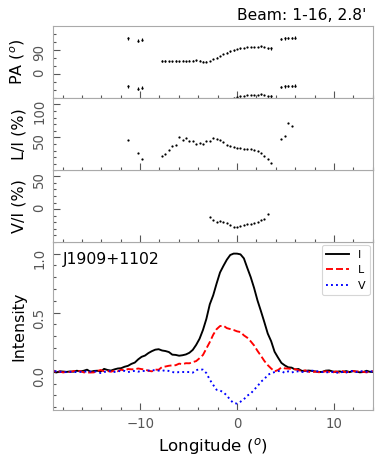}\\
  \caption{Polarization profiles of PSRs J1903+0135 and J1909+1102
    obtained from the data recorded in the 11 nearby beams and 4
    nearby beams, respectively, as illuminated in
    Figure~\ref{fig:Pol_comp-snap_beam} and listed in
    Table~\ref{table:Pol_comp}. The total intensity, linear
    polarization and circular polarization, their fractions, as well
    as position angle of linear polarization, all are plotted from the
    bottom to the top sub-panels for each profile. They are all
    consistent though with very different offsets for different beams,
    which demonstrates the excellent polarization characteristics of
    the 19-beam L-band receiver. }
  \label{fig:Pol_comp-snap}
\end{figure*}

\subsection{Polarization characteristics for the 19 beams}

In order to understand polarization characteristics of the 19 beams,
polarized pulse profiles obtained by the central beam without any
offset and by the outer beams with different offsets are carefully
examined. The FAST observation results are also compared with those
published in the literature.

First of all, is there any difference between the polarization
profiles obtained by the central beam and outer beams? In the PI
projects, we do observe some pulsars targeted by the central beam
without offset (see the list in Table~\ref{table:Pol_comp}), while the
covers with some of these pulsars are also tested in the GPPS survey
with a somehow ``random'' location offset in the outer beams.
Figure~\ref{fig:Pol_comp} shows such a comparison of pulse
polarization profiles for the cases for offset observations with the
beams of M01, M02, M03, M04, M10, M11, M12, M17 and M19. Apparently
there are some but not significantly different results from the offset
observations. The quantitative differences in PA, linear polarization
and circular polarization are obtained as listed in the first part of
Table~\ref{table:Pol_comp} for their averages and standard
deviations. The difference in PA can reach $6.4^\circ$ for an offset
of $2.2^\prime$ of a pulsar to the beam center of M02, and $4.3^\circ$
for an offset of $0.6^\prime$ for a pulsar from the beam center of
M03. The difference in the degree of linear polarization reaches 2.1\%
in an offset observation of $1.2^\prime$ from the beam center of
M19. The data scattering of the fractional linear polarization of PSR
J0631+1036 leads to a large standard deviation for the offset
observations of beam M02, probably due to the intrinsic data
scattering rather than measurements. The maximum difference in
fractional circular polarization is only 2.5\% from PSR J0711+0931
observed by the beam M04. We therefore conclude that polarization
characteristics of 19 beams, even observed with an offset as large as
$3'$, are so excellent that polarization profiles have less than 3\%
difference compared with measurements by the central beam M01 without
offset.

Second, we compare the polarization profiles obtained by data recorded
quasi-simultaneously by several beams in snapshot observations (see
e.g. Figure~\ref{fig:Pol_comp-snap_beam}). For 4 nearby beams with
various offsets, the polarization profiles in
Figure~\ref{fig:Pol_comp-snap} are excellently consistent with each
other, see the quantitative comparisons in Table~\ref{table:Pol_comp}.
If the observation of a pulsar of the nearest beam is chosen as the
references, the average deviations in PA, fractional linear
polarization and circular polarization are maximally of 3.5$^\circ$,
1.8\% and 2.4\% for beams within $3.0'$. Only for pulsars detected by
a very far beam with an offset more than $3.4'$, a slightly large
difference emerges with a large uncertainty due to the low S/N. That
is to day, the polarization response of these beams even in the very
offset positions are so consistent with each other.

Certainly, to understand polarization characteristics of the FAST
observations, we have to compare the polarization profiles with
published ones, at least one pulsar is chosen for each beam. For beam
M01, the polarized pulse profile PSR J1932+1059 obtained by FAST is
 consistent with the one observed at 1.4~GHz by
\citet{scw+84} and \citet{jhv+05}. The pulsars for the other beams are
also compared, such as PSR J0631+1036 \citep{jk18} to that from the
FAST beam M02, PSR J1932+1059 \citep{hr10} to that from the beam M03,
PSR J0711+0931 \citep{lzb+00} to that from the beam M04, PSR
J1904+0004 \citep{jk18} to that from the beam M06, PSRs J1954+2923
\citep{wcl+99, hr10} and J2008+2513 \citep{hdvl09} to those from the
beam M07, PSR J1932+2220 \citep{mr11} to that from the beam M10, PSR
J1935+1616 \citep{hdvl09,hr10} to that from the beam M11, PSR
J1909+1102 \citep{hr10} to that from the beam M16, PSR J1946+1805
\citep{hr10,mr11} to that from the beam M17, PSR J1857+0943
\citep{ovhb04} to that from the beam M18, and PSR J1842+1332
\citep{jk18} to that from the beam M19. All the FAST polarization
profiles are  consistent with the published ones,
albeit at slightly different observing frequencies.

%

\begin{table*}
  \centering     
  \caption{Polarization parameters of 682 pulsars (see notes for the columns at the end, -- {\it to be continued}. )}
  \label{table:psrs}
  \tabcolsep 2pt
  \footnotesize
\begin{tabular}{llrcccrrrrrrrrll}
    \hline  
    \hline
PSR        &   P     &   DM$^a$   & Date &  Mode & Beam   &  Offset  &  $W_{50}$ &  $W_{10}$ &  L/I & V/I  & $|V|$/I & RM &  Fig. & Note.  &  Ref.       \\
           &  (s)    &  &   &            &        &($^\prime$)&($^{\circ}$)&($^{\circ}$)& (\%) & (\%) & (\%)    &($\rm rad/m^2$)&      &       &          \\
(1)        &  (2)    &  (3)   &    (4)   &    (5)     &  (6)   &   (7)    & (8)      &  (9)     & (10) & (11) &  (12)   & (13)          &  (14) &  (15)  & (16)   \\
\hline
J0006+1834 & 0.693 &  11.4 & 20210110 & TR  & P1M01 & 0.0  & 39.9(28)& -       &100.0(35) & -6.6(31) &  7.5(31) & -19.4(4) & \ref{fig:intprof_HiLP}& hiL,S    &          \\
J0011+08   & 2.552 &  24.9 & 20210927 & SD  & P3M02 &$<$1.5&  1.8(7) &  7.4(7) & 39.9(38) & -6.0(34) &  6.1(34) & -17.4(27)& \ref{fig:intprof_CD}  & cd       &          \\
J0023+0923 & 0.0030&  14.3 & 20201120 & TR  & P1M01 & 0.0  & 23.1(14)& 85.6(14)& 24.6(31) &  1.8(31) &  6.6(31) &  -5.5(30)& \ref{fig:intprof_OM}  & ot       &          \\
J0050+03   & 1.366 &  26.5 & 20210117 & TR  & P1M01 &      & 10.0(14)& 14.1(14)& 36.5(33) & -8.5(32) & 12.3(32) & -25.6(22)& \ref{fig:intprof}     &          &          \\
J0058+4950 & 0.996 &  66.9 & 20201120 & TR  & P1M01 & 0.0  & 11.1(7) & 16.8(7) & 30.6(30) & 14.4(30) & 15.5(30) & -114.2(5)& \ref{fig:intprof_S}   & S        &          \\
\hline
 \end{tabular}
\end{table*}

\addtocounter{table}{-1}
\begin{table*}
  \centering     
  \caption{\it -- continued -- }
  \tabcolsep 2pt
  \footnotesize
  \begin{tabular}{llrcccrrrrrrrrll}
    
    \hline  
    \hline
PSR        &   P     &   DM$^a$   & Date &  Mode & Beam   &  Offset  &  $W_{50}$ &  $W_{10}$ &  L/I & V/I  & $|V|$/I & RM &  Fig. & Note.  &  Ref.       \\
           &  (s)    &  &   &            &        &($^\prime$)&($^{\circ}$)&($^{\circ}$)& (\%) & (\%) & (\%)    &($\rm rad/m^2$)&      &       &          \\
(1)        &  (2)    &  (3)   &    (4)   &    (5)     &  (6)   &   (7)    & (8)      &  (9)     & (10) & (11) &  (12)   & (13)          &  (14) &  (15)  & (16)   \\
\hline
J1544+4937 & 0.0021&  23.2 & 20201219 & TR  & P1M01 & 0.0  & 20.4(14)& 61.6(14)& 14.9(31) & -0.3(31) &  7.5(31) &  10.3(11)& \ref{fig:intprof}     &          &          \\
J1628+4406 & 0.181 &   7.3 & 20201026 & TR  & P1M01 & 0.0  &198.6(14)&205.5(14)& 36.9(30) & -0.6(30) &  4.5(30) &   1.6(4) & \ref{fig:intprof_IP}  & ip,hiL   &          \\
J1630+3734 & 0.0033&  14.1 & 20201216 & TR  & P1M01 & 0.0  &201.8(14)&237.8(14)& 36.4(30) &  1.5(30) &  5.5(30) &   1.1(2) & \ref{fig:intprof_WP}  & w,ot,hiL &          \\
J1635+2332 & 1.208 &  37.5 & 20201026 & TR  & P1M01 & 0.0  & 13.4(7) & 16.1(7) & 22.2(30) &  1.0(30) & 17.3(30) &  25.9(8) & \ref{fig:intprof}     &          &          \\
J1638+4005 & 0.767 &  33.4 & 20210201 & TR  & P1M01 & 0.0  &  5.9(7) & 11.2(7) & 18.1(33) &  0.8(32) &  3.0(32) &  17(4)   & \ref{fig:intprof}     &          &          \\
\end{tabular}
\end{table*}
\addtocounter{table}{-1}
\begin{table*}
  \centering     
  \caption{\it -- continued -- }
  \tabcolsep 2pt
  \footnotesize
  \begin{tabular}{llrcccrrrrrrrrll}
    
    \hline  
    \hline
PSR        &   P     &   DM$^a$   & Date &  Mode & Beam   &  Offset  &  $W_{50}$ &  $W_{10}$ &  L/I & V/I  & $|V|$/I & RM &  Fig. & Note.  &  Ref.       \\
           &  (s)    &  &   &            &        &($^\prime$)&($^{\circ}$)&($^{\circ}$)& (\%) & (\%) & (\%)    &($\rm rad/m^2$)&      &       &          \\
(1)        &  (2)    &  (3)   &    (4)   &    (5)     &  (6)   &   (7)    & (8)      &  (9)     & (10) & (11) &  (12)   & (13)          &  (14) &  (15)  & (16)   \\
\hline
J1838-0107 & 0.444 & 270.1 & 20210928 & SS  & P4M15 & 0.2  &  3.7(4) & 29.9(4) & 14.9(30) & -8.5(30) & 10.1(30) & 412.7(12)& \ref{fig:intprof_CD}  & cd       &        \\
J1838+1523 & 0.549 &  68.2 & 20210117 & TR  & P1M01 & 0.0  & 11.9(7) & 39.4(7) & 38.5(30) &-21.8(30) & 23.5(30) & 160.4(4) & \ref{fig:intprof_OM}  & ot?      &        \\
J1839-0223 & 1.266 & 336.8 & 20210816 & SS  & P4M18 & 1.5  & 12.0(7) & 15.4(7) & 21.0(33) & -9.8(33) & 10.2(33) & 147.1(34)& \ref{fig:intprof_CD}  & cd,S     &        \\
J1840+0214 & 0.797 & 182.4 & 20210616 & SS  & P3M11 & 0.7  &  4.5(7) &  7.6(7) & 27.5(31) & -5.3(31) &  6.7(31) & 108.1(10)& \ref{fig:intprof}     &          &        \\
J1841-0157 & 0.663 & 475.0 & 20210901 & SS  & P3M01 & 1.7  & 11.5(4) & 19.9(4) & 16.0(30) & -3.3(30) &  3.8(30) & 100.7(5) & \ref{fig:intprof}     &          &       \\
\end{tabular}
\end{table*}
\addtocounter{table}{-1}
\begin{table*}
  \centering     
  \caption{\it -- continued -- }
  \tabcolsep 2pt
  \footnotesize
  \begin{tabular}{llrcccrrrrrrrrll}
    
    \hline  
    \hline
PSR        &   P     &   DM$^a$   & Date &  Mode & Beam   &  Offset  &  $W_{50}$ &  $W_{10}$ &  L/I & V/I  & $|V|$/I & RM &  Fig. & Note.  &  Ref.       \\
           &  (s)    &  &   &            &        &($^\prime$)&($^{\circ}$)&($^{\circ}$)& (\%) & (\%) & (\%)    &($\rm rad/m^2$)&      &       &          \\
(1)        &  (2)    &  (3)   &    (4)   &    (5)     &  (6)   &   (7)    & (8)      &  (9)     & (10) & (11) &  (12)   & (13)          &  (14) &  (15)  & (16)   \\
\hline
J1851+1259 & 1.205 &  70.2 & 20210605 & SS  & P3M19 & 2.7  &  3.2(7) &  6.0(7) & 21.5(31) &  2.9(31) &  5.1(31) & 147(2)   & \ref{fig:intprof}     &          & 13,16    \\
J1852-0024g& 0.355 & 292.0 & 20200905 & TR  & P1M01 &$<$1.5&  6.5(7) & 14.0(7) & 24.7(32) &  1.4(32) &  9.5(32) &  95.9(24)& \ref{fig:intprof}     &          & 31       \\
J1852-0054g& 0.164 & 211.3 & 20210408 & TR  & P1M01 &$<$1.5& 28.7(28)& --      & 46.4(37) & -1.8(34) &  6.4(34) & 166(3)   & \ref{fig:intprof}     &          &          \\
J1852-0118 & 0.451 & 286.4 & 20211113 & TR  & P1M04 & 1.5  &142.0(7) &154.9(7) & 65.1(30) & -6.1(30) & 11.2(30) & 708.4(6) & \ref{fig:intprof_HiLP}& hiL,S    &          \\
J1852+0008 & 0.467 & 254.9 & 20200404 & TR  & P1M06 & 1.9  & 11.9(14)& 24.2(15)& 58.4(34) &  1.1(32) &  7.3(32) &  -1(1)   & \ref{fig:intprof}     &          &          \\
\end{tabular}
\end{table*}
\addtocounter{table}{-1}
\begin{table*}
  \centering     
  \caption{\it -- continued -- }
  \tabcolsep 2pt
  \footnotesize
  \begin{tabular}{llrcccrrrrrrrrll}
    
    \hline  
    \hline
PSR        &   P     &   DM$^a$   & Date &  Mode & Beam   &  Offset  &  $W_{50}$ &  $W_{10}$ &  L/I & V/I  & $|V|$/I & RM &  Fig. & Note.  &  Ref.       \\
           &  (s)    &  &   &            &        &($^\prime$)&($^{\circ}$)&($^{\circ}$)& (\%) & (\%) & (\%)    &($\rm rad/m^2$)&      &       &          \\
(1)        &  (2)    &  (3)   &    (4)   &    (5)     &  (6)   &   (7)    & (8)      &  (9)     & (10) & (11) &  (12)   & (13)          &  (14) &  (15)  & (16)   \\
\hline
J1859+00   & 0.559 & 420.0 & 20200513 & SS  & P3M15 &      & 34.8(7) & 61.9(7) & 24.4(30) & -4.2(30) &  4.9(30) & 581.6(4) & \ref{fig:intprof_OM}  & ot,S     &        \\
J1859+0126g& 0.957 & 529.2 & 20200815 & TR  & P1M01 &$<$1.5& 13.4(14)& --      & 57.4(38) &  6.4(37) &  8.4(37) & 358.0(12)& \ref{fig:intprof_S}   & S        &        \\
J1859+0430g& 0.336 & 785.4 & 20200402 & TR  & P1M01 &$<$1.5& 18.8(28)& --      & 24.5(39) & 17.9(38) & 19.6(37) &1554(9)   & \ref{fig:intprof}     &          &       \\
&        &       & 20200404 & TR  & P1M01 &        &         &         &        &         &         &         &          &                       &        \\
&        &       & 20200816 & TR  & P1M01 &        &         &         &        &         &         &         &          &                       &        \\
\end{tabular}
\end{table*}
\addtocounter{table}{-1}
\begin{table*}
  \centering     
  \caption{\it -- continued -- }
  \tabcolsep 2pt
  \footnotesize
  \begin{tabular}{llrcccrrrrrrrrll}
    
    \hline  
    \hline
PSR        &   P     &   DM$^a$   & Date &  Mode & Beam   &  Offset  &  $W_{50}$ &  $W_{10}$ &  L/I & V/I  & $|V|$/I & RM &  Fig. & Note.  &  Ref.       \\
           &  (s)    &  &   &            &        &($^\prime$)&($^{\circ}$)&($^{\circ}$)& (\%) & (\%) & (\%)    &($\rm rad/m^2$)&      &       &          \\
(1)        &  (2)    &  (3)   &    (4)   &    (5)     &  (6)   &   (7)    & (8)      &  (9)     & (10) & (11) &  (12)   & (13)          &  (14) &  (15)  & (16)   \\
\hline
J1906+0649 & 1.286 & 249.0 & 20200418 & SS  & P4M06 & 1.5  & 10.7(4) & 16.0(4) & 42.9(30) & 11.3(30) & 12.5(30) & -17.5(5) & \ref{fig:intprof_OM}  & ot,S     & 23     \\
J1906+0724 & 1.536 & 475.2 & 20220210 & SC  & P1M01 &      &  5.0(7) &    -    & 22.5(42) & -4.5(40) & 12.6(40) & 665(18)  & \ref{fig:intprof}     &          &        \\
J1906+0746 & 0.143 & 218.4 & 20200901 & SS  & P2M02 & 1.4  &  8.0(14)& 17.0(14)& 58.1(36) & -6.3(35) &  7.5(34) & 157.1(17)& \ref{fig:intprof_S}   & S        &        \\
J1906+1854 & 1.019 & 155.6 & 20210802 & SS  & P1M02 & 1.4  & 25.9(14)& 46.2(14)&  9.3(31) &  1.8(31) &  4.6(31) & 398(8)   & \ref{fig:intprof_OM}  & ot       & 23     \\
J1907+0249 & 0.351 & 261.0 & 20210617 & SS  & P3M15 & 0.6  & 19.3(7) & 30.9(7) & 29.4(30) &  0.4(30) &  4.0(30) & 334.3(9) & \ref{fig:intprof_HiLP}& hiL,ot   &        \\
\end{tabular}
\end{table*}
\addtocounter{table}{-1}
\begin{table*}
  \centering     
  \caption{\it -- continued -- }
  \tabcolsep 2pt
  \footnotesize
  \begin{tabular}{llrcccrrrrrrrrll}
    
    \hline  
    \hline
PSR        &   P     &   DM$^a$   & Date &  Mode & Beam   &  Offset  &  $W_{50}$ &  $W_{10}$ &  L/I & V/I  & $|V|$/I & RM &  Fig. & Note.  &  Ref.       \\
           &  (s)    &  &   &            &        &($^\prime$)&($^{\circ}$)&($^{\circ}$)& (\%) & (\%) & (\%)    &($\rm rad/m^2$)&      &       &          \\
(1)        &  (2)    &  (3)   &    (4)   &    (5)     &  (6)   &   (7)    & (8)      &  (9)     & (10) & (11) &  (12)   & (13)          &  (14) &  (15)  & (16)   \\
\hline
J1914+0625 & 0.878 & 204.8 & 20210507 & SS  & P4M04 &$<$1.5& 19.8(7) & 22.3(7) & 25.6(33) & -2.5(32) &  8.8(32) & -49.3(27)& \ref{fig:intprof}     &          &           \\
J1914+0631 & 0.693 &  58.0 & 20210507 & SS  & P3M09 & 1.3  &  7.2(4) & 11.6(4) & 21.8(31) & -5.7(30) &  6.3(30) & -98.3(15)& \ref{fig:intprof_OM}  & ot       &           \\
J1914+0838 & 0.440 & 290.6 & 20200829 & TR  & P1M15 &      &  3.5(7) &  8.5(7) &  9.3(30) &  8.8(30) & 10.5(30) & 357(2)   & \ref{fig:intprof}     &          &           \\
J1914+1029g& 2.485 &  65.9 & 20200821 & TR  & P1M01 &$<$1.5&  3.0(2) &  5.4(2) & 68.1(33) & -2.6(32) &  3.6(31) &  18.0(9) & \ref{fig:intprof_HiLP}& hiL      &           \\
&        &       & 20200914 & TR  & P1M01 &        &        &         &         &         &         &         &          &                       &    \\
\end{tabular}
\end{table*}
\addtocounter{table}{-1}
\begin{table*}
  \centering     
  \caption{\it -- continued -- }
  \tabcolsep 2pt
  \footnotesize
  \begin{tabular}{llrcccrrrrrrrrll}
    
    \hline  
    \hline
PSR        &   P     &   DM$^a$   & Date &  Mode & Beam   &  Offset  &  $W_{50}$ &  $W_{10}$ &  L/I & V/I  & $|V|$/I & RM &  Fig. & Note.  &  Ref.       \\
           &  (s)    &  &   &            &        &($^\prime$)&($^{\circ}$)&($^{\circ}$)& (\%) & (\%) & (\%)    &($\rm rad/m^2$)&      &       &          \\
(1)        &  (2)    &  (3)   &    (4)   &    (5)     &  (6)   &   (7)    & (8)      &  (9)     & (10) & (11) &  (12)   & (13)          &  (14) &  (15)  & (16)   \\
\hline
J1925+1720 & 0.075 & 223.3 & 20200404 & SS  & P1M09 & 1.2  & 14.7(28)& --      & 81.2(51) & 12.2(42) & 14.2(41) & 437(3)   & \ref{fig:intprof_HiLP}& hiL      & 30         \\
J1926+0431 & 1.074 & 101.8 & 20210702 & SS  & P2M19 & 1.3  &  4.5(2) &  9.1(2) & 16.2(30) &  0.8(30) &  5.2(30) & -21.0(6) & \ref{fig:intprof}     &          & 13,16  \\
J1926+0737 & 0.318 & 159.0 & 20201226 & SS  & P2M09 & 1.2  &171.5(7) &177.1(7) & 13.0(31) &  0.3(31) &  4.4(31) & 352(7)   & \ref{fig:intprof_IP}  & ip       &              \\
J1926+1434 & 1.324 & 211.4 & 20210317 & TR  & P1M15 & 1.5  &  3.3(7) & 40.3(7) & 30.4(30) & -7.3(30) &  9.9(30) & 252.9(7) & \ref{fig:intprof_CD}  & cd,ot,S  & 13,16,23  \\
J1926+1631g& 0.678 & 197.0 & 20200818 & TR  & P1M01 &$<$1.5& 12.1(7) & 14.9(7) & 18.1(31) &  4.7(31) &  8.3(31) & 555(2)   & \ref{fig:intprof}     &          &        \\
\end{tabular}
\end{table*}
\addtocounter{table}{-1}
\begin{table*}
  \centering     
  \caption{\it -- continued -- }
  \tabcolsep 2pt
  \footnotesize
  \begin{tabular}{llrcccrrrrrrrrll}
    
    \hline  
    \hline
PSR        &   P     &   DM$^a$   & Date &  Mode & Beam   &  Offset  &  $W_{50}$ &  $W_{10}$ &  L/I & V/I  & $|V|$/I & RM &  Fig. & Note.  &  Ref.       \\
           &  (s)    &  &   &            &        &($^\prime$)&($^{\circ}$)&($^{\circ}$)& (\%) & (\%) & (\%)    &($\rm rad/m^2$)&      &       &          \\
(1)        &  (2)    &  (3)   &    (4)   &    (5)     &  (6)   &   (7)    & (8)      &  (9)     & (10) & (11) &  (12)   & (13)          &  (14) &  (15)  & (16)   \\
\hline
J1941+2525 & 2.306 & 314.4 & 20200419 & SS  & P2M03 & 1.6  &  5.4(4) &  9.7(4) & 22.6(31) & -1.5(31) &  3.2(31) &-100.3(21)& \ref{fig:intprof_OM}  & ot       &         \\
J1941+4320 & 0.840 &  79.3 & 20201215 & TR  & P1M01 & 0.0  &  6.5(7) & 10.9(7) & 13.6(30) & -1.6(30) &  5.3(30) &  23.7(14)& \ref{fig:intprof}     &          &         \\
J1942+3941 & 1.353 &  95.5 & 20210801 & SS  & P2M01 & 1.3  & 12.0(4) & 28.2(4) & 14.0(30) & -5.3(30) & 10.4(30) &   1.7(5) & \ref{fig:intprof}     &          &         \\
J1943+0609 & 0.446 &  70.7 & 20210717 & SS  & P1M13 & 0.7  &  2.4(2) & 16.3(2) & 21.1(30) & -1.4(30) &  4.6(30) & -34.6(3) & \ref{fig:intprof_OM}  & ot       & 20,22  \\
J1943+28   & 0.737 & 231.6 & 20220508 & SS  & P3M09 &      &  7.0(7) & 12.4(7) & 26.1(32) & -2.5(31) &  6.7(31) &  37.0(16)& \ref{fig:intprof}     &          &         \\
\end{tabular}
\end{table*}
\addtocounter{table}{-1}
\begin{table*}
  \centering     
  \caption{\it -- continued --}
  \tabcolsep 2pt
  \footnotesize
  \begin{tabular}{llrcccrrrrrrrrll}
    
    \hline  
    \hline
PSR        &   P     &   DM$^a$   & Date &  Mode & Beam   &  Offset  &  $W_{50}$ &  $W_{10}$ &  L/I & V/I  & $|V|$/I & RM &  Fig. & Note.  &  Ref.       \\
           &  (s)    &  &   &            &        &($^\prime$)&($^{\circ}$)&($^{\circ}$)& (\%) & (\%) & (\%)    &($\rm rad/m^2$)&      &       &          \\
(1)        &  (2)    &  (3)   &    (4)   &    (5)     &  (6)   &   (7)    & (8)      &  (9)     & (10) & (11) &  (12)   & (13)          &  (14) &  (15)  & (16)   \\
\hline
J2008+3758 & 4.352 & 140.5 & 20211225 & SS  & P1M19 & 1.0  &  2.8(4) &  4.6(4) & 24.0(32) &  0.5(31) &  5.1(31) & 270(2)   & \ref{fig:intprof}     &          &        \\
J2009+3326 & 1.438 & 263.6 & 20200816 & TR  & P1M04 & 2.3  &  6.4(7) & 22.7(7) & 19.5(30) & -4.5(30) & 10.9(30) &   9.5(16)& \ref{fig:intprof_OM}  & ot,S     &        \\
J2010+2845 & 0.565 & 112.5 & 20210619 & SS  & P1M04 & 1.3  &  9.2(4) & 15.2(4) & 18.4(30) &-12.1(30) & 19.1(30) &-230.7(9) & \ref{fig:intprof}     &          &        \\
J2010+31   & 1.551 & 251.4 & 20221205 & SS  & P2M08 &      &  5.3(4) & 10.2(4) & 24.5(35) & -5.3(33) & 15.7(33) & -43.8(17)& \ref{fig:intprof}     &          &        \\
J2010+3230 & 1.442 & 371.8 & 20200812 & TR  & P1M03 & 2.1  &  7.4(4) & 12.8(4) & 32.4(31) & 12.7(31) & 14.8(31) &-644.7(15)& \ref{fig:intprof}     &          &        \\
\end{tabular}
\end{table*}
\addtocounter{table}{-1}
\begin{table*}
  \centering     
  \caption{\it -- end.}
  \tabcolsep 2pt
  \footnotesize
  \begin{tabular}{llrcccrrrrrrrrll}
    
    \hline  
    \hline
PSR        &   P     &   DM$^a$   & Date &  Mode & Beam   &  Offset  &  $W_{50}$ &  $W_{10}$ &  L/I & V/I  & $|V|$/I & RM &  Fig. & Note.  &  Ref.       \\
           &  (s)    &  &   &            &        &($^\prime$)&($^{\circ}$)&($^{\circ}$)& (\%) & (\%) & (\%)    &($\rm rad/m^2$)&      &       &          \\
(1)        &  (2)    &  (3)   &    (4)   &    (5)     &  (6)   &   (7)    & (8)      &  (9)     & (10) & (11) &  (12)   & (13)          &  (14) &  (15)  & (16)   \\
\hline
J2208+4056 & 0.636 &  11.8 & 20201018 & TR  & P1M01 & 0.0  & 16.3(7) &202.2(7) & 76.2(30) & -8.0(30) &  8.9(30) & -41.7(2) & \ref{fig:intprof_IP}  & ip,S,ot,hiL &      \\
J2208+4610 & 0.642 &  61.8 & 20211124 & SS  & P4M09 &      & 20.6(7) & 27.1(7) & 19.8(30) & -2.7(30) &  6.1(30) & -51.9(11)& \ref{fig:intprof_OM}  & ot       &         \\
J2209+22   & 1.777 &  45.4 & 20210129 & TR  & P1M01 &      &  3.4(7) &  7.4(7) & 23.2(30) & 19.2(30) & 20.4(30) & -90.1(6) & \ref{fig:intprof}     &          &         \\
J2215+5135 & 0.0026&  69.2 & 20210626 & SS  & P2M10 & 1.5  & 15.6(28)& --      & 49.4(38) & -4.3(36) &  2.0(34) & -53.5(12)& \ref{fig:intprof_WP}  & w        &         \\
J2219+4754 & 0.538 &  43.4 & 20210608 & SS  & P1M10 & 2.0  &  5.5(2) & 10.9(2) & 14.8(30) &  3.6(30) &  5.2(30) & -36.4(4) & \ref{fig:intprof}     &          & 13,25   \\
\end{tabular}
\tablenotes{
\item
  Columns (1)-(3): pulsar name, period and DM as identification of a
  pulsar. $^a$ DM has units of $\rm pc/cm^3$, $^\star$ period updated,
  $^\dagger$ DM updated by FAST observations compared to the pulsar
  catalog \citep{mhth05};
  Columns (4)-(7): FAST observation dates, observation modes, the
  beam name and the pulsar offsets from the beam center. Observation
  modes: TR -- tracking, SS -- snapshot, SD - snapshotdec, SC --
  swiftcalibration;
Columns (8)-(12): the profile properties: pulse widths $W_{50}$ and
$W_{10}$ at 50\% and 10\% the peak intensities, degree (and
uncertainty on the last digit) of linear, circular and absolute
circular polarization $L/I$, $V/I$ and $|V|/I$;
Column (13): FAST measured RM (and uncertainty), i.e. $\rm
RM_{ISM}=RM_{obs}-RM_{ion}$; columns (14)-(16): figure number for
polarization profile, notes and the references for profile comparison.
Notes: S: S-shaped position angle; ot: orthogonal modes; hiL: highly
linearly polarized; hiC: highly circularly polarized; cd: conal double
pulsar; ip: interpulse; w: wide profile; c: scattering.
References for profile comparison: 1 = \citet{man71}; 2 =
\citet{mhma78}; 3 = \citet{rb81}; 4 = \citet{scr+84}; 5 =
\citet{scw+84}; 6 = \citet{srs+86}; 7 = \citet{lm88}; 8 =
\citet{rsw89}; 9 = \citet{ts90}; 10 = \citet{bcw91}; 11 =
\citet{zcwl96}; 12 = \citet{vx97}; 13 = \citet{gl98}; 14 =
\citet{xkj+98}; 15 = \citet{vkk98}; 16 = \citet{wcl+99}; 17 =
\citet{nss01}; 18 = \citet{bcc+02}; 19 = \citet{wck+04}; 20 =
\citet{kjm05}; 21 = \citet{jkmg08}; 22 = \citet{wj08}; 23 =
\citet{hdvl09}; 24 = \citet{hr10}; 25 = \citet{mr11}; 26 =
\citet{ymv+11}; 27 = \citet{ckr+12}; 28 = \citet{dhm+15};
29 = \citet{jk18}; 30 = \citet{sjk+21}; 31 = \citet{hww+21}.
}

\end{table*}

%

\begin{figure*}
  \centering
  \includegraphics[angle=0,width = 0.33\textwidth] {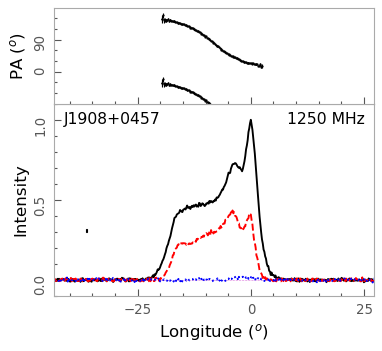}
  \includegraphics[angle=0,width = 0.33\textwidth] {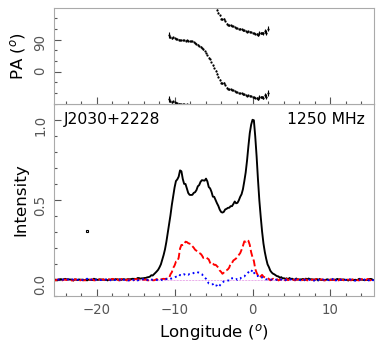}
  \includegraphics[angle=0,width = 0.33\textwidth] {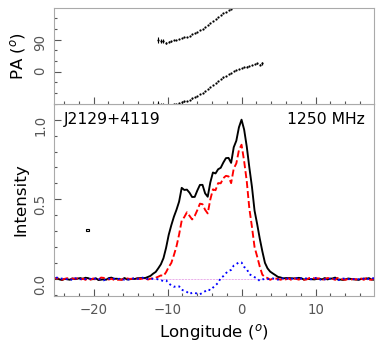}\\
  \caption{Polarization profiles with S-shaped PA curves for 3 pulsars
    as examples of 96 pulsars in Figure~\ref{fig:intprof_S} and also
    98 pulsars in other figures of Appendix A. For each pulsar, the
    total intensity, linear and circular polarization are represented
    by black solid, red dashed and blue doted lines in the bottom
    sub-panel. The left-hand circular polarization is defined to be
    positive. The bin size and $3\sigma$ are marked inside the
    sub-panel, here $\sigma$ is the standard deviation of off-pulse
    bins. In the top panel, dots with error-bar are measurements of
    polarization position angles for linear polarization intensity
    exceeding 3$\sigma$.
  }
  \label{fig:intprof_S4}
\end{figure*}

\section{FAST pulsar polarization profiles}

Polarization profiles of 682 pulsars are finally obtained from FAST
observations, as listed in Table~\ref{table:psrs}. The period and DM
are listed after the name, allowing identification of a pulsar with
parameters. Then FAST observation dates, FAST observation modes, the
beam name and the pulsar offsets from the beam center are given in
columns (4), (5), (6) and (7). The profile widths $W_{50}$ and
$W_{10}$ at 50\% and 10\% the peak intensities (only if the 10\% peak
intensities exceeding three times the standard deviation of off pulse
data) are listed in columns (8) and (9). Their uncertainties are
estimated as, $\sigma_{W,H}=t_b \sqrt{1+(\sigma_I/H)^2}$, by following
\citet{kg97} . Here, $t_b$ represents the time resolution of the phase
bins of a profile, $H$ is 0.5 for $W_{50}$, and 0.1 for $W_{10}$.  The
degrees of linear, circular and absolute circular polarization and
their uncertainties are listed in columns (10) - (12). We obtain new
RMs (i.e. $\rm RM = RM_{obs}-RM_{ion}$) for 402 pulsars as given in
column (13), which have been employed to investigate the magnetic
fields in the Galaxy by \citet{xhwy22}. The figure number for
polarization profile, and the short notes as well as references for
profile comparison are given in columns (14)-(16).

The observed polarization profiles exhibit diverse features, such as
those with S-shaped PA curves or orthogonal modes, profiles or
components with significant linear polarization or circular
polarization, profiles of conal double pulsars, profiles with
interpulse emission or emission occupying wide ranges of longitudes,
and those affected by interstellar scattering. These different
features are discussed in the following.

\begin{figure*}
  \centering
  \includegraphics[angle=0,width = 0.33\textwidth] {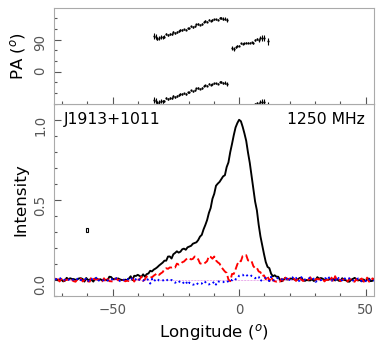}
  \includegraphics[angle=0,width = 0.33\textwidth] {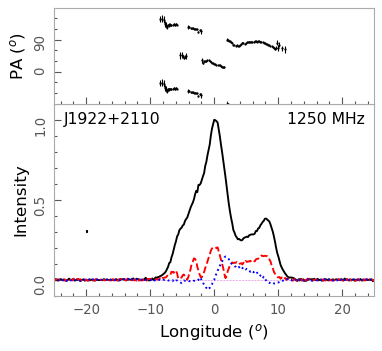}
  \includegraphics[angle=0,width = 0.33\textwidth] {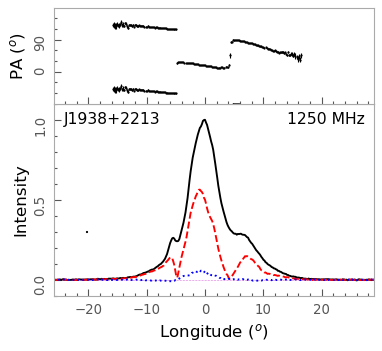}

  \caption{The same as Figure~\ref{fig:intprof_S4} but for
    polarization profiles with orthogonal modes for 3 pulsars as
    examples of 136 pulsars in Figure~\ref{fig:intprof_OM}, and also
    34 pulsars in other figures of the Appendix A.  }
  \label{fig:intprof_OM4}
\end{figure*}

\subsection{Polarization profile with a S-shaped PA curve}

The S-shaped PA curves are generally deemed to be related to the
orientation of the dipolar magnetic field line planes, and are
important indicators for pulsar radio emission generated from the
poles of dipolar magnetosphere. They are usually employed to determine
pulsar geometry as discussed in section 4.2, and can be naturally
reproduced by various emission mechanisms and propagation effects
\citep[e.g.][]{xlhq00, gan10, wwh12}.

Among the integrated pulse profiles of 682 pulsars observed by FAST,
we see the S-shaped PA curves for 195 pulsars, as shown for 3 example
pulsars in Figure~\ref{fig:intprof_S4} of 96 pulsars in
Figure~\ref{fig:intprof_S}, and 99 pulsars in other figures: 46
pulsars with orthogonal mode jumps in Figure~\ref{fig:intprof_OM}, 14
pulsars with high linear polarization in
Figure~\ref{fig:intprof_HiLP}, one pulsar with high circular
polarization in Figure~\ref{fig:intprof_HiCP}, 22 conal double pulsars
in Figure~\ref{fig:intprof_CD}, 13 pulsars with an interpulse in
Figure~\ref{fig:intprof_IP}, and 3 pulsars with wide pulses in
Figure~\ref{fig:intprof_WP}, as denoted by the ``Note'' column in
Table~\ref{table:psrs}.

These S-shaped PA curves are simple and represent the monotonic
variation of the PAs of a series of radiating sources, regardless of
the complexity of the source arrangement. This behavior is manifested
by the intensity profiles with a single component (e.g. PSRs
J0711+0931 and J1900+0227 in Figure~\ref{fig:intprof_S}), double
(e.g. PSRs J1312+1810 and J1931+1439 in Figure~\ref{fig:intprof_CD}),
triple (e.g. PSRs J1843+0119 in Figure~\ref{fig:intprof_S} and
J2113+4644 in Figure~\ref{fig:intprof_OM}), or multiple components
(e.g. PSRs J1908+0457 and J2129+4119 in Figure~\ref{fig:intprof_S4}),
and even the `partial cone' components (e.g. PSRs J1917+1353 and
J2013+3845 in Figure~\ref{fig:intprof_S}), as defined in
\citet{lm88}. More interesting is the S-shaped PA curves detected from
7 millisecond pulsars: PSRs J0605+3757, J1641+3627A, J1828+0625,
J1908+0128, J1921+0137, J2042+0246, and J2234+0944. It demonstrates
that the PA swings of MSPs are not very different from those of normal
pulsars.

\subsection{Polarization profiles with orthogonal modes}

Orthogonal modes have been observed as two populations of PAs of
individual pulses, and they manifest as a sudden $90^o$ jump in PA
curves of integrated pulse profiles, often accompanied by a great
reduction of linear polarization intensity. Among the polarization
profiles we observed by FAST, the PA jumps have been detected for 171
pulsars, as shown in Figure~\ref{fig:intprof_OM4} for 3 pulsars as
examples of 136 pulsars in Figure~\ref{fig:intprof_OM} and 35 pulsars
in other figures: 9 pulsars with highly linearly polarized components
in Figure~\ref{fig:intprof_HiLP}, 5 conal double pulsar in
Figure~\ref{fig:intprof_CD}, 6 pulsars with interpulse emission in
Figure~\ref{fig:intprof_IP}, 13 pulsars with wide profiles in
Figure~\ref{fig:intprof_WP}, and 2 pulsars affected by interstellar
scattering in Figure~\ref{fig:intprof_Scat}, as denoted by the
``Note'' column in Table~\ref{table:psrs}.

The orthogonal PA jumps can happen at one, or two, or multiple phases
within the pulse window. Among the total of  171
pulsars with orthogonal modes, 90 pulsars show one PA
jump, e.g. PSR J1913+1011 in
Figure~\ref{fig:intprof_OM4}. 60 pulsars exhibit two PA
jumps, e.g. PSR J1938+2213 in Figure~\ref{fig:intprof_OM4}. Multiple
jumps are detected for about 21 pulsars, e.g. PSR J1922+2110 in
Figure~\ref{fig:intprof_OM4}.

Apparently the PA jumps from the orthogonal modes can take place at
any phases inside the pulse window, e.g. at the profile center for PSR
J1913+1011 in Figure~\ref{fig:intprof_OM4}, or at the profile wings
for PSRs J1859+00 and J1946+1805 in
Figure~\ref{fig:intprof_OM}. Because of FAST sensitive observations,
weak emission has been detected in the wings far away from previously
known profiles, such as for PSRs J2113+4644 and J1935+1616 in
Figure~\ref{fig:intprof_OM}, which often exhibit significant
orthogonal modes.

Millisecond pulsars also exhibit orthogonal modes,
such as MSPs J0340+4130 and J1518+0204A in
Figure~\ref{fig:intprof_OM}, indicating similar magnetosphere
conditions for emission production and wave propagation as those for
normal pulsars \citep[e.g.][]{wwh14}.

Despite these 90$^o$ jumps, some non-orthogonal jumps have been
detected, e.g. for PSRs J1946+2244 and J2018+2839 in
Figure~\ref{fig:intprof_OM}. Such non-orthogonal jumps can be caused
by intrinsic mixture of emission of two modes near the transition
phase, e.g. the modes originating from different heights of pulsar
magnetosphere \citep[e.g.][]{xqh97}, or the suppositions of the
emission from different propagation paths as seen for pulsars with
strong interstellar scattering for PSRs J1852+0031 and J1853+0545
shown in Figure~\ref{fig:intprof_Scat}.

\begin{figure*}
  \centering
  \includegraphics[angle=0,width = 0.33\textwidth] {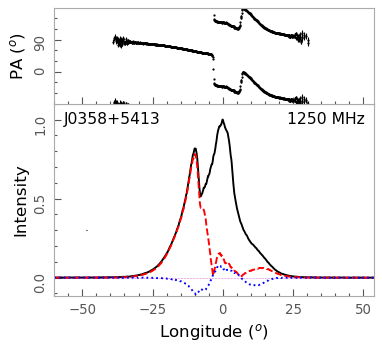}
  \includegraphics[angle=0,width = 0.33\textwidth] {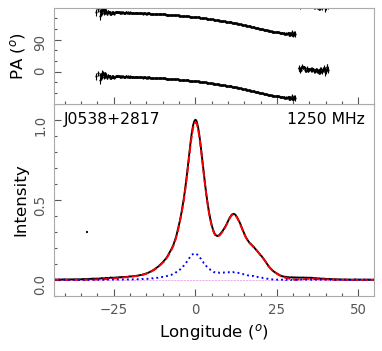}
  \includegraphics[angle=0,width = 0.33\textwidth] {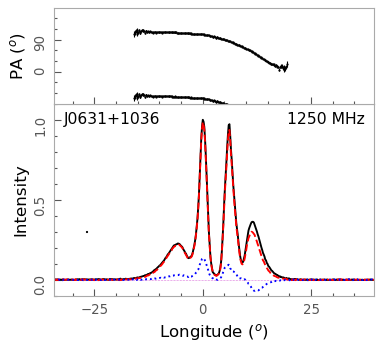}

  \caption{The same as Figure~\ref{fig:intprof_S4} but for profiles
    with highly linearly polarized emission for 3 pulsars as examples
    of 45 pulsars in Figure~\ref{fig:intprof_HiLP}, and also 28
    pulsars in other figures of the Appendix A. The highly linearly
    polarized emission can appear for the whole profile, the leading
    component, and/or the trailing component. }
  \label{fig:intprof_HiLP4}
\end{figure*}

\begin{figure*}
  \centering
  \includegraphics[angle=0,width = 0.33\textwidth] {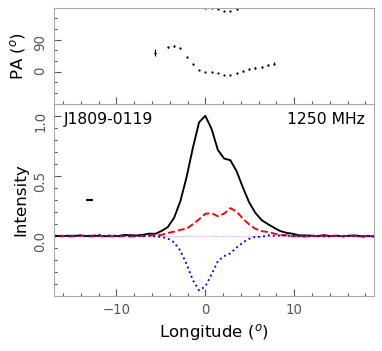}
  \includegraphics[angle=0,width = 0.33\textwidth] {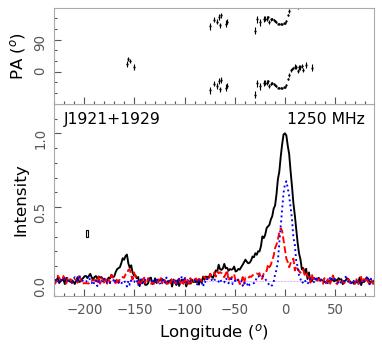}
  \includegraphics[angle=0,width = 0.33\textwidth] {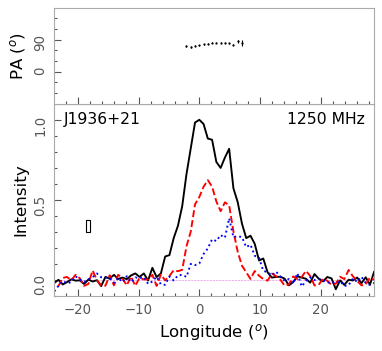}

  \caption{The same as Figure~\ref{fig:intprof_S4} but for profiles
    with highly circularly polarized emission for 3 pulsars as
    examples of 9 pulsars in Figure~\ref{fig:intprof_HiCP}, and also 7
    pulsars in other figures of the Appendix. The significant
    circularly polarized emission can be of left and/or right hand
    senses. }
  \label{fig:intprof_HiCP4}
\end{figure*}

\subsection{Polarization profiles with highly linearly polarized components}


Highly linearly polarized profiles or components have been detected
for 73 pulsars in our FAST observations, as shown in
Figure~\ref{fig:intprof_HiLP4} for 4 pulsars as examples of 45 pulsars
in Figure~\ref{fig:intprof_HiLP}, and also 28 pulsars in other
figures, including 13 pulsars with S-shaped position angles in
Figure~\ref{fig:intprof_S}, 2 pulsars with orthogonal modes in
Figure~\ref{fig:intprof_OM}, 9 pulsars with interpulse emission in
Figure~\ref{fig:intprof_IP}, 3 pulsars with wide profiles in
Figure~\ref{fig:intprof_WP} and one conal double pulsar in
Figure~\ref{fig:intprof_CD}, as denoted by the ``Note'' column in
Table~\ref{table:psrs}.

For most of these pulsars, highly linearly polarized emission is
detected for the whole pulse profiles with a polarization degree
larger than 70\%. These pulsars can have a single component such as
PSR J1838-01 in Figure~\ref{fig:intprof_HiLP}, or double components,
e.g., PSRs J0006+1834 in Figure~\ref{fig:intprof_HiLP} and J0538+2817
in Figure~\ref{fig:intprof_HiLP4}, or multiple components, e.g., PSR
J0631+1036 in Figure~\ref{fig:intprof_HiLP4}. For some pulsar, the
highly linearly polarized emission is only detected from the leading
component such as PSRs J0538+5413 in Figure~\ref{fig:intprof_HiLP4},
or the trailing component such as PSRs J1911+0939g and J1928+1809g in
Figure~\ref{fig:intprof_HiLP}, or both such as PSRs J1907+0249 and
J1957+2831 in Figure~\ref{fig:intprof_HiLP}. These highly linearly
polarized emission generally has flat PAs compared with those less
polarized ones.


One possible explanation for such highly linearly polarized
emission  is by considering the emission
process and propagation effects within pulsar magnetosphere. As
discussed by \citet{wh16},
the highly linearly polarization can appear in anywhere inside the
emission beam, e.g. the leading part, the trailing part or the entire
beam, depending on the plasma conditions in pulsar magnetosphere.
When the sight lines cut across the emission beam with different
impact angles, one can detect highly linearly polarized component at
different parts of pulse profiles.

\begin{figure*}
  \centering
  \includegraphics[angle=0,width = 0.33\textwidth] {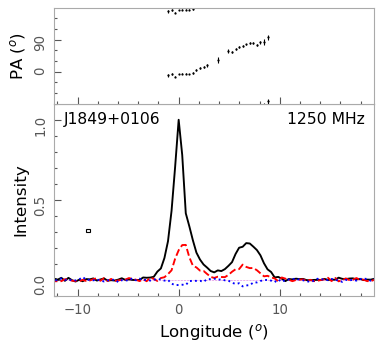}
  \includegraphics[angle=0,width = 0.33\textwidth] {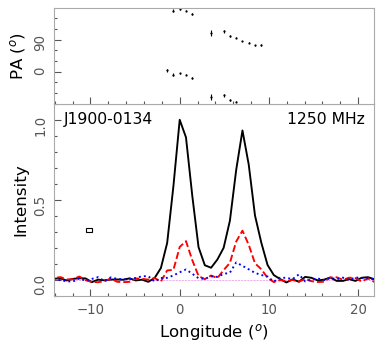}
  \includegraphics[angle=0,width = 0.33\textwidth] {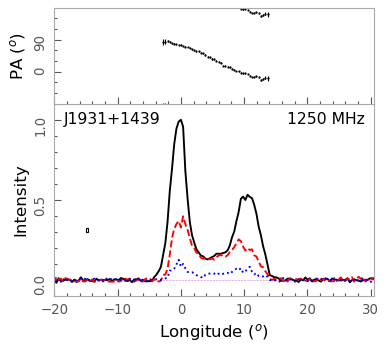}\\

  \caption{The same as Figure~\ref{fig:intprof_S4} but for 3 conal
    double pulsars as examples of 28 pulsars in
    Figure~\ref{fig:intprof_CD}.  }
  \label{fig:intprof_CD4}
\end{figure*}

\subsection{Polarization profiles with highly circularly polarized components}

Most pulsar polarization profiles have a low degree of circular
polarization \citep{hmxq98}. Among our FAST results, polarization
profiles of 16 pulsars have highly circularly polarized components
with the degree of circular polarization greater than 30\%, as shown
for 3 pulsars in Figure~\ref{fig:intprof_HiCP4} as examples of 9
pulsars in Figure~\ref{fig:intprof_HiCP} and PSRs J1853+0056 and
J2006+3102 with S-shaped PA curves in Figure~\ref{fig:intprof_S}, PSRs
J1855+0139g and J1907+0345 with highly linearly polarized emission in
Figure~\ref{fig:intprof_HiLP}, PSRs J1851+0118 and J1852+0056g with
interpulse emission in Figure~\ref{fig:intprof_IP}, and PSR J1855+0527
affected by interstellar scattering in
Figure~\ref{fig:intprof_Scat}. The strong circular polarization can be
of the left hand such as PSR J1921+1929 or the right hand such as PSR
J1809-0119 in Figure~\ref{fig:intprof_HiCP4}.

Circular polarization can be natural produced by curvature emission
mechanism in pulsar magnetosphere. For example, \citet{gan10} showed
that the circular polarization with a sense reversal correlates with
the swing of PA in curvature radiation. \citet{wwh12} explored the
polarization features for curvature radiation for the entire pulsar
emission beam, and demonstrated that the density gradient of the
relativistic particles result in net circular polarization, and the
sense or sense reversal can vary across the emission beam depending on
the density distribution of the relativistic particles and the
geometry with respect of the line of sight. Moreover, non symmetric
cyclotron absorption of emission in the higher magnetosphere can also
lead to circular polarization of pulsar emission \citep{wlh10}.

\begin{table}
  \centering
  \caption{Circular polarization of 28 conal-double pulsars.}
  \label{table:CP_conal-double}
  \tabcolsep 4pt
   \begin{tabular}{llcc}
    \hline
    \hline
PSR         & PA     & \multicolumn{2}{c}{Sense of V}  \\
            & swing    & Comp 1 & Comp 2    \\
\hline
J0011+08    & Increase & $-$   &  ?      \\
J0848+16    & Increase & ?   &  ?      \\
J1312+1810  & Increase & $-$   &  $-$      \\
J1538+2345  & Increase & $-$   &  $-$      \\
J1641+3627A & Increase & $-$   &  $-$      \\
J1833-0209  & Increase & $-$   &  $-$      \\
J1838+0044g & Increase & ?   &  ?      \\
J1839-0223  & Increase & $-$   &  $-$      \\
J1843-0211  & Increase & $-$   &  $-$      \\
J1849+0106  & Increase & $-$   &  $-$      \\
J1910+0728  & Increase & ?   &  +      \\
J1917+0834  & Increase & $-$   &  $-$      \\
J1919+1745  & Increase & +   &  +      \\
J1926+1434  & Increase & $-$   &  ?     \\
J1933+1304  & ?        & $-$   &  $-$      \\
J1933+5335  & Increase & $-$   &  +      \\
J1946+14    & Increase & +   &  +      \\
J2051+4434g & Increase & ?   &  ?      \\
\hline
J1832+27    & Decrease & +   &  +      \\
J1838-0107  & Decrease & $-$   &  ?     \\
J1854+0319  & Decrease & +   &  +      \\
J1856+0243g & Decrease & $-$   &  ?      \\
J1900-0134  & Decrease & +   &  +      \\
J1904+1011  & Decrease & +   &  +      \\
J1921+1540  & Decrease & $-$ &  $-$    \\
J1931+1439  & Decrease & +   &  +      \\
J1937+2544  & Decrease & ?   &  $-$    \\
J1938+14A   & Decrease & +   &  +      \\

\hline
\multicolumn{4}{l}{Note: ? = not very clear}\\
    \end{tabular}
\end{table}

\begin{figure*}
  \centering
  \includegraphics[angle=0,width = 0.33\textwidth] {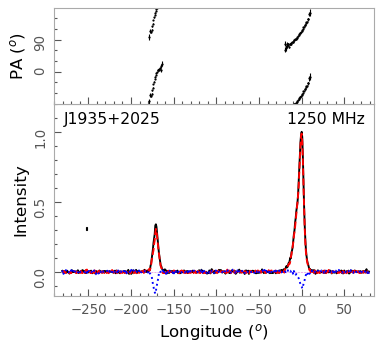}
  \includegraphics[angle=0,width = 0.33\textwidth] {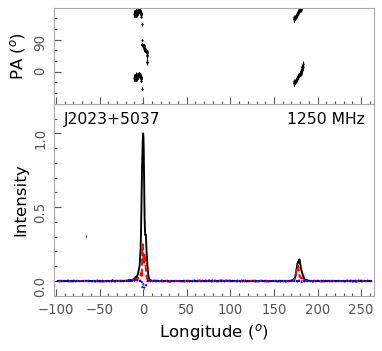}
  \includegraphics[angle=0,width = 0.33\textwidth] {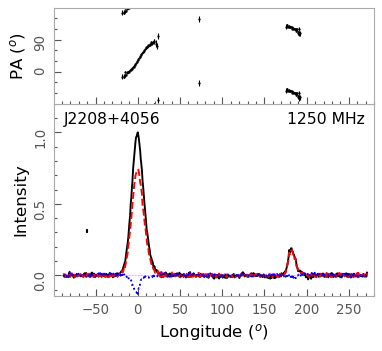}
  \caption{The same as Figure~\ref{fig:intprof_S4} but for 3 pulsars
    with interpulse emission as examples of 27 pulsars in
    Figure~\ref{fig:intprof_IP}.  }
  \label{fig:intprof_IP4}
\end{figure*}

\subsection{Conal-double pulsars}

Conal-double pulsars are defined as two distinct components
  likely from a conal beam. They generally exhibit S-shaped PA curves,
  as shown in Figure~\ref{fig:intprof_CD4} as examples.  . \citet{hmxq98} found that the
senses of PA swing and the senses of circular polarization are
related. The decrease of PA swing usually accompanies with the
left-hand ($+$ values) circular polarization, and the increase of PA
swing with the right-hand ($-$ values) circular polarization. Such a
sense correlation was strengthened by \citet{yh06} using a large
dataset.

Using the newly measured polarization profiles, we list the senses of
PA swing and circular polarization for 28 conal-double pulsars in
Table~\ref{table:CP_conal-double}. Among them, 18 pulsars exhibit
clear PA changes and the sense of circular polarization, in which 14
of them follow the correlation found by \citet{hmxq98}, while 4
pulsars (PSRs J1919+1745,J1946+14, J1933+5335 and J1921+1540 ) do
not. .

\begin{figure*}
  \centering
  \includegraphics[angle=0,width = 0.46\textwidth] {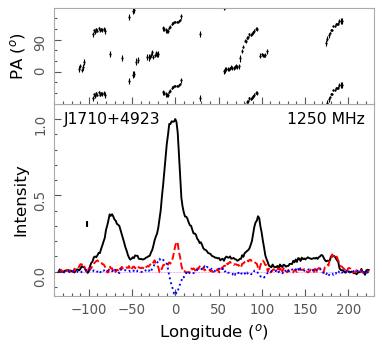}
  \includegraphics[angle=0,width = 0.46\textwidth] {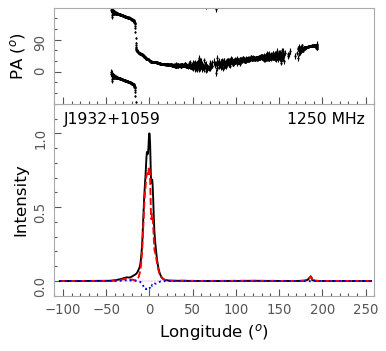}

  \caption{The same as Figure~\ref{fig:intprof_S4} but for two pulsars
    with extremely wide profiles as examples of 21 pulsars in
    Figure~\ref{fig:intprof_WP}. Note that PSR J1932+1059 has a very weak
    emission between the two peaks detected by FAST.}
  \label{fig:intprof_WP4}
\end{figure*}

\subsection{Polarization profiles with an interpulse component}

The interpulse represents a pulse component located at a longitude
separated by about 180$^\circ$ from the main one, which usually
indicates that the main pulse and interpulse come from the two
opposite poles of pulsar magnetosphere. Among the FAST pulsar
database, we get polarization profiles of 27 pulsars most likely to
have an interpulse component, as shown in Figure~\ref{fig:intprof_IP4}
for 3 pulsars as examples of the 27 ones in
Figure~\ref{fig:intprof_IP}. More than half of them have good quality
so that polarization profiles of the weak interpulse have been well
measured.  For PSRs J1909+0749 and J2208+4056 in
Figure~\ref{fig:intprof_IP4}, the main pulse and interpulse have the
different emission mode so that their PA curves have an offset of
$90^{\circ}$ in the rotating vector model fitting. The orthogonal mode
is clearly seen in the main pulse of PSR J2023+5037 in
Figure~\ref{fig:intprof_IP4}.

For these pulsars with interpulses, the highly linear polarization is
often detected for the main pulse such as PSR J1017+3011 in
Figure~\ref{fig:intprof_IP}, or for the interpulses such as PSRs
J1909+0749 and J1913+0832 in Figure~\ref{fig:intprof_IP}, or both such
as PSRs  J1935+2025, J2004+3429, J2032+4127 and
J2208+4056 in Figure~\ref{fig:intprof_IP}. The leading part of the
main pulse of PSR J1628+4406 in Figure~\ref{fig:intprof_IP} and the
trailing part of the main pulse of J1918+1541 in
Figure~\ref{fig:intprof_IP} are also highly linearly
polarized. Significant circular polarization is detected from the main
pulses of PSRs J1852+0056g and J2047+5029 or the interpulse of PSR
J1913+0832 in Figure~\ref{fig:intprof_IP}, and both the main pulse and
interpulse of PSR J1935+2025 in
Figure~\ref{fig:intprof_IP4}.

Based on high quality polarization profiles of 5 interpulse pulsars,
\citet{jk19} found that the polarization properties of main pulse and
interpulse are related, so that the signs of circular polarization
should be the same when the PA swings in the same trend. We get new
data for another 5 pulsars in Figure~\ref{fig:intprof_IP} which can be
used to verify the conclusion. The main pulse and interpulse of PSR
J1935+2025 have the same PA swing and the same sign of circular
polarization, and PSRs J1934+2352 and J2047+5029 have different PA
swings and the opposite signs for circular polarization. The
polarization data of these 3 pulsars are consistent with the
conclusion of \citet{jk19}. However, the PA swing of the main pulse
and interpulse for PSR  J2208+4056 is different,
while the circular polarization has the same sign.

\begin{figure*}
  \centering
  \includegraphics[angle=0,width = 0.33\textwidth] {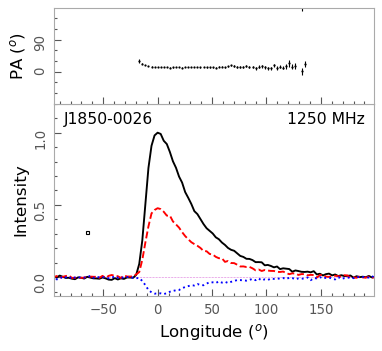}
  \includegraphics[angle=0,width = 0.33\textwidth] {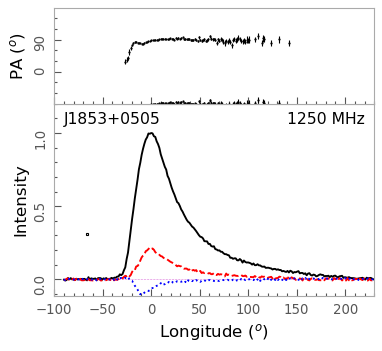}
  \includegraphics[angle=0,width = 0.33\textwidth] {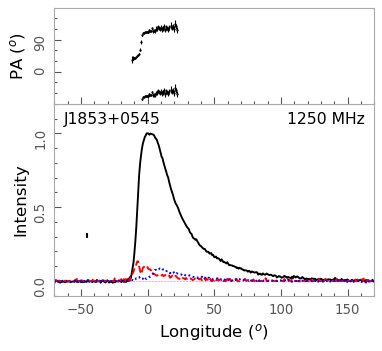}

  \caption{The same as Figure~\ref{fig:intprof_S4} but for 3 pulsars
    with scattering tails as examples of 22 pulsars in
    Figure~\ref{fig:intprof_Scat}.  }
  \label{fig:intprof_Scat4}
\end{figure*}

\subsection{Very wide pulse profiles}

Pulsar profiles generally occupy a few percent of a period,
i.e. 10$^\circ$ to 30$^\circ$ in the rotation phase. Sensitive
observations can reveal the weak emission components \citep[e.g. PSR
  J2124-3358 in][]{mh04}. From our FAST observations, we get very wide
profiles for 21 pulsars (see Figure~\ref{fig:intprof_WP}), occupying
the rotation phase by more than 180$^\circ$, as shown in
Figure~\ref{fig:intprof_WP4} for 2 pulsars as examples. Emission
occupies almost all rotation phase for PSR J1710+4923 in
Figure~\ref{fig:intprof_WP4}, and PSRs J1916+0748 and J2007+2722 in
Figure~\ref{fig:intprof_WP}. Most of the pulsars are millisecond
pulsars, but there do have 4 normal ones, PSRs J1851+0418, J1903+0925,
J1916+0748 and J1932+1059 in Figure~\ref{fig:intprof_WP}.

These pulsars exhibit diverse polarization features, regardless of
normal pulsars or millisecond ones. Some of them show S-shaped PA
curves, e.g., PSR J2234+0944 in Figure~\ref{fig:intprof_WP}. The
rotating vector model can fit some of PA curves very well. About half
of them show orthogonal modes, such as PSRs J0337+1715 and J1916+0748
in Figure~\ref{fig:intprof_WP}. The orthogonal modes and PA swing
  of PSR J1710+4923 in Figure~\ref{fig:intprof_WP4} regularly repeats 4
  times within one rotation. Highly linear polarized components
appear in PSRs J1630+3734 and J1709+2313 in
Figure~\ref{fig:intprof_WP}. Strong circular polarization is detected
for some components of PSRs J1835-0114 and J1955+2908 in
Figure~\ref{fig:intprof_WP}.

The profile morphology and polarization are very much complicated for
these very wide profiles. For normal pulsars, the wide profiles
  mean that inclination angles are small, and the sight line remains
  within the emission beam. Much more discrete radiating sources can be
  detected inside the beam, and hence the profiles have the
  complexity. This can be also the case for MSPs, but with much larger
  polar caps. It means that wide profiles are much more likely to be
  detected, even for large inclination angles. The wide profiles might
  also result from emission from both poles, such as PSR J1932+1059 in
  Figure~\ref{fig:intprof_WP}.

\begin{figure*}
  \centering
  \includegraphics[width = 0.47\textwidth] {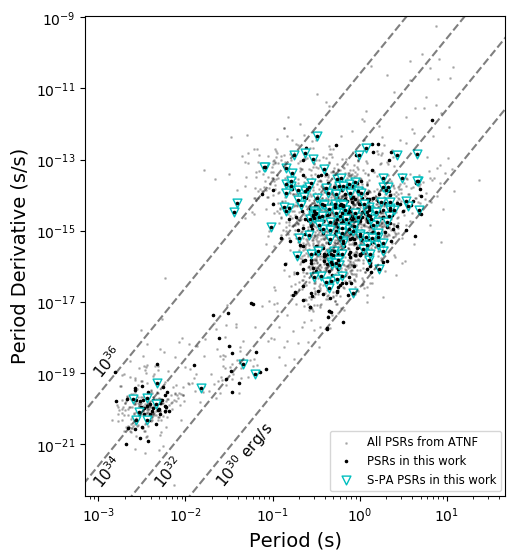}
  \includegraphics[width = 0.47\textwidth] {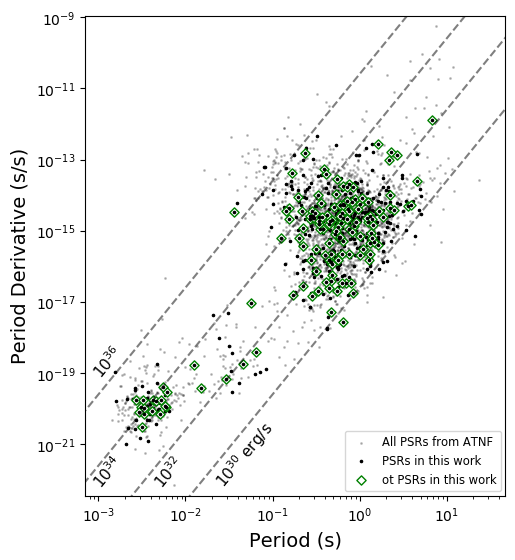}\\
  \includegraphics[width = 0.47\textwidth] {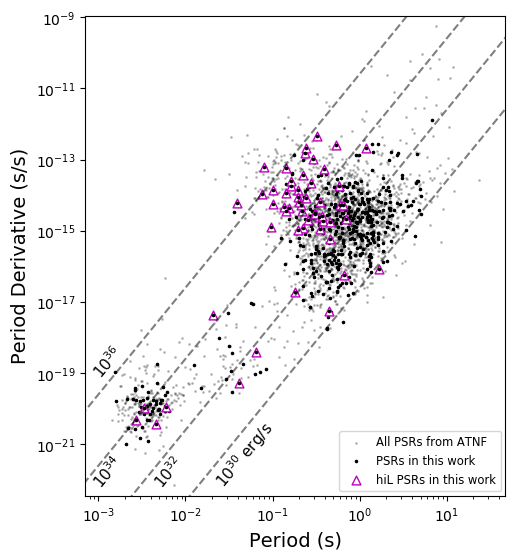}
  \includegraphics[width = 0.47\textwidth] {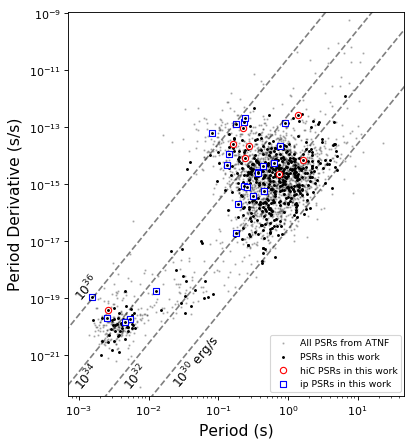}\\
  \caption{Pulsars with different polarization properties in the
      period and period derivative diagram. Grey dots are plotted for
      the known pulsars taken from the ATNF pulsar catalogue
    \citep{mhth05}, blank dots are for 502 of the 682 pulsars
      reported in the current work, which have both period and period
      derivative available. Pulsars with S-shaped PA curve are
      indicated by ``$\triangledown$'' in the top left panel, pulsars
      with orthogonal modes by the ``$\diamond$'' in the top right
      panel, pulsars with highly linear polarization by
      ``$\vartriangle$'' in the lower left panel, and pulsars with
      highly circular polarization by ``$\circ$'' and inter pulses by
      ``$\square$'' in the lower right panel. }
  \label{fig:P_Pdot}
\end{figure*}

\subsection{Profiles with scattering tails}

Distant pulsars with large dispersion measures tend to suffer from
interstellar scattering, showing not only a long tail of the pulse
profile but also a flat PA curve in the tail part \citep{lh03}.

In our FAST pulsar database, we observed a large number of pulsars
with scattering features, as presented by Jing et al (2023, in
preparing). Here we present the polarization profiles for 3 pulsars in
Figure~\ref{fig:intprof_Scat4} as examples of the 22 ones in
Figure~\ref{fig:intprof_Scat}. Apparently all polarization profiles
tend to have flat PA curves in the tail part, confirming the
conclusion first proposed by \citet{lh03}. In the leading wing before
the profile peak, diverse polarization features can be seen, such as
orthogonal modes for PSRs J1852+0031 and J1853+0545 in
Figure~\ref{fig:intprof_Scat}, which means no significant
depolarization from the scattered signals in these phase
bins. Circular polarization of these profiles generally has one
sense. Although affected by scattering, the circular polarization of
PSR J1855+0527 in Figure~\ref{fig:intprof_Scat} is the most prominent
among all the pulsars presented in this paper, reaching 45.5\%.

\subsection{Other Pulsars}

Polarization profiles of the other 298 pulsars are shown in
Figure~\ref{fig:intprof}, which are about 44$\%$ of all pulsars
reported in this paper. Most of our FAST observations give an
unprecedented S/N of profiles. The diverse polarization features may
be caused by complicated emission beam structures, such as multiple
emission patches from different parts or different heights in pulsar
magnetosphere.  Superposition of radio emission from the emission of
various patches leads to depolarization or complicated polarization
behaviors.

\begin{figure}
  \centering
  \includegraphics[width = 0.41\textwidth] {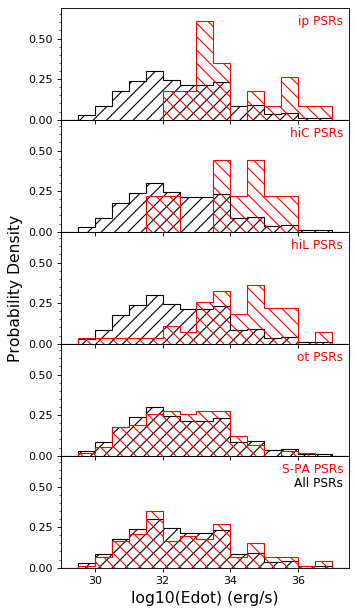}
  \caption{Distribution of pulsars with different properties against
    the Edot. Grey histogram is made for 502 pulsars in the current
    work with available Edot. From bottom to top panels, red steps are
    for pulsars with S-shaped PA curve, with orthogonal modes, with
    highly linear polarization or circular polarization, and with
    inter pulses.}
  \label{fig:Edot_hist}
\end{figure}

\section{Further understanding of pulsar profiles}

Diverse profile morphology and different polarization properties of
these pulsars need to be further understood in pulsar physics. Here,
we first do statistic analysis on profile widths, linear and circular
polarization. With the large sample of pulsars with S-shaped PA curves,
we then work on the emission geometry. Finally, frequency dependencies
of profile width, linear and circular polarization are also explored
for some pulsars with the sensitive FAST data.

\begin{figure}
  \centering
  \includegraphics[width = 0.49\textwidth] {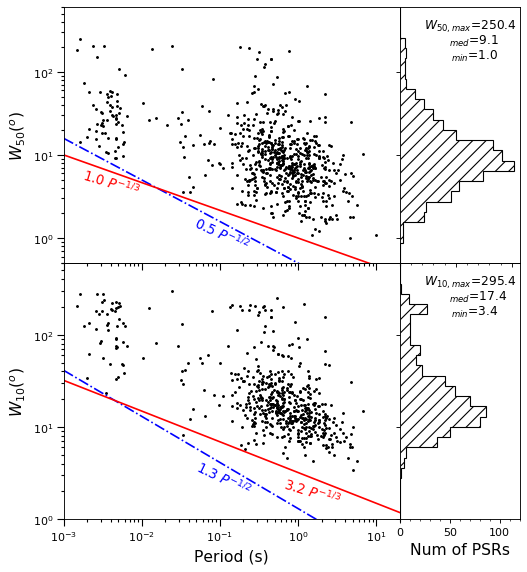}
  \caption{Distributions of profile widths $W_{10}$ and $W_{50}$ of
    682 pulsars. They have the lower boundaries with $P^{-1/2}$ and
    $P^{-1/3}$. Histograms of profile widths are shown in the right
    subpanels with indications of the maximum, median and minimum of
    pulse widths. }
  \label{fig:wp}
\end{figure}

\subsection{Statistics on pulse profile properties}

\subsubsection{Polarization properties versus fundamental parameters of pulsars}

To investigate if the profile properties are related to the period and
period derivative, we mark pulsars with different properties on the
period-period derivative diagram, as shown in
Figure~\ref{fig:P_Pdot}.  Pulsars with
S-shaped PA curves or orthogonal modes have Edots over a wide range
with no preference. These features are clearly demonstrated in the
histograms of Edot for different types of pulsars in
Figure~\ref{fig:Edot_hist}. MSPs exhibit all the types of polarization
features as the normal pulsars. Such results imply that pulsar
profiles are fundamentally related to the energy lose of rotating
pulsars.

\subsubsection{Statistics on profile width}

Our FAST observations provide accurate measurements of profile widths
for a large number of pulsars. The distributions of pulse profile
widths are obtained in Figure~\ref{fig:wp}. The typical widths are
9.1$^\circ$ for $W_{50}$ and 17.4$^\circ$ for $W_{10}$.

Pulse width in principle is related to the emission geometry and beam
width. The beam width should be the lower limit of pulse width in the
case of central cut of the line of sight on the beam. As shown in
Figure~\ref{fig:wp}, the lower boundary of pulse widths was plotted
with power-law relation, $0.5^\circ P^{-1/2}$ for $W_{50}$ or
$1.3^\circ P^{-1/2}$ for $W_{10}$, similar to that for core components
\citep{ran90} and/or conal components
\citep[e.g.][]{mgm12,mbm+16}. The scaling relations with $P^{-1/2}$
are understandable for the dipole opening angles of the magnetic field
lines by assuming a constant emission height
\citep[e.g.][]{big90,kxl+98}. Actually, pulsar emission at a given
frequency is not produced at the same height of pulsar
magnetosphere. With the data set of 682 pulsars, the lower boundary of
observed profile width $W_{10}$ follow $3.2^\circ P^{-1/3}$ or
$1.0^\circ P^{-1/3}$ for $W_{50}$, as shown in
Figure~\ref{fig:wp}. These dependencies have a slope less steeper than
$P^{-1/2}$, which meet with those from \citet{jka19}.

\subsubsection{Linear polarization}

\begin{figure}
  \centering
  \includegraphics[width = 0.49\textwidth] {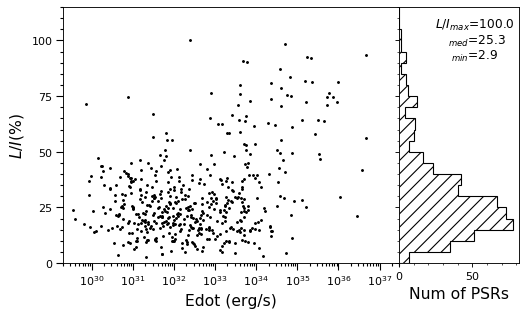}
  \caption{Distribution of the fractional linear polarization and its
    correlations with Edot and pulsar age.}
  \label{fig:Ldis}
\end{figure}

The pulsars have a fractional linear polarization of around 25.3\% as
observed by FAST at 1250~MHz, as shown in the right panel of
Figure~\ref{fig:Ldis}.

The young pulsars with large Edots tend to have a large fraction of
linear polarization at 4.9GHz and 1.4GHz \citep{vlk98, cmk01}, but not
at lower frequencies of 408~MHz and 774~MHz \citep{vlk98, hdvl09}. The
gradual transition of highly linear polarization is at Edot of about
$10^{34}-10^{35}$ erg/s \citep{wj08, mbm+16,jk18}. Our observation
results at 1250~MHz by FAST confirm the correlation and transition, as
shown in Figure~\ref{fig:Ldis}. The linear polarization is about 25\%
for pulsars with a low Edot. When Edot increases to about $10^{33}$
erg/s, more pulsars tend to have higher linear polarization. When Edot
further increases to about $10^{35}$ erg/s, the linear polarization is
more likely to be higher.

\begin{figure}
  \centering
  \includegraphics[width = 0.44\textwidth] {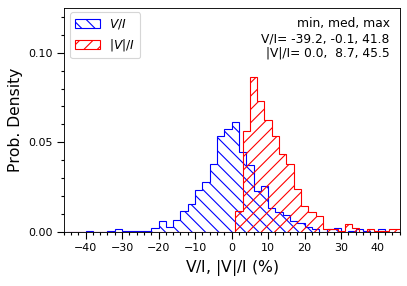}

  \caption{The distribution of the fraction of circular polarization
    observed by the FAST in this paper. }
  \label{fig:Vdis}
\end{figure}

\subsubsection{Circular polarization}

Pulsar radio emission at 1250~MHz usually has an absolute circular
polarization of about 8.7\%. There is no preference for the handness
of circular polarization, as shown by the symmetric distribution of
V/I in Figure~\ref{fig:Vdis}. No significant correlation of $|V|/I$
with pulsar period, period derivative, Edot and age has been found
from our observation data.

\begin{figure*}
  \centering
  \includegraphics[angle=0,width = 0.37\textwidth] {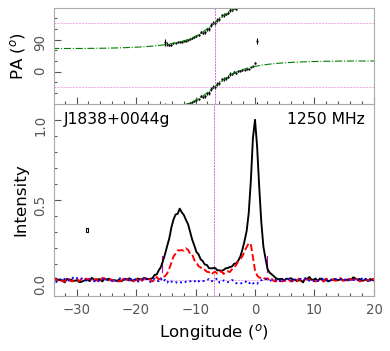}
  \includegraphics[angle=0,width = 0.4\textwidth] {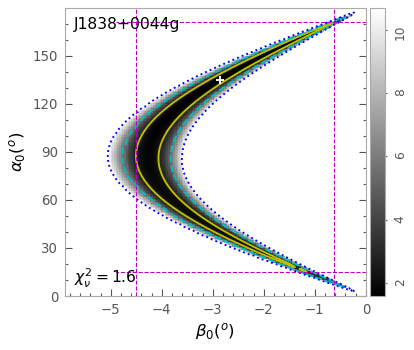}\\
  \includegraphics[angle=0,width = 0.37\textwidth] {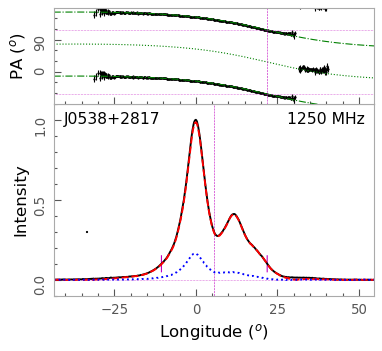}
  \includegraphics[angle=0,width = 0.4\textwidth] {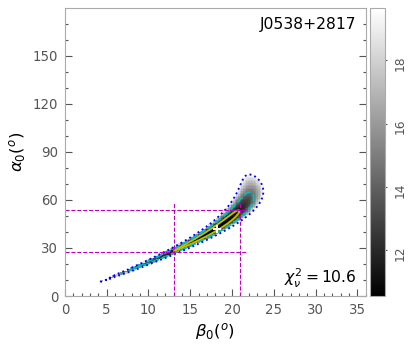}

  \caption{The RVM fittings for PSRs J1838+0044g and J0538+2817 as two
    examples. {\it Left plots:} Polarized pulse profiles with S-shaped
    PA curve and orthogonal modes. For each pulsar, the total
    intensity, linear and circular polarization are represented by
    black solid, red dashed and blue doted lines in the bottom
    sub-panel. The bin size and $3\sigma$ are marked inside the
    sub-panel, here $\sigma$ is the standard deviation of off-pulse
    bins. The short doted vertical lines represent the boundaries for
    10\% the peak intensity and the long vertical line for the profile
    center. In the top panel, dots with error-bar are measurements of
    polarization position angles for linear polarization intensity
    exceeding 3$\sigma$. The dash-doted line indicates the RVM
    solution, and the doted green one for the orthogonal mode with a
    $90^\circ$ shift. The vertical and horizonal doted lines in this
    subpanel are for the phase and position angle of the steepest
    position of the PA curve, i.e., $\phi_0$ and $\psi_0$. {\it Right
      plots:} $\chi^2$ distribution of the RVM fitting. For each
    pulsar, the position at the minimum of $\chi^2$ is indicated by
    ``+" with a value of $\chi^2_\nu$ marked in the panel. The solid,
    dashed and doted contour lines are for $\Delta \chi^2$=1.0, 4.0,
    and 9.0, respectively. Projection of the red line bounded region
    (1-sigma region) on the $\alpha_0$ and $\beta_0$ axes is indicated
    by the horizonal and vertical dashed lines.}
  \label{fig:PAGeo2}
\end{figure*}

\subsection{Pulsar geometry}

The geometry of a pulsar can be described by an inclination angle
$\alpha$ that is the angle between the magnetic axis and its rotation
axis, and an impact angle $\beta$ that is the minimum angle of a sight
line with respect to the magnetic axis. The line of sight then has an
angle of $\zeta = \alpha+\beta$ from the rotation axis. For the
geometry, one can find that the polarization position angle $\psi$
swings with respect to the rotation phase $\phi$ by following
\citep{rc69},
\begin{equation}
  \tan(\psi-\psi_{0})=
  \frac{\sin \alpha \sin(\phi-\phi_0)}{\sin(\zeta)\cos\alpha-\cos(\zeta)\sin\alpha\cos(\phi-\phi_0)}.
  \label{eq:rvm}
\end{equation}
This is called the rotating vector model (RVM). Here, $\phi_0$ and
$\psi_0$ represent the phase and position angle where the PA swing has
the largest gradient.

\subsubsection{Analysis methods}

Pulsar geometries can be derived through RVM fitting. Considering that
the PAs are defined to increase clockwise in the normal RVM but
counter-clockwise in practical observations, we therefore relate
$\alpha$ and $\beta$ in the model to the original $\alpha_0$ and
$\beta_0$ \citep{ew01} via $ \alpha = 180^{\circ}-\alpha_0$ and $\beta
= -\beta_0$.

Instead of fitting to the PA curves, two-dimensional fitting is
performed directly to Stokes Q and U by using PSRMODEL program of
PSRCHIVE. We first search a 61$\times$61 grid with $\alpha_0$
extending from $0^\circ$ to $180^\circ$, and $\beta_0$ extending from
$0^\circ$ to $60^\circ$ or $-60^\circ$ to $0^\circ$ depending the
sense of PA curve. That is to say, the increase PA corresponds to a
negative $\beta_0$, and vise versa. With the fitting result of each
grid point, a 2-D $\chi^2$ surface is plotted. According to which, the
region for $\beta_0$ and occasionally that for $\alpha_0$ are adjusted
for a second search over much finer 181$\times$181 grids. In
  general, the searching region for $\beta_0$ is reduced in the second
  iteration, but it is extended to $0^\circ$ to $90^\circ$ or
  $-90^\circ$ to $0^\circ$ for some pulsars with flat PA curves.
Then, the 2-D $\chi^2$ surface is obtained, which is smoothed to avoid
the lock down to $\alpha_0 \rightarrow 180^\circ$ or $0^\circ$ and
$\beta_0 \rightarrow 0^\circ$. Finally, a continuous region containing
the minimum of $\chi^2$ (i.e., $\chi_\nu^2$) is found, as shown in the
right plots of Figure~\ref{fig:PAGeo2}. The optimal $\alpha_0$ and
$\beta_0$ are then obtained for the minimum $\chi_\nu^2$. With the
$\alpha_0$ and $\beta_0$ fixed, a new fitting can give $\phi_0$ and
$\psi_0$.

Uncertainties of $\alpha_0$ and $\beta_0$ are estimated from the 2-D
$\chi^2$ surface. The regions lying at $\Delta \chi^2$=1, 4 and 9
above the minimum $\chi_\nu^2$ are indicated by the solid, dashed and
doted contour lines, respectively, as shown in
Figure~\ref{fig:PAGeo2}. Its projection on $\alpha_0$ and $\beta_0$
axes forms one-dimensional intervals. The intervals for $\Delta
\chi^2$=1, instead of the 2-D region, contain 68.3\% of normally
distributed data. These confidence intervals around the minimum are
employed to estimate the lower and upper boundaries for the
uncertainties in $\alpha_0$ and $\beta_0$. With these
tangential $\alpha_0$ and $\beta_0$ fixed in new RVM fittings, 4 pairs
(occasionally 2 or 3) of $\phi_0$ and $\psi_0$ are obtained. Their
maximum offsets from the optimal $\phi_0$ and $\psi_0$, i.e. the one
estimated at $\chi_\nu^2$, are used to estimate the lower and upper
boundaries for the uncertainties in $\phi_0$ and $\psi_0$.

During the fitting, $90^\circ$ discontinuities are inserted for
  the orthogonal modes. But the discontinuities are not of $90^\circ$
  for 12 pulsars, e.g. PSR J0358+5413 in Figure~\ref{fig:PAGeo}, and
  the small range of longitudes for the minor mode are unweighed.
The best fitted PA curves are represented by the dashed lines in the
PA subpanel of Figure~\ref{fig:PAGeo2}.

\subsubsection{The RVM solutions}

The so-obtained geometry parameters of all the 190 pulsars are listed
in Table~\ref{table:Geo_para}, with their fittings shown in
Figure~\ref{fig:PAGeo}. It is obvious that $\alpha_0$ can not be well
constrained for most of the pulsars with a value in the range of
$0^\circ$ to $180^\circ$, and only 69 of them have uncertainties less
than $90^\circ$, as listed in
Table~\ref{table:Geo_para}. Nevertheless, $\beta_0$ is much better
constrained.

\begin{figure*}
  \centering
  \includegraphics[angle=0,height = 0.18\textheight] {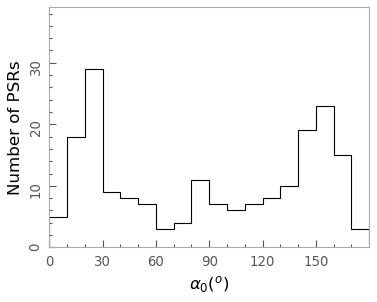}
  \includegraphics[angle=0,height = 0.18\textheight] {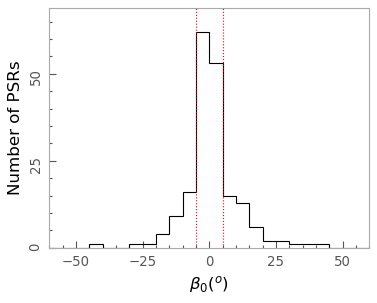}
  \includegraphics[angle=0,height = 0.18\textheight] {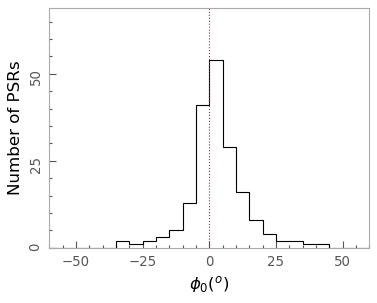}

  \caption{Histograms of the inclination angles $\alpha_0$, the impact
    angles $\beta_0$ and the phases of PA curves with the steepest
    gradients $\phi_0$ for the 190 pulsars.  }
  \label{fig:geo-hist}
\end{figure*}

Figure~\ref{fig:geo-hist} shows the distributions of $\alpha_0$,
$\beta_0$ and $\phi_0$ for these 190 pulsars. Apparently, magnetic
axes are most probably inclined at angles of about $25^\circ$ or
$155^\circ$ with respect to the rotation axes. There is also a good
number of orthogonal rotators, i.e., those around $\alpha_0 \sim
90^\circ$. Our sight lines most likely detect pulsars with
$|\beta_0|<5^\circ$, as shown by the significant overstep in the
histogram of $\beta_0$. Some pulsars have very flat PA curves,
  whose $\beta_0$ can not be well constrained. For example, it extends
  from $13^\circ$ to $90^\circ$ for PSR J1852-0118, $-89^\circ$ to
  $5^\circ$ for PSR J1916+0748 and $-90^\circ$ to $-18^\circ$ for PSR
  J2017+0603, as listed in Table~\ref{table:Geo_para} and shown
 in Figure~\ref{fig:PAGeo}. The minimum of $|\beta_0|$ is
only 0.053$^\circ$ for PSR J2113+4644 in Figure~\ref{fig:PAGeo},
indicating the sweep of sight line nearly across the magnetic axis. We
get $\beta_0$ only -0.002 for PSR J1856+09, but it has a large
uncertainty since the discontinuity in PA might be caused by orthogonal
emission modes.

%
\begin{table*}
  \centering
  \caption{Geometry parameters for 190 pulsars. -- {\it to be continued} --} 
  \label{table:Geo_para}
  \tabcolsep 3pt
  \footnotesize
  \begin{tabular}{llrrrrrrrrl}
    \hline
    \hline
    PSR    & P      & $\alpha_0$       & $\beta_0$         & $\phi_0$            & $\psi_0$            & $\chi^2_\nu$& $\rho$       & $r_{\rm geo}$      & $r_{\rm delay}$   &  Note\\
           & (s)    & ($^\circ$)        & ($^\circ$)         & ($^\circ$)           & ($^\circ$)           &       & ($^\circ$)         &  (km)             & (km)            &       \\
    (1)    &  (2)   &  (3)             &  (4)              &  (5)               &     (6)             &  (7)  &  (8)              &   (9)             &  (10)           & (11)    \\
\hline
J0006+1834 & 0.693 &   9.3            & 32.8              & -16.1               &  82.2               &  2.5 &  -                   &  -                 &   -                & l.un \\   
J0058+4950 & 0.996 &  21$_{-16}^{+60}$  &  1.7$_{-1.3}^{+3.0}$ &   3.7$_{-0.2}^{+0.2}$  & -17.2$_{-2.1}^{+1.9}$  &  7.0 &  3.5$_{-2.7}^{+6.0}$   &  82$_{-77}^{+516}$   &  231$_{-67}^{+70}$   &  \\
J0244+14   & 2.127 &  52$_{-47}^{+123}$ &  1.5$_{-1.3}^{+0.7}$ &  -1.4$_{-0.2}^{+0.3}$  &  88.3$_{-7.9}^{+4.7}$  &  1.4 &  3.5$_{-3.1}^{+1.1}$   & 176$_{-174}^{+121}$  &  neg               &  \\
J0358+5413 & 0.156 & 175              &  0.7              &   9.3               & -16.5                &169.2 &  1.7                &   3                &  401               & wo.ot, s.un \\
J0402+4825 & 0.512 & 159$_{-122}^{+17}$ & -2.4$_{-5.3}^{+1.9}$ &   8.6$_{-0.6}^{+1.2}$  & -16.7$_{-5.6}^{+6.7}$   &  3.6 &  4.0$_{-3.2}^{+7.5}$   &  55$_{-53}^{+394}$   & 192$_{-71}^{+126}$   &  \\
J0453+1559 & 0.045 &  23$_{-14}^{+38}$  &  0.7$_{-0.4}^{+0.8}$ &  -5.6$_{-0.1}^{+0.1}$  &  86.4$_{-0.7}^{+1.2}$   & 26.0 &  3.2$_{-1.9}^{+3.9}$   &   3$_{-3}^{+12}$     &  neg             &  \\
J0454+5543 & 0.340 &  8.0$_{-2.5}^{+9.5}$& 0.7$_{-0.2}^{+0.8}$ &   7.6$_{-0.1}^{+0.1}$  &  61.6$_{-0.4}^{+0.1}$   & 460.2 &  2.3$_{-0.7}^{+2.7}$  &  12$_{-6}^{+43}$     & 570$_{-1}^{+4}$    & wo.ot \\
J0457+23   & 0.504 &  20$_{-15}^{+61}$  &  2.7$_{-2.0}^{+5.2}$ &  16.7$_{-0.2}^{+1.2}$  &  70.6$_{-7.6}^{+0.9}$   &  5.8 &  7.3$_{-5.4}^{+13.0}$  & 178$_{-166}^{+1177}$  & 862$_{-191}^{+227}$ &  \\
J0538+2817 & 0.143 &  42$_{-14}^{+11}$  & 18.2$_{-5.0}^{+2.6}$ &  21.8$_{-1.6}^{+1.7}$  & -62.5$_{-4.2}^{+3.5}$   & 10.6 & 22.0$_{-6.0}^{+3.1}$   & 460$_{-218}^{+141}$   & 482$_{-46}^{+50}$  &  \\
J0543+2329 & 0.246 & 154              &  2.5               &  26.1                &  68.3                 &  1.7 &  5.7              &   54                & 1087            & l.un  \\
J0605+3757 & 0.0027&  11              & -2.3               &  31.9                & -87.2                 &  0.9 &  -                &  -                  &  -              & no.w, l.un\\
J0608+1635 & 0.945 & 131$_{-126}^{+43}$ &  2.5$_{-2.3}^{+2.5}$ &  -3.3$_{-0.4}^{+1.6}$  & -85$_{-25}^{+4}$        &  0.7 &  4.9$_{-4.3}^{+2.7}$   & 151$_{-149}^{+212}$  & neg              &  \\
J0613+3731 & 0.619 &  18$_{-13}^{+70}$  &  2.6$_{-1.8}^{+5.8}$ &  -3.5$_{-1.1}^{+0.6}$  & -76.9$_{-4.5}^{+7.5}$   & 15.5 &  4.2$_{-3.0}^{+8.8}$   &  71$_{-65}^{+620}$   & neg               &  \\
J0631+1036 & 0.287 &  16$_{-8}^{+9}$   &  2.3$_{-1.0}^{+1.1}$  &  11.2$_{-0.1}^{+0.1}$  &  54.8$_{-1.2}^{+0.8}$   & 35.2 &  4.3$_{-2.0}^{+2.2}$   &  35$_{-24}^{+45}$   &  508$_{-6}^{+9}$    &  \\
J0659+1414 & 0.384 &  17$_{-8}^{+17}$  &  5.7$_{-2.7}^{+4.9}$  &  12.8$_{-0.3}^{+0.8}$  &  17.2$_{-2.9}^{+1.0}$   &  5.6 &  7.5$_{-3.5}^{+6.5}$   & 143$_{-103}^{+355}$  & 1172$_{-23}^{+67}$  & wo.ot \\
J0711+0931 & 2.428 & 136$_{-119}^{+31}$ & -2.8$_{-1.4}^{+1.9}$  &   0.5$_{-0.2}^{+0.3}$  &  13.7$_{-3.0}^{+4.3}$   &  3.7 &  3.6$_{-2.4}^{+1.7}$  & 206$_{-183}^{+238}$  &  309$_{-113}^{+153}$ &  \\
J0944+4106 & 2.229 & 133$_{-31}^{+40}$  &  0.20$_{-0.17}^{+0.05}$& -3.7$_{-0.1}^{+0.1}$  & -12.8$_{-0.3}^{+0.4}$   & 41.5 &  3.8$_{-3.2}^{+1.3}$  & 217$_{-211}^{+172}$  &  neg               &  \\
J1017+3011 & 0.452 &  12$_{-9}^{+101}$  & -1.5$_{-51.0}^{+1.0}$ &  139$_{-33}^{+15}$    &  23$_{-1}^{+105}$       &  2.0 &  -                  &  -                 &  -               &  \\
J1236-0159 & 3.597 & 120$_{-113}^{+51}$ &  1.7$_{-1.5}^{+0.4}$  &  -4.3$_{-0.1}^{+0.2}$  &  20.9$_{-4.3}^{+1.4}$   &  8.1 &  5.1$_{-4.4}^{+0.9}$  & 630$_{-617}^{+231}$  &  neg               &  \\
J1312+1810A& 0.033 &  91$_{-84}^{+58}$  & -7.4$_{-0.7}^{+6.3}$  &  16.0$_{-0.7}^{+0.6}$  & -66.8$_{-3.5}^{+3.8}$   &  1.6 & 21.5$_{-18.9}^{+2.1}$ & 102$_{-100}^{+10}$   & 13.3$_{-13.2}^{+13.1}$ &  \\
J1538+2345 & 3.449 & 171$_{-19}^{+6}$   & -0.4$_{-0.8}^{+0.3}$  &  -3.0$_{-0.1}^{+0.1}$  &  55.5$_{-0.5}^{+0.5}$   & 44.6 &  0.9$_{-0.6}^{+1.7}$ &  17$_{-15}^{+134}$    & neg               &  \\
J1641+3627A& 0.010 & 166$_{-164}^{+13}$ & -3.9$_{-17.2}^{+3.9}$ &   5.9$_{-2.5}^{+6.0}$  &  60$_{-7}^{+18}$        &  1.0 &   9$_{-8}^{+26}$     &   5$_{-5}^{+75}$     & 25$_{-6}^{+13}$     &  \\
J1657+3304 & 1.570 &  28$_{-24}^{+147}$ & -1.5$_{-2.1}^{+1.3}$  &  -0.2$_{-0.3}^{+0.8}$  &  -18.1$_{-5.1}^{+13.9}$  &  2.2 &  2.7$_{-2.3}^{+3.3}$ &  73$_{-71}^{+300}$   & neg                &  \\
J1736+05   & 0.999 & 167$_{-165}^{+12}$ & -0.9$_{-4.0}^{+0.9}$  &  -2.0$_{-0.3}^{+0.9}$  &   89.1$_{-2.4}^{+13.0}$  &  2.9 &  1.3$_{-1.2}^{+5.1}$ &  13$_{-11}^{+263}$   & neg                &  \\
J1800-0125 & 0.783 & 125$_{-120}^{+51}$ & 11.3$_{-10.4}^{+4.0}$ &   4.5$_{-3.0}^{+2.5}$  &   56$_{-10}^{+12}$      &  1.2 &  14$_{-13}^{+5}$     & 1078$_{-1070}^{+849}$ & 617$_{-495}^{+405}$  &  \\
J1806+2819 & 0.015 &  33$_{-29}^{+143}$ & 13.5$_{-11.7}^{+26.4}$ &  19.9$_{-4.8}^{+6.5}$ &   25.1$_{-11.7}^{+5.6}$  &  1.9 &  29$_{-26}^{+26}$    &  83$_{-82}^{+218}$   &  3$_{-3}^{+21}$      &  \\
J1807+04   & 0.798 &  21$_{-16}^{+81}$  & -0.8$_{-1.4}^{+0.6}$  &   0.7$_{-0.2}^{+0.1}$  &  36.6$_{-3.0}^{+2.6}$   & 16.9 &  2.3$_{-1.7}^{+4.1}$  &  28$_{-26}^{+189}$   & neg                &  \\
J1807+0756 & 0.464 &  17$_{-13}^{+158}$ &  2.2$_{-2.1}^{+8.8}$  &  -0.8$_{-2.7}^{+2.0}$  &  -49$_{-14}^{+18}$      &  3.1 &  3.9$_{-3.0}^{+11.2}$ &  46$_{-44}^{+648}$   & neg                &  \\
J1808+00   & 0.425 & 150$_{-76}^{+21}$  &  3.7$_{-2.6}^{+3.8}$  &   1.1$_{-0.6}^{+0.3}$  &  -49.1$_{-1.6}^{+3.9}$  &  7.1 &  6.2$_{-4.3}^{+6.6}$  & 108$_{-97}^{+355}$   & 192$_{-54}^{+30}$    &  \\
J1809+17   & 2.066 & 166$_{-164}^{+12}$ &  1.2$_{-1.0}^{+6.1}$  &   2.1$_{-2.7}^{+0.4}$  &   74$_{-3}^{+28}$       &  3.4 &  1.5$_{-1.3}^{+6.7}$  &  31$_{-30}^{+895}$   & 1213$_{-1156}^{+179}$ &  \\
J1823-0154 & 0.759 &  36$_{-23}^{+65}$  &  0.8$_{-0.5}^{+0.6}$  &   1.0$_{-0.1}^{+0.1}$  &  -9.5$_{-1.4}^{+0.7}$   & 13.4 &  2.3$_{-1.4}^{+1.5}$  &  25$_{-22}^{+47}$    & 190$_{-5}^{+8}$      & wo.ot \\
J1824-0127 & 2.499 & 151$_{-144}^{+21}$ & -1.1$_{-1.4}^{+0.8}$  &   4.8$_{-0.2}^{+0.1}$  &  -50.2$_{-3.0}^{+1.2}$  &  2.7 &  2.6$_{-2.0}^{+2.8}$  & 116$_{-109}^{+378}$   & 822$_{-98}^{+50}$    & wo.ot \\
J1828+0625 & 0.0036&  13$_{-10}^{+65}$  &  4.4$_{-3.2}^{+18.2}$ &   2.8$_{-2.2}^{+9.2}$  &  -71.5$_{-20.7}^{+4.3}$  &  1.5 &  16$_{-12}^{+50}$    &  6.3$_{-6.0}^{+97.8}$ & 8.9$_{-1.1}^{+1.3}$   &  \\
J1832+27   & 0.631 & 146$_{-142}^{+32}$ &  4.2$_{-3.9}^{+4.6}$  &   4.3$_{-2.0}^{+4.2}$  &  -65$_{-34}^{+14}$      &  1.2 &  5.3$_{-4.9}^{+5.4}$  & 116$_{-115}^{+359}$   & 130$_{-130}^{+554}$  &  \\
J1833-0209 & 0.583 & 146$_{-141}^{+30}$ & -1.7$_{-1.9}^{+1.4}$  &  -4.1$_{-0.5}^{+0.5}$  &  -49.3$_{-6.1}^{+5.8}$   &  1.6 &  4.7$_{-4.1}^{+3.8}$ &  87$_{-86}^{+194}$    &  neg               &  \\
J1834+10   & 1.172 &  17$_{-11}^{+60}$  & -2.1$_{-4.6}^{+1.3}$  &  -8.0$_{-1.8}^{+0.3}$  &  -43.8$_{-6.6}^{+1.5}$   &  6.1 &  4.2$_{-2.7}^{+10.1}$& 135$_{-117}^{+1451}$   & 532$_{-439}^{+90}$   &  \\
J1837+0053 & 0.473 & 123$_{-89}^{+50}$  &-11.2$_{-6.4}^{+9.8}$  & -30.9$_{-4.8}^{+1.9}$  &  -76$_{-15}^{+5}$       &  1.4 &  35$_{-30}^{+5}$     & 3786$_{-3701}^{+1173}$ &  neg               &  \\
J1838+0044g& 2.203 & 135$_{-120}^{+36}$ &  -2.9$_{-1.6}^{+2.2}$  &  -6.8$_{-0.2}^{+0.3}$  &  -42.2$_{-1.1}^{+1.6}$  &  1.6 &  7.0$_{-5.4}^{+2.8}$ &  709$_{-674}^{+691}$   &  34$_{-34}^{+145}$   &  \\
J1839-0223 & 1.266 & 153$_{-151}^{+25}$ &  -2.6$_{-4.7}^{+2.3}$  &  -7.2$_{-0.9}^{+1.7}$  &   20$_{-7}^{+12}$      &  1.0 &  4.5$_{-4.1}^{+6.1}$  &  167$_{-166}^{+772}$  &  neg              &  \\
J1841+0912 & 0.381 & 114$_{-58}^{+32}$  &   2.3$_{-0.9}^{+0.3}$  &  0.12$_{-0.03}^{+0.03}$& -18.5$_{-0.5}^{+0.7}$   & 43.0 &  6.5$_{-2.5}^{+0.7}$  &  105$_{-66}^{+23}$    &  17$_{-5}^{+5}$     & wo.ot \\
J1842+0257 & 3.088 &  44$_{-36}^{+127}$ &   3.8$_{-3.1}^{+2.3}$  &  -0.2$_{-0.3}^{+0.4}$  & -41.8$_{-3.8}^{+3.1}$   &  5.4 &  5.4$_{-4.3}^{+2.7}$  &  590$_{-568}^{+730}$  &  neg               &  \\
J1842+1332 & 0.471 & 164$_{-14}^{+12}$  &  -43$_{-47}^{+33}$    &   21$_{-55}^{+2}$      & -86$_{-24}^{+1}$        &  2.4 &  50$_{-39}^{+50}$    & 7910$_{-7487}^{+23330}$&  336$_{-336}^{+221}$ &   \\
J1843-0211 & 2.027 & 158$_{-65}^{+18}$  &  -0.8$_{-1.7}^{+0.7}$  &   7.7$_{-0.2}^{+0.6}$  &  46.4$_{-2.2}^{+2.2}$   &  1.6 &  3.5$_{-2.8}^{+5.7}$  &  160$_{-155}^{+966}$  &  512$_{-87}^{+250}$  &  \\
J1843+0119 & 1.267 &  28$_{-25}^{+148}$ &  -0.8$_{-1.2}^{+0.7}$  &   5.2$_{-0.3}^{+0.2}$  &   2.7$_{-5.4}^{+2.7}$   &  1.7 &   -                 &   -                 &  -                & no.w \\
J1843-0000 & 0.880 &  34$_{-12}^{+13}$  &   3.3$_{-1.1}^{+0.9}$  &   0.1$_{-0.1}^{+0.1}$  &   2.5$_{-0.2}^{+0.3}$   & 27.4 &  5.4$_{-1.8}^{+1.6}$  &  171$_{-94}^{+115}$   &  298$_{-9}^{+12}$   &   \\
J1844+00   & 0.460 &  89$_{-77}^{+41}$  &  12.6$_{-9.0}^{+2.3}$  &  22.8$_{-4.1}^{+1.8}$  &   7$_{-13}^{+20}$      &  5.9 & 19$_{-14}^{+2}$       & 1056$_{-986}^{+182}$  & 1304$_{-390}^{+176}$ &  \\
J1846+0051 & 0.434 & 147$_{-142}^{+26}$ &   5.8$_{-4.9}^{+6.2}$  &   2.6$_{-2.4}^{+0.8}$  &  80$_{-4}^{+12}$       &  0.9 &  8$_{-7}^{+8}$        &  177$_{-172}^{+559}$  &  302$_{-229}^{+106}$ &  \\
J1847+0614g& 1.662 & 144$_{-138}^{+30}$ &   2.3$_{-1.9}^{+2.1}$  &   2.2$_{-0.7}^{+0.3}$  &  98.1$_{-3.4}^{+8.4}$   &  2.0 &  4.3$_{-3.5}^{+3.4}$   &  202$_{-195}^{+448}$  &  352$_{-263}^{+164}$ &  \\
J1848+0604 & 2.218 &  32$_{-28}^{+141}$ &  -1.4$_{-1.5}^{+1.2}$  &   1.0$_{-0.5}^{+0.2}$  &  55.1$_{-8.9}^{+3.1}$   &  4.0 &  2.7$_{-2.3}^{+2.6}$   &  103$_{-101}^{+294}$  &  336$_{-225}^{+102}$ &  \\
J1848+0826 & 0.328 &  57$_{-51}^{+37}$  &  -8.8$_{-3.6}^{+7.7}$  &   7.5$_{-3.6}^{+2.0}$  &  52$_{-17}^{+10}$      &  1.1 &  18$_{-16}^{+5}$       &  676$_{-666}^{+411}$  &  244$_{-246}^{+145}$ &  \\
J1848+12   & 0.754 & 153$_{-30}^{+20}$  & -11$_{-12}^{+8}$      &  24.6$_{-0.7}^{+1.4}$  &  82.0$_{-1.2}^{+2.9}$   &  4.2 &  23$_{-17}^{+19}$      & 2715$_{-2506}^{+6376}$ & 1385$_{-162}^{+258}$ & \\
J1849+0106 & 1.832 & 151$_{-145}^{+22}$ &  -1.6$_{-2.2}^{+1.2}$  &   4.5$_{-0.5}^{+0.2}$  &  42.8$_{-5.4}^{+1.5}$   &  2.7 &   3.0$_{-2.3}^{+3.3}$  &  108$_{-103}^{+368}$  &  389$_{-182}^{+110}$  &  \\
J1849+0127 & 0.542 &  94$_{-87}^{+81}$  &  -5.1$_{-0.8}^{+4.7}$  &   2.0$_{-0.4}^{+0.2}$  & -32.1$_{-4.2}^{+2.5}$   &  2.1 &  10.9$_{-9.9}^{+0.2}$  &  424$_{-421}^{+16}$   &  neg              & \\
J1849+0340g& 1.666 &  86$_{-82}^{+90}$  &   1.6$_{-1.5}^{+0.3}$  &   6.8$_{-0.1}^{+0.2}$  &  97.3$_{-3.1}^{+1.1}$   &  3.2 &   9.1$_{-8.5}^{+0.1}$  &  915$_{-910}^{+12}$   &  289$_{-166}^{+174}$ & \\
J1850+1335 & 0.345 &  82$_{-51}^{+53}$  &  11.6$_{-5.6}^{+0.5}$  &   0.5$_{-0.7}^{+0.9}$  & -80.2$_{-4.2}^{+3.6}$   &  2.4 &  12.9$_{-6.1}^{+0.5}$  &  381$_{-275}^{+27}$   &  neg               & \\
J1851-0029 & 0.518 & 159$_{-153}^{+17}$ &  -2.4$_{-5.6}^{+1.9}$  &   0.5$_{-0.9}^{+1.1}$  & -49.3$_{-6.8}^{+8.7}$   &  4.1 &   4.4$_{-3.5}^{+8.2}$  &   67$_{-64}^{+482}$   &  neg               & \\
J1851-0114 & 0.953 &  17$_{-13}^{+157}$ &   1.4$_{-1.1}^{+5.1}$  &  -6.6$_{-1.1}^{+0.2}$  & -18$_{-9}^{+16}$        &  2.8 &   3.1$_{-2.4}^{+8.0}$  &   61$_{-58}^{+711}$  &  neg               & \\

\hline

  \end{tabular}
  \tablenotes{
  \item
    Notes: neg: negative phase shift; ip: interpulse; wo.ot: without
    orthogonal mode in fitting; no.w: no $W_{10}$; l.un: large
    uncertainty;
  \item 
    s.un: small uncertainty.  }
\end{table*}
  
\addtocounter{table}{-1}
\begin{table*}
  \centering     
  \caption{\it -- continued -- }
  \tabcolsep 2pt
  \footnotesize
  \begin{tabular}{llrrrrrrrrl}
    \hline
    \hline
    PSR    & P      & $\alpha_0$       & $\beta_0$           &   $\phi_0$           & $\psi_0$             &$\chi^2_\nu$& $\rho$        & $r_{\rm geo}$       & $r_{\rm delay}$    &  Note\\
           & (s)    & ($^\circ$)        &($^\circ$)           & ($^\circ$)            & ($^\circ$)            &       &($^\circ$)          &  (km)             & (km)             &        \\
    (1)    &  (2)   &  (3)             &  (4)                &  (5)                 &     (6)              &  (7)  &  (8)              &   (9)             &  (10)            & (11)           \\
    \hline
J1852-0118 & 0.451 &  27$_{-18}^{+11}$  &  90$_{-77}^{+0}$      &  66$_{-28}^{+69}$     & -62$_{-58}^{+6}$        &  1.7 &   -                  &   -               &   -               &  \\
J1853+0056 & 0.275 &  26$_{-23}^{+152}$ &   5.2$_{-4.8}^{+10.6}$ &   6.5$_{-0.6}^{+5.1}$  &  77$_{-35}^{+5}$        &  1.2 &   8.7$_{-8.1}^{+12.7}$ &  137$_{-136}^{+698}$  & 614$_{-83}^{+302}$  &  \\    
J1853+0259 & 0.585 &  73$_{-64}^{+77}$  &  -5.9$_{-0.8}^{+4.8}$  & -10.4$_{-0.3}^{+0.4}$  & -43.7$_{-1.4}^{+2.8}$   &  2.1 &    -                 &  -                  &   -             & no.w\\
J1853+0427 & 1.320 & 154$_{-51}^{+18}$  &  -1.1$_{-1.5}^{+0.7}$  & -10.0$_{-0.1}^{+0.1}$  & -47.6$_{-1.2}^{+0.8}$   &  4.8 &   5.4$_{-3.7}^{+6.5}$  &  257$_{-231}^{+992}$  & neg               &  \\
J1854+0306 & 4.557 &  85$_{-78}^{+88}$  &  -2.0$_{-0.2}^{+1.8}$  &   0.6$_{-0.4}^{+0.2}$  & -83.4$_{-9.8}^{+5.1}$   &  0.2 &   3.6$_{-3.2}^{+0.1}$  &  394$_{-388}^{+25}$   & 227$_{-227}^{+349}$ & \\
J1854+0319 & 0.628 &  30$_{-25}^{+146}$ &   8.4$_{-7.2}^{+12.0}$ & -11.5$_{-11.0}^{+2.9}$  &  88$_{-9}^{+42}$       &  1.6 &  12$_{-11}^{+14}$      &  621$_{-612}^{+2150}$ &  neg              &  \\
J1855+0139g& 0.444 &  17$_{-15}^{+162}$ &  11.5$_{-11.0}^{+68.0}$&  -5$_{-41}^{+11}$      & -59$_{-134}^{+15}$      &  0.7 &    -                 &   -                &   -               & no.w\\
J1855+0700 & 0.258 &  93$_{-2}^{+2}$   &  -3.3$_{-2.6}^{+0.5}$  &   1.8$_{-3.1}^{+0.4}$   &  56$_{-41}^{+8}$       &  2.7 &    -                 &   -                &   -               & ip \\
J1856+0102 & 0.620 &  25$_{-20}^{+73}$  &  -0.4$_{-0.6}^{+0.3}$  &  -0.3$_{-0.2}^{+0.1}$  &  34.4$_{-1.7}^{+1.3}$   &  2.6 &   3.4$_{-2.7}^{+4.7}$  &   49$_{-47}^{+223}$   &  neg              & \\
J1856+0243g& 0.546 &  22$_{-18}^{+154}$ &   4.2$_{-3.5}^{+9.5}$  &  -1.0$_{-4.0}^{+1.0}$  &  58$_{-5}^{+20}$       &  1.4 &   6.4$_{-5.4}^{+11.7}$ &  147$_{-144}^{+1030}$  &  neg              &  \\
J1856+0245 & 0.080 &  16$_{-10}^{+50}$  &   15$_{-10}^{+66}$    &  19$_{-3}^{+31}$       & -67$_{-31}^{+2}$       &  1.2 &   -                 &  -                  &  -                & no.w\\
J1856+0912 & 2.170 &   6$_{-3}^{+30}$   & -0.002$_{-0.013}^{+0.001}$& -2.2$_{-0.01}^{+0.01}$ & 19.2$_{-0.5}^{+0.4}$   & 16.7 &   0.5$_{-0.2}^{+2.2}$ &   3$_{-3}^{+104}$    & 369$_{-30}^{+30}$    & \\
J1857+0057 & 0.356 &  94$_{-80}^{+65}$  &  -9.5$_{-0.5}^{+7.1}$  &  2.7$_{-0.5}^{+0.5}$   &  54.2$_{-2.4}^{+2.5}$   &  1.2 &   20$_{-16}^{+1}$     &  963$_{-913}^{+48}$  & 277$_{-55}^{+57}$    & \\
J1858-02   & 1.462 &  75$_{-68}^{+96}$  &  -1.2$_{-0.1}^{+1.0}$  &  3.3$_{-0.1}^{+0.1}$   &  22.1$_{-3.1}^{+1.1}$   &  1.0 &   4.0$_{-3.5}^{+0.1}$  &  154$_{-152}^{+10}$  & 216$_{-48}^{+41}$    & \\
J1858+0239 & 0.197 & 158$_{-155}^{+18}$ &  -6.6$_{-14.8}^{+5.6}$ &  2.3$_{-2.8}^{+4.7}$   &  59$_{-9}^{+13}$        &  1.8 &   9.5$_{-8.3}^{+16.9}$ &  117$_{-115}^{+790}$ & 401$_{-121}^{+197}$  & \\
J1858+0241 & 4.693 &  48$_{-44}^{+128}$ &  -3.2$_{-1.7}^{+2.8}$  &  1.4$_{-1.2}^{+1.2}$   &  25$_{-15}^{+15}$       &  1.6 &   4.4$_{-3.9}^{+2.0}$ &  591$_{-586}^{+668}$  & 315$_{-315}^{+1187}$  & \\
J1858+0310g& 0.372 & 144$_{-33}^{+32}$  &  -71$_{-19}^{+66}$    &  61$_{-17}^{+98}$      & 116$_{-6}^{+69}$        &  1.0 &   -                 &   -                 &  -                & no.w\\
J1859+00   & 0.559 & 158$_{-34}^{+17}$  &  -23$_{-41}^{+18}$    & -16$_{-16}^{+3}$       & -13$_{-14}^{+4}$        &  1.3 &   28$_{-21}^{+42}$    & 2866$_{-2712}^{+15426}$&  neg              &  \\
J1859+0126g& 0.957 & 107$_{-103}^{+68}$ &  -12.0$_{-3.2}^{+11.1}$&  0.5$_{-3.1}^{+2.3}$    & -14$_{-14}^{+10}$       &  0.7 &   -                 &    -                &  -               & no.w\\
J1900-0134 & 1.832 &  49$_{-37}^{+118}$ &   2.6$_{-1.9}^{+1.1}$  &   3.6$_{-0.5}^{+0.4}$   & -52.0$_{-3.5}^{+3.8}$   &  2.4 &   5.2$_{-3.8}^{+1.7}$  &  331$_{-305}^{+253}$  & neg               &  \\
J1900+0227 & 0.374 &  25$_{-21}^{+151}$ &   4.4$_{-3.6}^{+8.2}$  &   2.1$_{-1.3}^{+1.3}$   &  41.4$_{-5.1}^{+6.4}$   &  1.5 &   7.4$_{-6.3}^{+10.7}$ &  136$_{-133}^{+677}$  & 241$_{-106}^{+105}$  & \\
J1900+0634 & 0.389 &  21$_{-16}^{+147}$ &   0.6$_{-0.4}^{+1.4}$  &  -3.1$_{-0.2}^{+0.1}$   & -33$_{-5}^{+4}$        &  3.4 &   1.9$_{-1.4}^{+3.4}$  &   9.3$_{-8.8}^{+64.0}$ &  neg              & \\
J1901+0020g& 0.214 &  86$_{-81}^{+91}$  &  10.5$_{-10.0}^{+2.5}$ &   0.6$_{-2.9}^{+5.4}$   & -29$_{-34}^{+15}$       &  1.1 &   14$_{-13}^{+2}$     &  270$_{-269}^{+84}$   &  83$_{-83}^{+243}$   & \\
J1901+0124 & 0.318 &  39$_{-37}^{+139}$ &  -4.1$_{-4.8}^{+3.8}$  &   7.0$_{-2.4}^{+0.3}$   &  88$_{-21}^{+2}$        &  1.3 &   6.5$_{-6.1}^{+5.7}$ &   89$_{-88}^{+222}$   & 316$_{-159}^{+25}$   & \\
J1901+0234 & 0.885 &  70$_{-67}^{+107}$ &  12.5$_{-12.0}^{+7.5}$ &  -4.3$_{-7.7}^{+5.7}$   &  38$_{-24}^{+39}$       &  0.8 &   14$_{-13}^{+7}$     & 1152$_{-1150}^{+1439}$ &  neg              &  \\
J1901+0320 & 0.636 & 145$_{-93}^{+31}$  &  13.0$_{-11.3}^{+12.4}$& -20.4$_{-7.8}^{+5.2}$   & -19$_{-12}^{+20}$       &  1.0 &   19$_{-16}^{+19}$    & 1480$_{-1458}^{+4592}$  &  neg             & \\
J1901+0413 & 2.662 & 144$_{-131}^{+28}$ &  -3.0$_{-2.6}^{+2.3}$  &   6.6$_{-0.5}^{+0.5}$   & -53.6$_{-3.8}^{+3.7}$   &  1.8 &   6.0$_{-4.6}^{+4.2}$  &  642$_{-604}^{+1193}$  & 991$_{-304}^{+304}$  & \\
J1901+0510 & 0.614 &  82$_{-80}^{+95}$  &   24$_{-23}^{+27}$    &  40$_{-18}^{+6}$        & -11$_{-42}^{+38}$      &  1.4 &  38$_{-37}^{+20}$      & 5936$_{-5927}^{+8016}$ & 4064$_{-2312}^{+753}$ &  \\
J1901+0511 & 4.600 & 139$_{-132}^{+36}$ &  -1.5$_{-1.0}^{+1.3}$  &  -2.5$_{-0.4}^{+0.3}$   &  49.9$_{-6.9}^{+5.4}$   &  1.6 &   2.6$_{-2.3}^{+1.5}$ &  209$_{-205}^{+302}$   & neg                &  \\
J1901+0621 & 0.831 &  20              &  25.5               &  -12.2               &  -0.7                &  0.7 &  32                 &   5693               &  1928             & l.un \\
J1902+0723 & 0.487 & 142$_{-82}^{+34}$  &   3.5$_{-3.1}^{+2.6}$  &   7.4$_{-0.4}^{+0.4}$   &  14.5$_{-3.2}^{+3.7}$  &  1.6 &   8$_{-7}^{+6}$       &  214$_{-211}^{+397}$   & 136$_{-59}^{+57}$    & \\
J1903+0851g& 1.231 &  96$_{-92}^{+80}$  &   3.1$_{-2.9}^{+0.5}$  &  -1.5$_{-0.6}^{+0.9}$   & -16$_{-12}^{+8}$       &  0.5 &   -                 &    -                 &  -                & no.w\\
J1903+2225 & 0.651 & 163$_{-160}^{+14}$ &  -2.1$_{-7.0}^{+1.6}$  &  -4.1$_{-0.8}^{+5.0}$   &  29$_{-5}^{+35}$       &  1.5 &   3.2$_{-2.3}^{+8.4}$  &  43$_{-40}^{+532}$     & neg               & \\
J1904+0800 & 0.263 &  98$_{-94}^{+78}$  &  -5.4$_{-2.4}^{+5.1}$  &  10.7$_{-0.9}^{+1.1}$   & -71$_{-10}^{+12}$      &  2.4 &  12.1$_{-11.3}^{+1.2}$ &  254$_{-253}^{+52}$    & 245$_{-52}^{+60}$    & \\
J1904+1011 & 1.856 &  25$_{-15}^{+40}$  &   3.7$_{-2.2}^{+4.8}$  &  12.0$_{-0.2}^{+0.2}$   & -51.7$_{-0.8}^{+0.7}$   &  2.6 &   9$_{-5}^{+10}$     &  888$_{-736}^{+3097}$   & 314$_{-148}^{+150}$  &  \\
J1905+0616 & 0.989 & 133$_{-123}^{+37}$ &   3.9$_{-2.9}^{+1.8}$  &   1.2$_{-0.6}^{+0.2}$   &  51.8$_{-1.6}^{+6.5}$   &  9.2 &  4.7$_{-3.6}^{+2.1}$  &  145$_{-137}^{+160}$    & 237$_{-123}^{+36}$   & \\
J1905+0709 & 0.648 &  58$_{-41}^{+51}$  &  -9.6$_{-1.8}^{+6.2}$  &  -0.4$_{-0.2}^{+0.2}$   & -17.2$_{-1.1}^{+0.9}$   &  1.7 & 16.8$_{-11.1}^{+3.5}$ & 1209$_{-1071}^{+548}$   & 415$_{-36}^{+36}$    & \\
J1906+0414 & 1.043 &  29$_{-25}^{+145}$ &   2.5$_{-2.1}^{+3.5}$  &  10.9$_{-0.6}^{+0.8}$   & -70.0$_{-6.4}^{+5.0}$   &  1.1 &  6.7$_{-5.7}^{+7.0}$  &  308$_{-302}^{+991}$    & 720$_{-150}^{+191}$  & \\
J1906+0641 & 0.267 &  42$_{-28}^{+29}$  &  -7.3$_{-2.8}^{+4.7}$  &  -3.9$_{-0.4}^{+0.2}$   &   5.2$_{-1.9}^{+1.0}$   & 30.5 & 13.0$_{-8.4}^{+5.8}$  &  299$_{-261}^{+329}$    & 45$_{-23}^{+13}$    & \\
J1906+0649 & 1.286 & 154$_{-152}^{+24}$ &  -12$_{-23}^{+11}$     &  -0.1$_{-3.6}^{+23.8}$  & -70$_{-8}^{+69}$       &  2.6 &  12$_{-11}^{+23}$    & 1270$_{-1267}^{+8982}$   &  neg               & wo.ot \\
J1906+0746 & 0.144 &  28$_{-25}^{+148}$ &  -6.0$_{-9.6}^{+5.6}$  &  0.0$_{-6.6}^{+11.9}$   &  68$_{-28}^{+64}$       &  0.8 &   7$_{-6}^{+11}$     &    46$_{-45}^{+251}$    &  102$_{-102}^{+361}$ & \\
J1907+0631 & 0.323 & 103$_{-99}^{+73}$  &  18.5$_{-17.5}^{+8.0}$ &  5.0$_{-8.7}^{+9.9}$    &  25$_{-34}^{+25}$       &  1.3 &   -                &    -                  &  -               &  no.w\\
J1907+0709g& 0.344 & 150$_{-144}^{+25}$ &  -1.1$_{-1.5}^{+0.9}$  & -4.3$_{-0.2}^{+0.3}$    &  20.0$_{-4.5}^{+6.1}$   &  1.5 &  5.3$_{-4.4}^{+5.1}$  &   65$_{-63}^{+181}$     &  neg              & \\
J1908+0128 & 0.0047&  32              &   37                 &  11.6               & -77                  &  0.7 & 49.4                 &    76                &  21               & l.un \\
J1908+0457 & 0.846 & 104$_{-37}^{+46}$  &   5.2$_{-2.4}^{+0.2}$  & -8.9$_{-0.1}^{+0.1}$    &  86.2$_{-0.9}^{+1.0}$   &  1.4 & 12.3$_{-6.1}^{+0.5}$  &  847$_{-630}^{+74}$     &  neg              & \\
J1908+0500 & 0.291 & 150$_{-146}^{+27}$ &   1.7$_{-1.5}^{+2.8}$  &  2.2$_{-0.8}^{+0.2}$    & -83$_{-6}^{+16}$        &  6.4 &  2.3$_{-2.0}^{+3.2}$  &   10$_{-9}^{+48}$      & 185$_{-47}^{+10}$   & \\
J1909+0616 & 0.755 &  28$_{-23}^{+147}$ &   4.2$_{-3.3}^{+6.1}$  & -5.2$_{-0.7}^{+1.4}$    & -32.4$_{-6.8}^{+4.5}$    &  1.2 &  6.7$_{-5.6}^{+8.0}$  &  227$_{-220}^{+854}$   & 254$_{-119}^{+217}$  & \\
J1909+0749 & 0.237 &  82              &   15.6              &  180.1               &  -5.0                 &  0.7 &   -                 &  -                   &   -               & ip \\
J1909+1205 & 1.229 & 112$_{-107}^{+63}$ &  -3.5$_{-1.0}^{+3.2}$  & 10.8$_{-0.4}^{+0.6}$    &  -9.9$_{-3.1}^{+2.4}$   &  0.9 & 12.5$_{-11.4}^{+0.6}$  & 1278$_{-1267}^{+133}$  & 369$_{-141}^{+175}$   & \\
J1910+0728 & 0.325 & 145$_{-134}^{+26}$ &  -5.0$_{-4.8}^{+3.6}$  & -0.9$_{-0.8}^{+0.6}$    &  21.4$_{-3.4}^{+3.8}$   &  3.4 &  8.3$_{-6.0}^{+6.3}$   &  150$_{-138}^{+312}$   & 254$_{-53}^{+38}$    &  \\
J1912+1036 & 0.409 &  31$_{-28}^{+147}$ &  12.0$_{-11.5}^{+24.5}$ &  11.6$_{-8.2}^{+12.8}$  &  0$_{-52}^{+19}$       &  1.5 & 15$_{-14}^{+24}$       &  566$_{-564}^{+3501}$  & 988$_{-702}^{+1095}$  & \\
J1912+2525 & 0.622 & 167$_{-161}^{+9}$  &  -0.4$_{-1.8}^{+0.3}$  &  2.4$_{-0.2}^{+0.3}$    &  36.6$_{-4.8}^{+10.8}$  &  4.3 & 1.0$_{-0.7}^{+3.5}$    &    4$_{-3}^{+79}$     & 289$_{-23}^{+39}$    & \\
J1913+0657 & 1.257 &  28$_{-22}^{+143}$ &  -0.4$_{-0.7}^{+0.3}$  &   -1.2$_{-0.3}^{+0.2}$  &  47.8$_{-11.1}^{+3.8}$  &  2.8 & 1.8$_{-1.4}^{+2.0}$    &   27$_{-26}^{+94}$    & 141$_{-99}^{+65}$    & \\
J1913+0832 & 0.134 & 85.3$_{-1.0}^{+0.5}$&   9.3$_{-1.2}^{+2.1}$ & -161.2$_{-0.6}^{+1.4}$  & -42.6$_{-2.7}^{+0.4}$   &  1.8 &  -                   &   -                 &  -                 & ip  \\
J1913+0904 & 0.163 & 162$_{-107}^{+16}$ &  -1.3$_{-3.4}^{+1.2}$  &   13.2$_{-0.6}^{+1.1}$  &  93.7$_{-5.6}^{+15.1}$  &  1.3 & 2.9$_{-2.6}^{+6.4}$    &   9.1$_{-9.0}^{+85}$  & 337$_{-20}^{+37}$    & \\
J1913+1011 & 0.035 &  26$_{-24}^{+147}$ & -10.7$_{-18.0}^{+9.8}$ &  -20.0$_{-1.5}^{+8.0}$  & -57.0$_{-3.1}^{+17.8}$  &  1.0 &  13$_{-12}^{+22}$      &   39$_{-38}^{+244}$   & neg                &  \\
J1913+1050 & 0.190 &  72$_{-65}^{+15}$  &  15.6$_{-13.6}^{+12.4}$ &  2.3$_{-5.5}^{+2.6}$   & -52.2$_{-11.3}^{+17.9}$  & 1.2  &   -                  &  -                  &   -               & ip\\
J1913+3732 & 0.851 &  13.5$_{-5.0}^{+0.0}$& -0.6$_{-0.0}^{+0.2}$ &   2.1$_{-0.1}^{+0.1}$   &   9.6$_{-0.1}^{+0.1}$   & 815.1 &  1.5$_{-0.5}^{+0.0}$   &  12$_{-7}^{+0}$      &  61$_{-5}^{+5}$     &   \\
J1914+0219 & 0.457 &  63$_{-30}^{+44}$  &   6.0$_{-2.4}^{+0.9}$  &   4.3$_{-0.1}^{+0.1}$   & -67.0$_{-0.9}^{+0.7}$   &  8.7 & 10.1$_{-3.9}^{+1.1}$   & 309$_{-191}^{+73}$    &  28$_{-10}^{+13}$   &   \\
J1914+1122 & 0.600 &  35$_{-24}^{+129}$ &   2.2$_{-1.4}^{+1.8}$  &  -3.8$_{-0.1}^{+0.2}$   &  67.8$_{-2.6}^{+1.6}$   &  2.0 &  4.7$_{-3.1}^{+3.4}$   &  86$_{-77}^{+173}$    &  63$_{-34}^{+42}$   &  \\

\hline
  \end{tabular}
\end{table*} 
\addtocounter{table}{-1}
\begin{table*}
  \centering     
  \caption{\it -- continued -- }
  \tabcolsep 2pt
  \footnotesize
  \begin{tabular}{llrrrrrrrrl}
    \hline
    \hline
    PSR    & P      & $\alpha_0$        & $\beta_0$         &   $\phi_0$         & $\psi_0$                &$\chi^2_\nu$& $\rho$         & $r_{\rm geo}$       & $r_{\rm delay}$    & Note\\
           & (s)    & ($^\circ$)        & ($^\circ$)         & ($^\circ$)          & ($^\circ$)               &       & ($^\circ$)          &  (km)             & (km)             &        \\
    (1)    &  (2)   &  (3)             &  (4)              &  (5)              &     (6)                   &  (7)  &  (8)               &   (9)             &  (10)            & (11)                 \\
    \hline
J1915+0738 & 1.542 & 162$_{-126}^{+15}$ &  -0.4$_{-1.1}^{+0.4}$  &   2.9$_{-0.2}^{+0.2}$   & -10.0$_{-3.4}^{+5.6}$   & 20.3 & 0.8$_{-0.7}^{+1.8}$    &   6.1$_{-6.0}^{+61.8}$ & 832$_{-56}^{+62}$   & \\
J1915+1410 & 0.297 & 118$_{-108}^{+54}$ &   1.6$_{-1.3}^{+0.5}$  &  11.5$_{-0.7}^{+0.4}$   &  21.2$_{-3.2}^{+5.9}$   &  0.5 & 12.3$_{-10.4}^{+1.8}$  & 296$_{-289}^{+90}$     & 174$_{-50}^{+40}$   & \\    
J1915+1647 & 1.616 & 168$_{-38}^{+10}$  &   0.4$_{-0.3}^{+1.0}$  &  2.78$_{-0.02}^{+0.02}$ & -23.7$_{-0.4}^{+0.4}$   & 17.9 &  1.1$_{-0.9}^{+3.0}$   &  13.3$_{-12.9}^{+168.8}$ & neg              &  \\
J1916+0748 & 0.541 & 162$_{-6}^{+16}$   &  -85$_{-4}^{+80}$     &   39$_{-29}^{+10}$      & -87.8$_{-7.3}^{+1.7}$   &  1.5 &  -                  &  -                   &  -               & \\
J1916+1023 & 1.618 &  43$_{-36}^{+132}$ &   7.1$_{-6.1}^{+4.7}$  & -6.0$_{-3.9}^{+1.0}$    &  71.0$_{-4.3}^{+19.6}$  &  1.2 & 11.3$_{-10.0}^{+5.6}$  & 1374$_{-1354}^{+1702}$ &  neg               & \\
J1917+0834 & 2.129 & 128$_{-59}^{+31}$  &  -0.9$_{-0.3}^{+0.5}$  & -3.2$_{-0.1}^{+0.1}$    & -40.5$_{-0.4}^{+0.6}$   &  5.9 &  4.4$_{-2.4}^{+1.2}$  & 274$_{-218}^{+162}$    & 111$_{-26}^{+34}$    &  \\
J1917+1353 & 0.194 &  19$_{-5}^{+0}$    &   2.2$_{-0.6}^{+0.0}$  &  5.9$_{-0.1}^{+0.1}$    & -31.9$_{-0.1}^{+0.2}$   &102.5 &  3.4$_{-0.9}^{+0.0}$  &  15$_{-7}^{+0}$       & 188$_{-2}^{+2}$      & \\
J1918+1444 & 1.181 & 165$_{-13}^{+2}$   &  -0.3$_{-0.2}^{+0.1}$  &  0.26$_{-0.01}^{+0.01}$ & -68.1$_{-0.1}^{+0.1}$   &211.9 &  1.0$_{-0.1}^{+0.8}$   &   8$_{-2}^{+19}$      & 15$_{-9}^{+9}$       & \\
J1918+1541 & 0.370 &  98$_{-7}^{+34}$   & -27$_{-63}^{+12}$     & -186.1$_{-2.8}^{+4.1}$  &  69$_{-23}^{+20}$       &  1.5 &  -                  &   -                 &  -                 & ip  \\
J1919+1745 & 2.081 & 127$_{-107}^{+42}$ &  -3.1$_{-1.0}^{+2.3}$  &   4.3$_{-0.1}^{+0.2}$   &  38.5$_{-1.5}^{+1.1}$   &  1.8 &  6.1$_{-4.6}^{+1.5}$  & 508$_{-478}^{+284}$    &  neg                &  \\
J1920+1040 & 2.215 & 135$_{-129}^{+39}$ &  -6.4$_{-4.5}^{+5.4}$  & -0.6$_{-0.8}^{+1.7}$    &  72.8$_{-4.6}^{+11.0}$  &  2.9 &  8.0$_{-6.9}^{+4.6}$  & 948$_{-927}^{+1402}$   & neg                 & \\
J1921+0137 & 0.0024& 116              &  28                 & -128                & -32                  &  1.0 &  -                  &   -                &  -                  & ip?   \\
J1921+0812 & 0.210 & 152$_{-149}^{+26}$ &  -2.4$_{-10.8}^{+2.2}$ &  8.7$_{-0.1}^{+3.7}$   &  -5.5$_{-20.2}^{+16.6}$ &  1.3 &  2.8$_{-2.5}^{+10.7}$  &  11$_{-10}^{+243}$   & 358$_{-5}^{+164}$     &  \\
J1921+1419 & 0.618 &  29$_{-21}^{+142}$ &   6.0$_{-4.4}^{+7.4}$  &  6.2$_{-0.4}^{+3.5}$   & -44.6$_{-16.4}^{+2.0}$  &  0.9 &  8.6$_{-6.2}^{+9.2}$  & 303$_{-279}^{+988}$   & 422$_{-65}^{+451}$    &  \\
J1921+1540 & 0.143 & 140$_{-134}^{+35}$ &   4.1$_{-3.5}^{+3.3}$  &  5.0$_{-1.3}^{+1.1}$   &  90.2$_{-8.4}^{+10.9}$  &  1.2 &  8.0$_{-6.9}^{+5.4}$  &  61$_{-60}^{+110}$    & neg                &  \\
J1922+1733 & 0.236 &  12$_{-5}^{+4}$   &   1.2$_{-0.5}^{+0.4}$   &  0.2$_{-0.1}^{+0.1}$   &  28.4$_{-0.3}^{+0.2}$  &261.0 &  2.3$_{-1.0}^{+0.8}$  &   8$_{-5}^{+6}$       & 246$_{-3}^{+4}$     &  \\
J1922+2018 & 1.172 &   7$_{-2}^{+16}$   &  -0.3$_{-0.7}^{+0.1}$  & -3.5$_{-0.1}^{+0.1}$   &  76.9$_{-0.5}^{+0.6}$  & 21.3 &  1.2$_{-0.3}^{+2.7}$  &  11$_{-5}^{+104}$     & 238$_{-34}^{+35}$    &  \\
J1923+1706 & 0.547 &   8$_{-5}^{+55}$   &  -0.2$_{-0.9}^{+0.1}$  &  2.2$_{-0.1}^{+0.1}$   & -48.8$_{-0.9}^{+1.2}$  & 12.3 &  1.1$_{-0.7}^{+6.5}$  &   5$_{-4}^{+208}$     &  neg               &  \\
J1923+4243 & 0.595 & 140$_{-77}^{+23}$  &  -2.3$_{-1.4}^{+1.2}$  &  2.0$_{-0.1}^{+0.1}$   &  74.1$_{-1.7}^{+0.8}$  & 51.8 &  5.2$_{-2.8}^{+2.8}$  & 106$_{-84}^{+144}$    & 123$_{-18}^{+10}$     &  \\
J1926+1434 & 1.324 & 158$_{-47}^{+17}$  &  -1.6$_{-3.2}^{+1.2}$  & 18.3$_{-0.7}^{+0.3}$   &  66.3$_{-1.5}^{+1.6}$  &  2.1 &  7.9$_{-6.1}^{+11.7}$ & 551$_{-521}^{+2835}$   & 643$_{-199}^{+95}$   & wo.ot  \\
J1926+1648 & 0.579 &  52$_{-48}^{+118}$ & -10.7$_{-4.3}^{+9.8}$  &  6.0$_{-3.5}^{+1.4}$   & 109.9$_{-14.8}^{+6.7}$  &  1.1 & 11.6$_{-10.6}^{+4.6}$ & 512$_{-509}^{+488}$   & 830$_{-425}^{+170}$   &  \\
J1926+1857g& 0.278 & 138$_{-135}^{+39}$ &  11$_{-10}^{+13}$      &  3.6$_{-8.7}^{+6.8}$   &  13$_{-36}^{+28}$      &  0.7 &   -                &    -                &  -                & no.w\\
J1926+2016 & 0.299 &  28$_{-23}^{+147}$ &   5.2$_{-4.3}^{+7.8}$  &  4.5$_{-2.1}^{+3.1}$   &  -7.7$_{-15.8}^{+9.9}$  &  1.7 &  6.8$_{-5.6}^{+8.8}$  &  90$_{-88}^{+386}$    & 171$_{-130}^{+194}$  &  \\
J1927+1852 & 0.482 & 119$_{-111}^{+55}$ &  13.3$_{-11.9}^{+3.3}$ &  6.7$_{-6.1}^{+1.6}$   & -26.3$_{-5.9}^{+24.0}$  &  5.1 & 15.8$_{-14.1}^{+3.7}$ & 795$_{-785}^{+422}$   & 59$_{-59}^{+164}$    &  \\
J1927+1856 & 0.298 & 168$_{-45}^{+8}$   & -3.3$_{-10.0}^{+2.3}$  & 15.8$_{-0.4}^{+3.1}$   & -56.0$_{-2.0}^{+8.4}$   &  2.4 &  5.7$_{-3.9}^{+16.2}$ &  65$_{-58}^{+886}$    & 393$_{-32}^{+191}$  &  \\
J1928+1809g& 0.294 & 112              & -15.5               & -8.8                &  10.1                &  1.2 &  -                  &   -                 &  -               & ip  \\
J1930+1722 & 1.609 &  27$_{-21}^{+128}$ &   1.7$_{-1.3}^{+2.3}$  &  5.7$_{-0.4}^{+0.5}$   &  77.0$_{-5.4}^{+3.9}$   &  3.0 &  4.5$_{-3.5}^{+5.3}$  & 220$_{-208}^{+818}$   & 313$_{-300}^{+321}$ &  \\
J1931+1439 & 1.779 &  80$_{-65}^{+84}$  &   5.8$_{-4.3}^{+0.4}$  &  5.6$_{-0.3}^{+0.2}$   &  31.5$_{-1.1}^{+2.3}$   &  1.4 & 10.1$_{-7.4}^{+0.3}$  & 1206$_{-1119}^{+69}$  &  92$_{-92}^{+67}$  &  \\
J1932+1500 & 1.864 &  84$_{-78}^{+88}$  &   1.8$_{-1.6}^{+0.2}$  &  4.2$_{-0.1}^{+0.1}$   &  63.7$_{-2.0}^{+1.5}$   &  1.7 &  5.9$_{-5.3}^{+0.1}$  & 434$_{-429}^{+11}$   & 193$_{-70}^{+76}$   &  \\
J1932+2220 & 0.144 &  23$_{-18}^{+99}$  &  -2.5$_{-4.7}^{+2.0}$  &  6.6$_{-1.1}^{+0.3}$   & -26.1$_{-9.4}^{+6.1}$   &  3.7 &  3.3$_{-2.6}^{+6.0}$  &  11$_{-10}^{+71}$    & 137$_{-32}^{+8}$    &  \\
J1934+2352 & 0.178 & 101$_{-2}^{+4}$    & -12.7$_{-5.2}^{+4.5}$  &  8.0$_{-1.9}^{+1.0}$  &  75.9$_{-13.1}^{+12.8}$  &  0.8 &  -                 &  -                 &  -                &  \\
J1935+2025 & 0.080 &  87$_{-1}^{+1}$    &  -6.2$_{-0.6}^{+0.7}$  &-174.3$_{-0.7}^{+0.3}$ & -40.9$_{-4.2}^{+0.7}$   &  4.0 &  -                  &  -                 &  -                & ip \\
J1937-00   & 0.240 &  28$_{-26}^{+150}$ &  -3.7$_{-16.3}^{+3.3}$ &  7.5$_{-9.0}^{+0.8}$   &  55$_{-51}^{+3}$       &  1.7 &  5.1$_{-4.7}^{+16.3}$ &  42$_{-41}^{+689}$   & 473$_{-449}^{+43}$    &  \\
J1937+2544 & 0.200 &  18$_{-15}^{+63}$  &   1.6$_{-1.3}^{+3.3}$  & 14.4$_{-0.4}^{+1.0}$   &  81.5$_{-8.9}^{+2.5}$  &  1.8 &  5.3$_{-4.4}^{+11.0}$ &  38$_{-37}^{+317}$   & 215$_{-17}^{+42}$    & wo.ot \\
J1938+0650 & 1.121 &  21$_{-16}^{+153}$ &  -0.5$_{-1.2}^{+0.4}$  &  0.7$_{-0.2}^{+0.2}$   &  44.0$_{-8.4}^{+7.4}$  &  2.1 &  1.5$_{-1.1}^{+2.7}$  &  16$_{-15}^{+113}$   & 193$_{-54}^{+44}$    &  \\
J1938+14A  & 1.661 & 147$_{-68}^{+22}$  &   1.0$_{-0.7}^{+0.9}$  &  3.5$_{-0.2}^{+0.3}$   & -73.9$_{-2.1}^{+1.1}$  & 16.3 &  3.0$_{-1.9}^{+2.6}$  &  99$_{-87}^{+243}$   & neg               &  \\
J1938+2213 & 0.166 & 133$_{-32}^{+41}$  &   8.1$_{-6.9}^{+2.7}$  & 12.4$_{-0.9}^{+0.8}$   & -33.2$_{-6.5}^{+3.8}$  &  7.6 & 10.2$_{-8.7}^{+3.7}$  & 114$_{-112}^{+98}$   & 376$_{-33}^{+28}$   &  \\
J1938+2659 & 0.883 & 146.5            &  -0.7               &  -6.9                & -84.4               &  4.2 &  4.6                & 125               & neg               & \\
J1939+10   & 2.308 & 151$_{-97}^{+18}$  &  -1.5$_{-1.9}^{+0.9}$  &  5.2$_{-0.2}^{+0.3}$   &  -7.6$_{-1.7}^{+2.6}$   & 14.7 &  3.5$_{-2.1}^{+3.7}$  & 190$_{-160}^{+602}$  & neg               &  \\
J1939+2449 & 0.645 &  15$_{-3}^{+6}$    &  -1.6$_{-0.6}^{+0.3}$  &  1.5$_{-0.1}^{+0.1}$   &  -4.2$_{-0.1}^{+0.1}$  & 15.1  &  2.7$_{-0.5}^{+1.0}$  &  31$_{-11}^{+29}$    & neg               & wo.ot \\
J1945+1211 & 4.756 & 147$_{-60}^{+21}$  &   5.5$_{-3.3}^{+4.6}$  & 12.9$_{-0.9}^{+0.1}$   &  44.9$_{-0.4}^{+4.4}$   &  9.8  &  9.5$_{-5.9}^{+8.8}$ & 2848$_{-2433}^{+7745}$ & 1684$_{-928}^{+156}$ &  \\
J1946+1805 & 0.440 & 158$_{-47}^{+17}$  & -16$_{-15}^{+12}$      & -21.8$_{-12.0}^{+3.1}$ & -33$_{-25}^{+6}$       & 44.4  &  19$_{-14}^{+17}$    & 1019$_{-963}^{+2793}$  & neg               &   \\
J1946+2535 & 0.515 & 152$_{-147}^{+23}$ &  -4.1$_{-5.7}^{+3.4}$   & -0.3$_{-2.8}^{+0.7}$   &   4.6$_{-19.0}^{+4.8}$ &  1.9  &  5.4$_{-4.5}^{+6.6}$  &  100$_{-98}^{+392}$   & neg               &  \\
J1950+05   & 0.455 &  64$_{-58}^{+111}$ & -18.8$_{-5.0}^{+17.0}$  &  8.8$_{-4.7}^{+3.1}$   &  25$_{-13}^{+10}$      &  1.6  &  24$_{-22}^{+5}$     & 1753$_{-1733}^{+874}$ & 400$_{-400}^{+334}$ &  \\
J1952+3252 & 0.039 &  19$_{-16}^{+114}$ &  -6.7$_{-25.9}^{+5.9}$  & 27.4$_{-16.5}^{+0.7}$  &   7$_{-41}^{+11}$      &  2.1  &  9$_{-8}^{+30}$      &  22$_{-21}^{+384}$    & 259$_{-136}^{+7}$  &  \\
J1957+2831 & 0.307 &  28$_{-26}^{+150}$ &  14$_{-13}^{+28}$      & -13.5$_{-14.8}^{+9.0}$  & -69$_{-36}^{+10}$      &  1.0  &  15$_{-14}^{+28}$    & 465$_{-464}^{+3372}$  & neg              & \\
J1959+3620 & 0.406 & 154$_{-37}^{+15}$  & -16.8$_{-22.2}^{+9.6}$  & -14.3$_{-1.6}^{+2.1}$  & -14.7$_{-1.8}^{+3.1}$  &  1.9  &  26$_{-15}^{+27}$    & 1868$_{-1511}^{+5765}$ & neg              &  \\
J2002+1637 & 0.276 & 112$_{-109}^{+66}$ &  14.7$_{-14.3}^{+8.0}$  &  4.9$_{-17.5}^{+2.0}$  &  -16$_{-7}^{+85}$      &  1.4  &  17$_{-16}^{+8}$     & 532$_{-531}^{+587}$   & 260$_{-260}^{+117}$ &  \\
J2005+3552 & 0.307 &  27$_{-18}^{+71}$  &   8.0$_{-5.4}^{+11.0}$  & -9.1$_{-2.8}^{+0.2}$   & 48.5$_{-0.5}^{+8.2}$   &  1.8  &  11$_{-7}^{+13}$     & 228$_{-202}^{+877}$   & neg               &  \\
J2006+3102 & 0.163 & 147$_{-82}^{+26}$  &  -6.9$_{-7.8}^{+5.2}$   & -3.4$_{-2.3}^{+0.9}$  & -62.3$_{-6.7}^{+4.8}$   &  2.0  &  12$_{-11}^{+6}$     & 156$_{-155}^{+208}$   & 250$_{-38}^{+43}$   &  \\
J2008+2513 & 0.589 &  14$_{-7}^{+32}$  &  -1.0$_{-2.0}^{+0.6}$   & 10.6$_{-0.4}^{+0.2}$   & 38.0$_{-2.5}^{+2.3}$    & 52.4 &  2.2$_{-1.1}^{+4.3}$  &  18$_{-14}^{+147}$    & 907$_{-43}^{+20}$   &  \\
J2009+3326 & 1.438 & 152$_{-147}^{+24}$ &  -2.7$_{-4.3}^{+2.3}$   &  9.9$_{-0.7}^{+0.7}$   & -20.6$_{-5.7}^{+5.0}$  &  1.0  &  6.2$_{-5.2}^{+7.0}$ & 365$_{-356}^{+1295}$   & 1461$_{-282}^{+279}$ & wo.ot \\
J2013+3845 & 0.230 &  52$_{-40}^{+32}$  &  20.1$_{-13.1}^{+3.3}$  & 30.6$_{-5.4}^{+2.5}$   &  90$_{-21}^{+15}$     &  1.8  & 26.7$_{-18.0}^{+2.5}$ & 1086$_{-970}^{+209}$   & 1545$_{-259}^{+119}$ &  \\
J2017+0603 & 0.0028& 139$_{-29}^{+31}$  &  -87$_{-3}^{+69}$      &  17$_{-58}^{+20}$      &  6$_{-34}^{+25}$      &  2.0  &   -                 & -                   &  -               &  \\
J2018+2839 & 0.557 & 159$_{-3}^{+3}$    &  -3.0$_{-0.4}^{+0.3}$  & -3.8$_{-0.1}^{+0.3}$   &  51.9$_{-0.5}^{+2.4}$  & 561.0 &  4.1$_{-0.5}^{+0.5}$  &  61$_{-13}^{+17}$     & neg               &  \\
J2022+5154 & 0.529 & 140$_{-30}^{+10}$  &  -9.1$_{-4.4}^{+1.9}$   & -1.7$_{-0.5}^{+0.1}$   &  20.6$_{-2.4}^{+0.6}$  & 11.9  & 10.8$_{-2.2}^{+5.0}$  & 407$_{-152}^{+461}$  & 226$_{-59}^{+12}$   &  \\
J2027+4557 & 1.099 & 102$_{-20}^{+18}$  &   1.1$_{-0.1}^{+0.1}$   &  8.9$_{-0.1}^{+0.1}$   &  66.7$_{-0.2}^{+0.3}$  & 226.8 & 10.0$_{-1.2}^{+0.2}$  & 722$_{-159}^{+36}$   & 187$_{-10}^{+10}$   &  \\
J2030+2228 & 0.630 & 143$_{-13}^{+14}$  &   0.6$_{-0.2}^{+0.2}$   & -5.0$_{-0.1}^{+0.1}$   &  12.1$_{-0.1}^{+0.1}$  & 14.1  &  4.1$_{-1.5}^{+1.1}$  &  71$_{-41}^{+44}$    &  neg              &  \\
J2032+4127 & 0.143 &  54              &  40.5                &  17.7               &  32.4                &  1.1  &    -               &   -                &  -                & ip, ot?  \\
J2036+2835 & 1.358 &  57$_{-35}^{+89}$  &   1.8$_{-1.0}^{+0.4}$   &  0.9$_{-0.1}^{+0.1}$   &  -43.8$_{-1.6}^{+0.8}$ & 11.9  &  3.8$_{-2.1}^{+0.7}$  & 131$_{-105}^{+53}$    & 221$_{-13}^{+22}$   &  \\

\hline
  \end{tabular}
\end{table*} 
\addtocounter{table}{-1}
\begin{table*}
  \centering     
  \caption{\it -- continued -- }
  \tabcolsep 2pt
  \footnotesize
  \begin{tabular}{llrrrrrrrrl}
    \hline
    \hline
    PSR    & P      & $\alpha_0$        & $\beta_0$         &   $\phi_0$         & $\psi_0$                &$\chi^2_\nu$& $\rho$         & $r_{\rm geo}$       & $r_{\rm delay}$    & Note\\
           & (s)    & ($^\circ$)        & ($^\circ$)         & ($^\circ$)          & ($^\circ$)                &        & ($^\circ$)          &  (km)             & (km)             &        \\
    (1)    &  (2)   &  (3)             &  (4)              &  (5)               &     (6)                   &  (7)  &  (8)               &   (9)             &  (10)            & (11)             \\
    \hline
J2042+0246 & 0.0045& 166$_{-77}^{+11}$  &  -2.6$_{-20.4}^{+2.2}$  &  6.5$_{-5.1}^{+2.3}$   &  39$_{-26}^{+10}$      &  1.5  &    -               &  -                  &  -                &  \\    
J2043+2740 & 0.096 & 121$_{-88}^{+50}$  &   8.3$_{-7.0}^{+3.4}$   &  13.1$_{-2.8}^{+0.7}$  & -14.5$_{-7.9}^{+17.1}$ &  2.3  & 10.9$_{-9.1}^{+3.6}$  &  76$_{-74}^{+59}$     & 207$_{-55}^{+14}$   &  \\
J2044+4614 & 1.392 &  68$_{-30}^{+25}$  &   2.5$_{-1.0}^{+0.4}$   &  14.4$_{-0.1}^{+0.2}$  &  30.1$_{-1.8}^{+1.7}$  &  9.0  & 16.8$_{-5.6}^{+1.1}$  &2585$_{-1434}^{+364}$  & 1662$_{-69}^{+72}$  & wo.ot \\    
J2047+5029 & 0.445 & 143$_{-134}^{+32}$ &  -1.5$_{-1.2}^{+1.2}$   & -1.4$_{-0.4}^{+0.3}$   & -28.6$_{-4.3}^{+3.4}$  &  3.2  &  -                  &   -                 &  -               &  \\
J2051+4434g& 1.303 & 160$_{-27}^{+13}$  &  -9.1$_{-12.5}^{+5.6}$  & -26.5$_{-0.4}^{+0.5}$  &  38.6$_{-0.7}^{+0.7}$  &  3.1  & 20$_{-13}^{+23}$      & 3481$_{-3012}^{+12445}$& neg               &  \\
J2105+07   & 3.746 & 149$_{-63}^{+22}$  &   1.6$_{-1.0}^{+1.5}$   &  -3.3$_{-0.2}^{+0.2}$  & -58.7$_{-2.6}^{+3.6}$  &  7.4  &  3.9$_{-2.7}^{+3.8}$  & 375$_{-340}^{+1105}$   & 313$_{-243}^{+210}$ &  \\
J2113+4644 & 1.014 & 179.5            &   0.053              &  -1.9               & -84.0               &73014  &  0.3                &   0.6                & 461              & s.un \\
J2122+2426 & 0.541 &  47$_{-35}^{+103}$ & -13.3$_{-5.3}^{+9.5}$   &  -1.0$_{-0.4}^{+2.1}$  &  35.3$_{-1.4}^{+6.6}$  &  1.7  & 15$_{-11}^{+7}$       & 847$_{-778}^{+892}$   & neg               &  \\
J2129+1210A& 0.110 & 123$_{-54}^{+47}$  &  -1.8$_{-0.6}^{+1.5}$   &   0.6$_{-0.1}^{+0.1}$  &  45.7$_{-1.4}^{+2.2}$  & 11.8  &  9.4$_{-7.5}^{+1.7}$  &  65$_{-62}^{+26}$     &  72$_{-3}^{+3}$     &  \\
J2129+4119 & 1.687 & 104$_{-27}^{+30}$  &  -3.5$_{-0.2}^{+1.0}$   &  -3.8$_{-0.1}^{+0.1}$  & -35.7$_{-0.4}^{+0.6}$  &  3.6  &  7.7$_{-1.9}^{+0.2}$  & 654$_{-287}^{+26}$    &  27$_{-27}^{+30}$    &   \\
J2138+4911 & 0.696 &  25$_{-19}^{+147}$ &   4.5$_{-3.3}^{+7.5}$   &   2.8$_{-1.3}^{+0.5}$  & -46.2$_{-2.5}^{+7.3}$  &  3.0  &  5.2$_{-3.8}^{+8.1}$  & 123$_{-114}^{+681}$    & 435$_{-194}^{+71}$   &  \\
J2208+4056 & 0.636 & 120$_{-11}^{+46}$  & -12.8$_{-1.5}^{+9.3}$   &   0.8$_{-0.3}^{+0.3}$  &  36.2$_{-1.1}^{+1.1}$  &  1.5  &  -                  & -                   &  -                & ip?  \\
J2340+08   & 0.303 & 161$_{-149}^{+15}$ &   4.5$_{-3.7}^{+13.1}$   &  -0.3$_{-3.1}^{+3.0}$  & -46$_{-12}^{+14}$     &  0.8  &  7.5$_{-5.9}^{+19.4}$  & 112$_{-107}^{+1337}$  & neg               &   \\

\hline

\hline
  \end{tabular}

\end{table*} 

%

The offsets of the PA curves with respect to profile peaks are
demonstrated by the histogram of $\phi_0$ in
Figure~\ref{fig:geo-hist}. PA curves lag the profile peaks for 120 of
the 190 pulsars.  The
majority of pulsars exhibiting the lag can be explained by the
underlying physics of rotation induced abberation, which leads the
profiles to be shifted to an early phases and the PAs lately
\citep[e.g.][]{bcw91,wwh12}. For some pulsars, the offsets are larger
than $60^\circ$, e.g., PSR J1852-0118 in Figure~\ref{fig:PAGeo}. It
means that the emission regions are far away from the meridional
plane.

\begin{figure}
  \centering
  \includegraphics[angle=0,width = 0.42\textwidth] {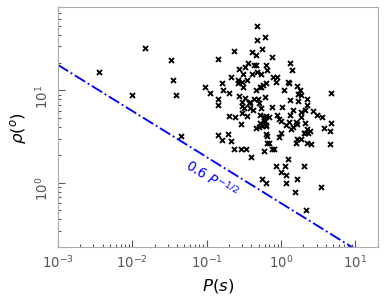}
  \caption{The beam radius $\rho$ in Table~\ref{table:Geo_para}
    against pulsar period. The lower boundary is indicated by a
    power-law $0.6 P^{-1/2}$. }
  \label{fig:rho-p}
\end{figure}

\begin{figure*}
  \centering
  \includegraphics[angle=0,width = 0.33\textwidth] {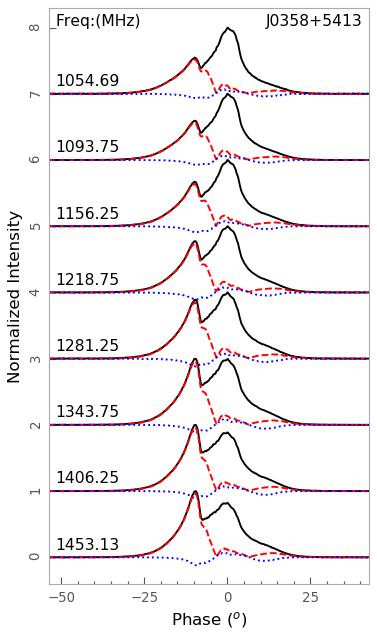}
  \includegraphics[angle=0,width = 0.33\textwidth] {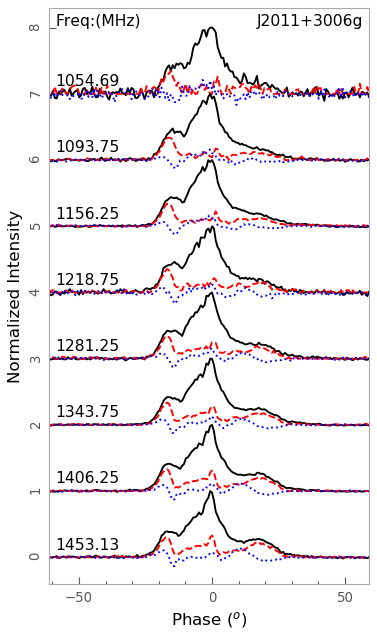}
  \includegraphics[angle=0,width = 0.33\textwidth] {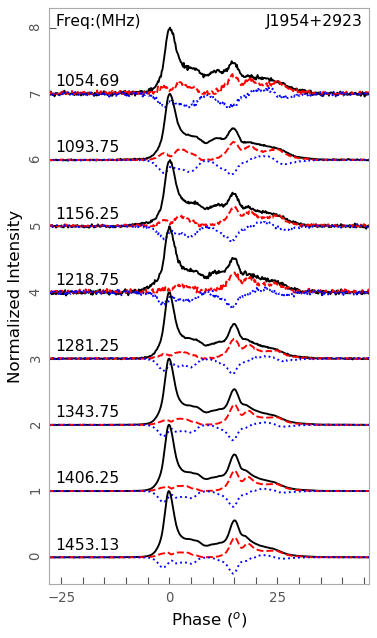}\\
  \includegraphics[angle=0,width = 0.33\textwidth] {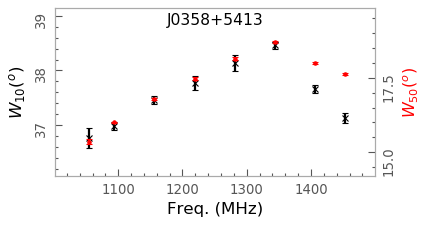}
  \includegraphics[angle=0,width = 0.33\textwidth] {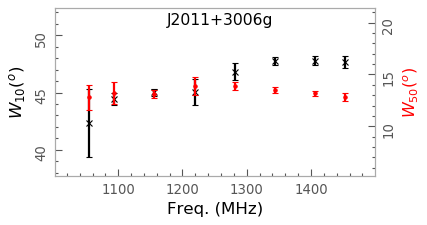}
  \includegraphics[angle=0,width = 0.33\textwidth] {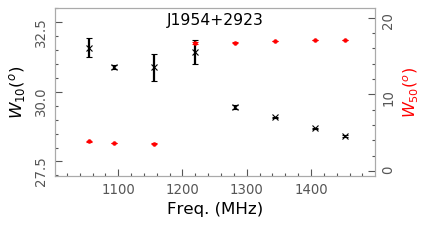}\\
  \caption{The frequency evolution of polarization profiles ({\it
      upper panels}) for PSRs J0358+5413, J2011+3006g and J1954+2923,
    and the variations of pulse width $W_{10}$ and $W_{50}$ ({\it
      lower panels}). The evolution and variation are caused by the
    different spectral indexes for different profile components.  }
  \label{fig:profWD-inc}
\end{figure*}

\subsubsection{Emission heights}

With these geometry parameters and the measured profile widths, beam
radii $\rho$ are estimated for the pulsars, $\rho = 2
\arcsin[\sin^2(W_{10}/4) \sin\alpha \sin( \zeta
  )+\sin^2(\beta/2)]^{1/2}$, as listed in column (8) of
Table~\ref{table:Geo_para}. Its correlation with pulsar period is
shown in Figure~\ref{fig:rho-p}. It is apparent that the beam radii
vary a lot even for pulsars with similar periods, as caused by the
emission originated from different heights of pulsar magnetosphere.
However, there is a clear low boundary for the beam radii versus
period, which can be described by a power-law of $0.6 P^{-1/2}$.

Pulsar radio emission is generated at a finite height above the polar
cap. By assuming dipole magnetosphere and a circular emission beam,
\citet{kg97} found that
\begin{equation}
  r_{\rm geo} \simeq R_\star P (\rho/1.24^\circ)^2 s^{-2},
  \label{eq:geo-height}
\end{equation}
here $R_\star$ is the neutron star radius. The parameter $s$ describes
the relative location of the footpoint of a magnetic field line. It is
within the range of 0 to 1, with $s=0$ for the magnetic axis and $s=1$
for the polar cap boundary. Simply assuming that the profile edges at
10\% the peak intensity originate from the beam boundary, $r_{\rm
  geo}$ can be estimated, as listed in column (9) of
Table~\ref{table:Geo_para}. 29 ones of them are not determined due to
the lack of measurable $W_{10}$.

As caused by the rotation induced aberration, a pulse profile is
shifted towards an early rotation phase, while the PA curve is shifted
laterlly \citep[e.g.][]{bcw91, wwh12}. The phase offset between a
pulse profile and its PA curve reads, $\Delta \phi=\phi_0-\phi_{\rm
  prof}$. Here, $\phi_{\rm prof}$ represents the central phase of the
profile. We take the midpoint of the longitudes defining $W_{10}$ as
the profile center. The emission height, $r_{\rm delay}$, is then
estimated as being,
\begin{equation}
  r_{\rm delay} = \Delta \phi R_{LC}/4,
  \label{eq:dphi-height}
\end{equation}
here, $R_{LC}=c P/2 \pi$ is the light-cylinder radius. $r_{\rm delay}$
for these pulsars are listed in column (10) of
Table~\ref{table:Geo_para}. The abberation effect cannot be figured out
for 84 pulsars, with 55 ones having $\Delta \phi<0$ and 29 ones
lacking midpoint of pulse profiles.

108 pulsars have measurements for both $r_{\rm geo}$ and $r_{\rm
  delay}$. They are consistent within the uncertainties for 75
ones. $r_{\rm geo}$ is smaller than $r_{\rm delay}$ for 26 pulsars. It
is reasonable, since $r_{\rm geo}$ represents only the lower limit of
the emission height, which is calculated with s=1 in
Equation~\ref{eq:geo-height}. If the profile edges originate from the
inner magnetic field lines, i.e., $s<1$, the so estimated $r_{\rm
  geo}$ will be larger. While $r_{\rm geo}$ is larger than $r_{\rm
  delay}$ for 6 pulsars, J1841+0912, J1901+0621, J1908+01,J1914+0219,
J2027+4557 and J2129+4119. It results from the small phase offset
$\Delta \phi$ in Equation~\ref{eq:dphi-height}. These small phase
offsets together with those negative ones might be caused by the
asymmetry of emission regions around the meridional plane. Moreover,
if propagation effects dominate, the phase offsets will also be
negative \citep{wlh10}.

   \begin{figure*}
     \centering
     \includegraphics[angle=0,width = 0.32\textwidth] {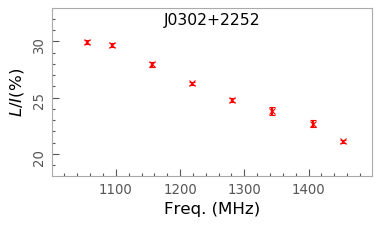}
     \includegraphics[angle=0,width = 0.32\textwidth] {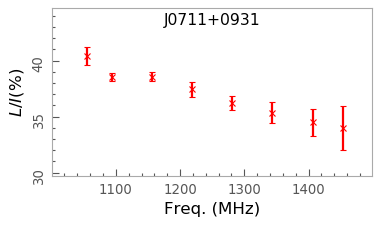}
     \includegraphics[angle=0,width = 0.32\textwidth] {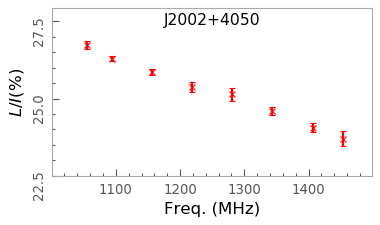}\\
     \includegraphics[angle=0,width = 0.32\textwidth] {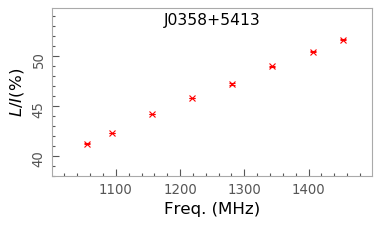}
     \includegraphics[angle=0,width = 0.32\textwidth] {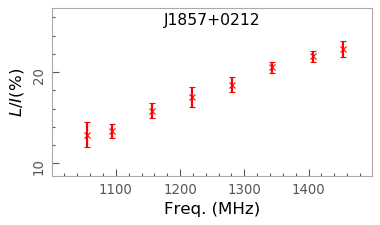}  
     \includegraphics[angle=0,width = 0.32\textwidth] {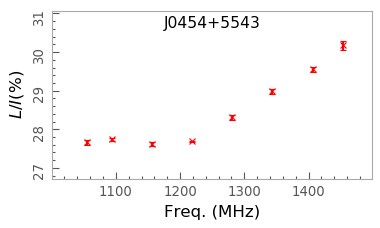}\\
     \includegraphics[angle=0,width = 0.32\textwidth] {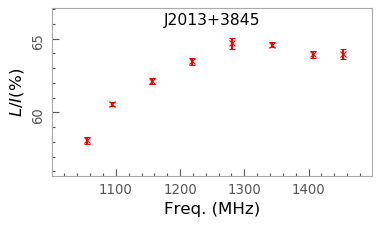}  
     \includegraphics[angle=0,width = 0.32\textwidth] {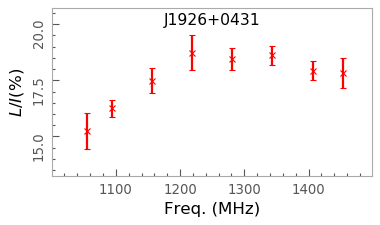}
     \includegraphics[angle=0,width = 0.32\textwidth] {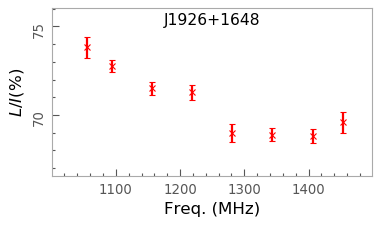} 
       \caption{Frequency dependencies of the fractional linear polarization, $L/I$. }
     \label{fig:L_freq}
   \end{figure*}  
\begin{figure*}
  \centering
  \includegraphics[angle=0,width = 0.32\textwidth] {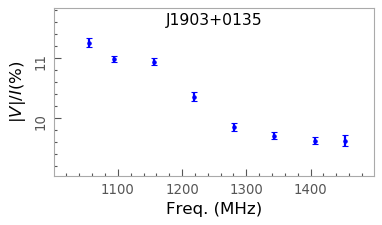}
  \includegraphics[angle=0,width = 0.32\textwidth] {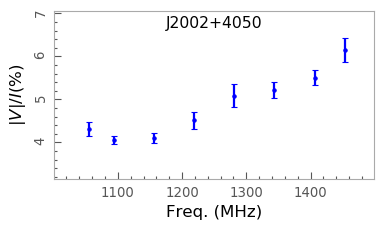}  
  \includegraphics[angle=0,width = 0.32\textwidth] {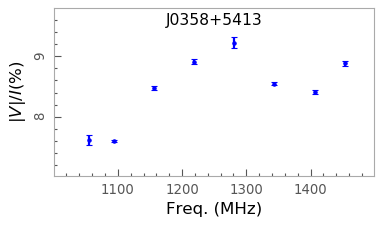}
  \caption{Frequency dependencies of the fractional circular
    polarization, $|V|/I$. }
  \label{fig:V_freq}
\end{figure*}

\subsection{Frequency evolution of pulse profiles}

Pulse profiles can evolve with frequency by several reasons. It can be
caused by the geometric change of emission region. The
radius-to-frequency mapping \citep{rs75, cor78} tells that pulsar
emission of a lower frequency is generated from a higher altitude in
pulsar magnetosphere with a beam of broader opening angle. Therefore
pulse profiles should evolve with frequency, and the pulse width
decreases with frequency as described by a power-law relation,
$W=a\nu^k+W_0$ \citep{th91}. Simulations show that the index $k$ in
this radius-frequency-mapping depends on the density and energy
distributions of relativistic particles within pulsar magnetosphere
\citep{whw13}. In many pulsars, however, one can see that profiles may
evolve together with the emerge of new components
\citep[e.g.][]{nsk+15}.

For some bright pulsars observed by the FAST, we examine the frequency
evolution of polarization profiles. Profiles of most pulsars do not
show clear evolution in such a small frequency range of 1.0 -- 1.5~GHz
of our FAST observations. However, we do see the evolution of a few
pulsars, such as PSRs J0358+5413, J2011+3006g and J1954+2923 in
Figure~\ref{fig:profWD-inc}. The evolution is caused by the
enhancement of leading component of PSR J0358+5413 and trailing
component of J2011+3006g towards higher frequency, while the tail of
PSR J1954+2923 is diminishing at higher frequency. These evolution can
lead to the variation of pulse width $W_{10}$ and/or $W_{50}$ with
frequency. Therefore, the different spectral indexes of different
profile components should be investigated to understand the frequency
evolution of pulse profiles.

We have also examined the frequency evolution of linear polarization
and circular polarization for the pulse profile we observed. Again,
for most pulsars within the observation frequency range of
1.0-1.5~GHz, significant evolution is not found for almost all
pulsars. However, we do see the evolution of linear polarization
percentage of several pulsars (see Figure~\ref{fig:L_freq}). For
example, The significant decreases of linear polarization percentage
have been observed for PSRs J0302+2252, J0711+0931 and J2002+4050,
while the percentage of linear polarization increase with frequency
for PSRs J0358+5413 and J1857+0212. More complicated cases are also
found for PSRs J0454+5542, J2013+3845, J1926+0431 and J1926+1648.
By jointly considering the curvature emission mechanism and
propagation effects within pulsar magnetosphere, the fractional linear
polarization should decreases with the increase of observing frequency
\citep{wwh15}. Because emission at a higher frequency is usually
generated from a lower altitude of pulsar magnetosphere, where the
pulsar rotation induced the separation of the two orthogonal modes,
the O-mode and X-mode, is less significant and depolarization is
serious.  When the two orthogonal modes are almost separated above
certain heights of pulsar magnetosphere, the fractional linear
polarization remains almost unchanged with frequency.  When the
emission is generated from the lower magnetosphere, O-mode refraction
becomes serious, which reduces the depolarization for the emission at
higher frequencies \citep{wwh15}. Hence, the fractional linear
polarization will increase with frequency.

We noticed that circular polarization of a small number of pulsar
profiles evolve with frequency, as shown in Figure~\ref{fig:V_freq}.
The circular polarization can increase, or decrease or vary
irregularly with frequency.  The physical processes for such
complicated behaviors are unclear.

\section{Conclusions}

We report in this paper the database for polarization profiles of 682
pulsars observed by FAST with unprecedent sensitivity. Polarization
profiles for about 460 pulsars are obtained for the first time. The
profiles exhibit diverse features. About 194 pulsars have S-shaped PA
curves and 171 pulsars exhibit orthogonal modes. The profiles can be
highly linearly polarized for its leading and/or trailing or the whole
part, and a few of them have high circular polarization.
28 pulsars are probably ``conal double'', whose sense of circular
polarization interplays with their PA variations.
Some 27 pulsars show interpulse emission, and 21 pulsars have
extremely wide profiles.
Profiles of about 22 pulsars are seriously affected by interstellar
scattering.

At 1250~MHz, the typical width of pulsar profiles is about
$17.4^\circ$ for $W_{10}$. In general, profile widths increase with
the decrease of pulsar period. Young pulsars with a large Pdot and
Edot tend to have a higher fraction of linear polarization.

For 190 pulsars with either S-shaped PA curves or the ones superposed
with orthogonal modes, geometry parameters are investigated through
RVM fitting. The inclination angle, $\alpha_0$, has a value in a wide
range from about $0^\circ$ to $180^\circ$. Pulsars are most likely to
be detected within an impact angle of $|\beta_0|<5^\circ$ by sight
lines. The emission heights estimated from these phase lags are
generally larger than or consistent with the geometric ones.

\section*{Data availability}

All the polarized pulse profiles presented in this paper are available
on the webpage http://zmtt.bao.ac.cn/psr-fast/.

\begin{acknowledgements}
FAST is a Chinese national mega-science facility built and operated by
the National Astronomical Observatories, Chinese Academy of
Sciences.
J. L. Han is supported by the National Natural Science
Foundation of China (No. 11988101 and 11833009).
P. F. Wang is supported by the National Key R\&D Program of China
(No. 2021YFA1600401 and 2021YFA1600400), National Natural Science
Foundation of China (No. 11873058, 12133004) and the National SKA
program of China (No. 2020SKA0120200).
J. Xu is supported by the National Natural Science Foundation of
China (No. U2031115).

\end{acknowledgements}

\bibliographystyle{raa}
\bibliography{psr_geometry}



\appendix

\section{Pulsar profiles}

We have obtained the polarization profiles of 682 pulsars. Instead of
the example pulsars in the main text, here we present all
profiles. They are classified into 9 categories as in Section
3. Figure~\ref{fig:intprof_S} is for 96 pulsars with S-shaped PA
curves, Figure~\ref{fig:intprof_OM} for 136 pulsars with orthogonal
modes, Figure~\ref{fig:intprof_HiLP} for 45 pulsars with highly linear
polarization, Figure~\ref{fig:intprof_HiCP} for 9 pulsars with highly
circular polarization, Figure~\ref{fig:intprof_CD} for 28 conal double
pulsars, Figure~\ref{fig:intprof_IP} for 27 pulsars with inter pulses,
Figure~\ref{fig:intprof_WP} for 21 pulsars with very wide profiles,
Figure~\ref{fig:intprof_Scat} for 22 pulsars with obvious scattering
tails, and Figure~\ref{fig:intprof} for the other 298 pulsars.

\begin{figure*}
  \centering  
  
  \caption{Polarization profiles with S-shaped PA curves for 96
    pulsars here and 35 pulsars in other figures, including 7 pulsars
    with high linear polarization in Figure~\ref{fig:intprof_HiLP}, 19
    conal double pulsars in Figure~\ref{fig:intprof_CD}, 8 pulsars
    with an interpulse in Figure~\ref{fig:intprof_IP}, and one pulsar
    with a wide pulse in Figure~\ref{fig:intprof_WP}. For each pulsar,
    the total intensity, linear and circular polarization are
    represented by solid, dashed and doted lines in the bottom
    sub-panel. The left-hand circular polarization is defined to be
    positive. The bin size and $3\sigma$ are marked inside the
    sub-panel, here $\sigma$ is the standard deviation of off-pulse
    bins. In the top panel, dots with error-bar are measurements of
    polarization position angles for linear polarization intensity
    exceeding 3$\sigma$ line. -- {\it to be continued}-- }
  \label{fig:intprof_S}
\end{figure*}

\begin{figure*}
  \centering
      
  \caption{The same as Figure~\ref{fig:intprof_S} but for polarization
    profiles with orthogonal modes for 136 pulsars in this figure and
    for 33 pulsars in other figures (8 pulsars with highly linearly
    polarized components in Figure~\ref{fig:intprof_HiLP}, 5 conal
    double pulsars in Figure~\ref{fig:intprof_CD}, 6 pulsars with
    interpulse emission in Figure~\ref{fig:intprof_IP}, 12 pulsars
    with wide profiles in Figure~\ref{fig:intprof_WP}, and 2 pulsars
    affected by interstellar scattering in
    Figure~\ref{fig:intprof_Scat}).-- {\it to be continued}--
  }
  \label{fig:intprof_OM}
\end{figure*}

\begin{figure*}
  \centering

 \caption{The same as Figure~\ref{fig:intprof_S} but for highly
   linearly polarized profiles of 45 pulsars in this figure and for 28
   pulsars in other figures (13 pulsars with S-shaped position angles
   in Figure~\ref{fig:intprof_S}, 2 pulsars with orthogonal modes in
   Figure~\ref{fig:intprof_OM}, 9 pulsars with interpulse emission in
   Figure~\ref{fig:intprof_IP}, 3 pulsars with wide profiles in
   Figure~\ref{fig:intprof_WP}, and one conal double pulsar in
   Figure~\ref{fig:intprof_CD}, as listed in
   Table~\ref{table:psrs}). The highly linearly polarized emission can
   appear for (a) the whole profile, (b) the leading component, (c)
   the trailing component, and (d) both the leading and trailing
   components. -- {\it to be continued}-- }
  \label{fig:intprof_HiLP}
\end{figure*}

\begin{figure*}  
  \centering

  \caption{The same as Figure~\ref{fig:intprof_S} but for highly
    circularly polarized profiles ($|V|/I>30\%$) of 9 pulsars here and
    7 pulsars in other figures (PSRs J1853+0056 and J2006+3102 with
    S-shaped PA curves in Figure~\ref{fig:intprof_S}, PSRs J1855+0139g
    and J1907+0345 with highly linearly polarized emission in
    Figure~\ref{fig:intprof_HiLP}, PSRs J1851+0118 and J1852+0056g
    with interpulse emission in Figure~\ref{fig:intprof_IP}, and PSR
    J1855+0527 affected by interstellar scattering in
    Figure~\ref{fig:intprof_Scat}).  }
  \label{fig:intprof_HiCP}
\end{figure*}

\begin{figure*}
  \centering
  
  \caption{The same as Figure~\ref{fig:intprof_S} but for polarized
    pulse profiles of 28 conal double pulsars. -- {\it to be continued}.
  }
  \label{fig:intprof_CD}
\end{figure*}

\begin{figure*}
  \centering
  
  \caption{Polarized pulse profiles of 27 pulsars with interpulse
     emission. See the keys in Figure~\ref{fig:intprof_S}.  -- {\it to be continued}. }
  \label{fig:intprof_IP}
\end{figure*}

\begin{figure*}
  \centering

  \caption{The same as Figure~\ref{fig:intprof_S} but for very wide
    profiles for 21 pulsars. PSRs J1851+0418, J1903+0925, J1916+0748
    and J1932+1059 are normal pulsars, while others are millisecond
    pulsars. -- {\it to be continued}.}
  \label{fig:intprof_WP}
\end{figure*}

\begin{figure*}
  \centering

  \caption{The same as Figure~\ref{fig:intprof_S} but for 22 pulsars with obvious scattering tails. -- {\it to be continued}.
  }
  \label{fig:intprof_Scat}
\end{figure*}
 
\begin{figure*}
  \centering

  \caption{Polarized pulse profiles of the other 298 pulsars. See the
    keys in Figure~\ref{fig:intprof_S}. -- {\it to be continued} --  }
  \label{fig:intprof}
\end{figure*}

\section{The RVM solutions}

The RVM solutions of the fittings are presented for all the 190
pulsars. They are arranged in pulsar name, whose corresponding
geometry parameters are listed in Table~\ref{table:Geo_para}.


\begin{figure*}
  \centering

  \caption{RVM fitting to the PA curves of 190 pulsars. See the
    keys in Figure~\ref{fig:PAGeo2}. -- {\it to be continued} --  }
  \label{fig:PAGeo}
\end{figure*}


\label{lastpage}

\end{document}